\documentclass[usenatbib]{mn2e}
\usepackage[dvips]{graphicx}
\usepackage{amssymb,amsmath}

\title[Baryons and the orbital structure of haloes]{Influence of baryons on the orbital structure of dark matter haloes}
\author[S.E. Bryan et al.]
{
{\parbox{\textwidth}{S. E. Bryan,$^{1,2}$\thanks{E-mail:sb526@le.ac.uk}
S. Mao,$^{1,3}$
S. T. Kay,$^{1}$
J. Schaye,$^{4}$
C. Dalla Vecchia $^{4,5}$ and 
C. M. Booth$^{4}$}}\vspace{0.4cm}\\
\parbox{\textwidth}{$^{1}$Jodrell Bank Centre for Astrophysics, School of
  Physics and Astronomy, The University of Manchester, Manchester M13 9PL,
  U.K.\\
$^{2}$Department of Physics \& Astronomy, University of Leicester, Leicester  LE1 7RH. \\
$^{3}$National Astronomical Observatories of China, Chinese Academy of
  Sciences, 20A Datun Road, Beijing, China 100012. \\
$^{4}$Leiden Observatory, Leiden University, Postbus 9513, 2300 RA Leiden, The
Netherlands.\\
$^{5}$Max Planck Institute for Extraterrestrial Physics, Giessenbachstra\ss{}e 1, 85748 Garching, Germany.\\
}
}

\begin{document}

\def\aj{AJ}					
\def\araa{ARA\&A}				
\def\apj{ApJ}					
\def\apjl{ApJL}					
\def\apjs{ApJS}					
\def\apss{Astrophysics and Space Science}
\def\capsp{Comments on Astrophysics and Space Physics}
\def\aap{A\&A}					
\def\aapr{A\&A~Rev.}				
\def\aaps{A\&AS}				
\def\azh{AZh}					
\def\baas{BAAS}					
\def\jrasc{JRASC}				
\def\memras{MmRAS}				
\def\mnras{MNRAS}					
\def\pasp{PASP}					
\def\pasj{PASJ}					
\def\qjras{QJRAS}				
\def\skytel{S\&T}				
\def\solphys{Sol.~Phys.}			
\def\sovast{Soviet~Ast.}			
\def\ssr{Space~Sci.~Rev.}			
\def\zap{ZAp}					
\def\na{New Astronomy}				
\def\iaucirc{IAU~Circ.}				
\def\aplett{Astrophys.~Lett.}			
\def\apspr{Astrophys.~Space~Phys.~Res.}		
\def\bain{Bull.~Astron.~Inst.~Netherlands}	
\def\memsai{Mem.~Soc.~Astron.~Italiana}		

\def\ao{Appl.~Opt.}				

\def\pra{Phys.~Rev.~A}				
\def\prb{Phys.~Rev.~B}				
\def\prc{Phys.~Rev.~C}				
\def\prd{Phys.~Rev.~D}				
\def\pre{Phys.~Rev.~E}				
\def\prl{Phys.~Rev.~Lett.}			
\def\nat{Nature}				
\def\fcp{Fund.~Cosmic~Phys.}			
\def\gca{Geochim.~Cosmochim.~Acta}		
\def\grl{Geophys.~Res.~Lett.}			
\def\jcp{J.~Chem.~Phys.}			
\def\jgr{J.~Geophys.~Res.}			
\def\jqsrt{J.~Quant.~Spec.~Radiat.~Transf.}	
\def\nphysa{Nucl.~Phys.~A}			
\def\physrep{Phys.~Rep.}			
\def\physscr{Phys.~Scr}				
\def\planss{Planet.~Space~Sci.}			
\def\procspie{Proc.~SPIE}			
\def\rpp{Rep.~Prog.~Phys.}			
\let\astap=\aap
\let\apjlett=\apjl
\let\apjsupp=\apjs
\let\applopt=\ao
\let\prep=\physrep


\date{Accepted ...... Received ...... ; in original form......   }

\pagerange{\pageref{firstpage}--\pageref{lastpage}} \pubyear{2002}
\maketitle
\label{firstpage}
\begin{abstract}
We explore the dynamical signatures imprinted by baryons on dark matter haloes
during the formation process using the OverWhelmingly Large Simulations (OWLS), a set
of state-of-the-art high-resolution cosmological hydrodynamical simulations.
We present a detailed study of the effects of the implemented feedback
prescriptions on the orbits of dark matter particles, stellar particles and
subhaloes, analysing runs with no feedback, with stellar feedback and with
feedback from supermassive black holes.  We focus on the central regions
(0.25$r_{200}$) of haloes with virial masses
$\sim 6 \times 10^{13} ( \sim7 \times 10^{11})$\,$h^{-1}$M$_\odot$ at
$z = 0$$(2)$.  We also investigate how the orbital
content (relative fractions of the different orbital types) of these haloes depends on several key parameters such as their mass,
redshift and dynamical state.  The results of spectral analyses of the orbital
content of these simulations are compared, and the change in fraction of box,
tube and irregular orbits is quantified.  Box orbits are found to dominate
the orbital structure of dark matter haloes in cosmological simulations.
There is a strong anticorrelation between the fraction of box orbits and the central baryon
fraction.  While radiative cooling acts to reduce the fraction of box orbits,
strong feedback implementations result in a similar orbital distribution to
that of the dark matter only case.  The orbital content described by the
stellar particles is found to be remarkably similar to that drawn from the
orbits of dark matter particles, suggesting that either they have forgotten
their dynamical history, or that subhaloes
bringing in stars are not biased significantly with respect to the main
distribution.  The orbital content of the subhaloes is in broad agreement with
that seen in the outer regions of the particle distributions.

\end{abstract}

\begin{keywords}
galaxies: haloes - galaxies: clusters: general - galaxies: evolution - galaxies: kinematics and dynamics - methods: numerical - cosmology: theory
\end{keywords}

\section{Introduction}
\label{orbitsintro}

Dark matter structure formation is well understood within the standard cosmological model.  Haloes are thought to form hierarchically, through the merging and accretion of smaller systems.  As such, there should be observational signatures of these merging processes in the resulting remnants, providing dynamical information about their formation histories.  We investigate the orbital content of dark matter haloes in order to explore what signatures may result.  

Dark matter haloes formed in a $\Lambda$cold dark matter ($\Lambda$CDM) cosmology appear to share a nearly
universal internal morphology; they have density profiles that are well
described by the Navarro, Frenk \& White (hereafter NFW; \citeyear{bib:Navarro96},
  \citeyear{bib:Navarro97}) profile and pseudo-phase space densities with a
constant ($\alpha\sim 1.9$) power-law slope (\citealt{bib:Taylor01,
  bib:Dehnen05, bib:Barnes06, bib:Ludlow11}).  There is a universal relation between the
radial density profile slope and the velocity anisotropy within the inner
region of dark matter haloes
\citep{bib:Hansen06, bib:Navarro10}  and the velocity distribution function is found to have
a universal shape (\citealt{bib:Hansenetal06}).  Dark matter haloes are
thought to have spin distributions that are reasonably well characterised by a log-normal
distribution \citep{bib:Bullock01, bib:Bailin05, bib:Bett07, bib:Maccio08} and are thought to be triaxial (\citealt{bib:Frenk88, bib:Dubinski91, bib:Warren92, bib:Cole96, bib:Jing02, bib:Bailin05, bib:Allgood06, bib:Maccio06, bib:Bett07, bib:Jeeson-Daniel11}). Here we explore the underlying orbital distribution of dark matter haloes evolved within the standard cosmological model.

Of course, a complete understanding of structure formation requires careful
consideration of the baryonic component, particularly in the inner region of
haloes.  The baryonic physics processes involved are largely uncertain and
have become the focus of galaxy formation studies.  It is, however, well
established that the condensation of baryons to the centres of dark matter
haloes tends to result in the halo becoming more spherical or axisymmetric,
and that this is a direct consequence of the impact of baryons on the orbital
content of the halo (see, for example, \citealt{bib:Gerhard85, bib:Dubinski94,
  bib:Barnes96, bib:Merritt99, bib:Jesseit05, bib:Debattista08,
  bib:Valluri10}).  In this paper we consider a series of simulations with
varying implementations of baryonic physics to systematically explore the
effect of baryons on the orbital content of the haloes.  We do so by performing a spectral analysis on the snapshots of the simulation, assuming that the potential remains static during an orbit.

Orbits are divided into groups, or families, according to the phase space they cover.  Families of regular orbits have similar morphologies because they conserve similar integrals of motion.  These isolating integrals restrict the region of phase space available to an orbit.  Each lowers, by one, the dimensionality of the region available to the orbit.  As such, an orbit is shaped by its isolating integrals.  Axisymmetric potentials have two classical integrals of motion: energy $E$ and the $z$-component of the angular momentum $L_z$ and a third non-classical isolating integral (\citealt{bib:Lindblad33}, \citealt{bib:Contopoulos60}, \citealt{bib:Binney82}).  However, it is well-known that many elliptical galaxies are not axisymmetric (\citealt{bib:Franx91}).   In more relevant triaxial systems, there is a second non-classical integral that is likely to play a role in shaping the structure of the system (\citealt{bib:Schwarzschild79}).  For this reason numerical simulations provide an important tool in understanding the orbital content of such systems.  

Within simple generic triaxial models, regular orbits can be divided into two main families: box and tube orbits.  Tube orbits are further divided according to their orientation into major- and minor-axis tubes (\citealt{bib:Schwarzschild79}; \citealt{bib:Statler87}).  Box orbits are free to pass close to the centre of the potential and their orbit-averaged angular momentum is zero.  While box orbits show no sense of rotation, tube orbits tend to rotate around the centre of the system, avoiding the centre.  Box orbits are fundamentally important, as they are thought to be responsible for conveying information from the central regions of a halo to the outer parts of the system and are required to support the triaxial halo.  

Several authors have investigated the orbital content of analytic potentials and remnants of simulated disc mergers. 
An initial attempt to classify the orbital content of simulated merger remnants was conducted by \cite{bib:Barnes92} who simulated a small sample of merging encounters between equal mass disc galaxies.  Orbits were classified according to changes in the sign of the angular momentum vector.  The shapes and kinematic properties of the remnants were found to be related to the initial spin vectors and other encounter parameters.  By including gas dynamics in these merging galaxies, \cite{bib:Barnes96} showed the dramatic effect gas can have on the structure of the resulting remnant.  Torques experienced during the merger act to remove angular momentum from the gas, causing it to flow inwards to form a central mass concentration.  They found the depth of the potential well to be highly correlated with the stellar kinematics and that gas acts to destabilise box orbits (as discussed in \citealt{bib:Dubinski94}).  This causes minor-axis tubes to become dominant and results in a more oblate remnant.

\cite{bib:Jesseit05}  studied a statistical sample of disc galaxy mergers, using the automated spectral classification of \cite{bib:Carpintero98} to quantify the orbital content of the resulting remnants.  They found that the most abundant orbital classes were box and minor-axis tube orbits.  While the inner regions of the simulated remnants were dominated by box orbits, tube orbits became more important at intermediate radii.  They also found that the ratio of these two classes of orbits played a role in determining the basic properties of the remnant.  Minor-axis-tube-dominated haloes were found to be discy, while those dominated by box orbits were boxy.  Major-axis tubes were found to be dominant in prolate remnants.  Again, it was noted that gas affects the fraction of box orbits, causing an increase in the population of minor-axis tubes.

\cite{bib:Debattista08} studied the impact of growing a central disc on the
orbital content of a halo.  They find that while the central concentration
does result in rounder, more radially anisotropic haloes, the halo's shape is
essentially returned to its original state if the disc is artificially `evaporated'.  This indicates that the character of the orbits is not generally changed by the central mass concentration, the box orbits are not destroyed but simply become rounder in line with the potential.  This is also considered in \cite{bib:Valluri10} who explore the orbital evolution induced by baryonic condensation in triaxial haloes.  They find that the evolution depends on the radial distribution of the baryonic component, and that a massive compact central mass will result in the scattering of a large fraction of both box and long-axis tube orbits even at fairly large pericentric distances.

A comprehensive study of the orbital structure of 1:1 merger remnants can be
found in \cite{bib:Hoffman10}.  Mergers between equal mass discs at varying
initial gas fractions (ranging from 0 to 40 per cent) were simulated, taking
into account both star formation and feedback.  They showed that, by varying
the fraction of gas in a merger, a wide range of kinematic structures can be
produced.  The remnants formed in these simulations are typically
prolate-triaxial.  The central regions are dominated by box orbits, while tube
orbits dominate further out.  The inclusion of gas acts to decrease the
fraction of stellar particles on box orbits in the central region, replacing
them with minor-axis tubes.  The remnants were found to become progressively
more oblate as the gas fraction is increased.  Outside of 1.5 $R_e$ (where
$R_e$ is the effective radius) the remnants are found to be largely unaffected by the addition of gas.

This work aims to extend the previous work on the effect of baryons on orbital structure by comparing several models for the feedback implementation within realistic cosmological simulations. 

 The outline of the paper is as follows.  In Section \ref{orbitsims} we briefly review the simulations used for this study, describing the different baryonic physics implemented and the halo sample extracted for analysis.  In Section \ref{orbitsmethod} we describe the method used to define the orbital content of the haloes.  Our main results are presented in Section \ref{orbitsresults} and we finish with a summary of our conclusions in Section \ref{orbitsconclusions}.  Resolution issues/convergence tests and the effect of halo definition are discussed in Appendix A.   
  
\section{Numerical Simulations}
\label{orbitsims}

\begin{table*}
\caption[]{A list of the OWLS runs used in this analysis.  We list the identifier of the simulation (as in \citealt{bib:Schaye10}) and comment on the subgrid physics implemented in each case.  Further information can be found in Section \ref{orbitsims}.} 
\centering 
\begin{tabular}{ l l} 
\hline
Name & Description\\
\hline 
DMONLY &  Dark matter only run\\
NOSN\_NOZCOOL & No feedback, cooling assumes primordial abundances\\
REF & Weak stellar feedback with metal cooling\\
WDENS & Strong stellar feedback with metal cooling\\
AGN &  Weak stellar feedback and AGN feedback with metal cooling\\
\hline 
\end{tabular}
\label{table:simtypes} 
\end{table*}

The haloes used for this analysis were extracted from the OverWhelmingly Large
Simulations (OWLS).  OWLS is a set of high-resolution cosmological simulations
run with varying implementations of the subgrid physics.  For detailed
information about these simulations we refer the reader to
\cite{bib:Schaye10}.  Here we discuss only briefly the pertinent details of
the subset of simulations used for our analysis.  The cosmological parameters
were taken from the third year {\it Wilkinson Microwave Anisotropy
  Probe (WMAP3)} results \citep{bib:wmap3}, with: $\Omega_m$ = 0.238,
$\Omega_\Lambda$ = 0.762, $\Omega_b$ = 0.0418,\,$h$ = 0.73, \mbox{$n$ = 0.95}
and $\sigma_8$= 0.74.  The primordial baryonic mass fraction of hydrogen
(helium) is assumed to be 0.752 (0.248).  For all runs, cosmological initial
conditions were constructed using a linear power spectrum based on a transfer
function generated with {\sc{cmbfast}} \citep{bib:Seljak96}.  The initial
positions and velocities were computed from a glass-like state
\citep{bib:White96} using the \cite{bib:Zeldovich70} approximation.

The simulations were run using a modified version of {\small GADGET-3}
\citep{bib:Gadget2, bib:Springel08} to follow the evolution of $512^3$ dark matter particles
and $512^3$ gas particles in cubes of comoving lengths 25 and
100\,$h^{-1}$Mpc.  In the dark matter only run the particle mass is $7.65
\times 10^{6}$ and $4.93 \times 10^{8}$\,$h^{-1} \mbox{M}_{\odot}$ in the 25
and 100\,$h^{-1}$ Mpc boxes respectively.  The baryon runs follow the baryonic
component with smooth particle hydrodynamics (SPH), where the number of neighbours $N_{\rm ngb}$ for the SPH interpolation was set to 48.  The mass of the particles in the baryon runs is divided between the gas and dark matter particles according to the universal baryon fraction, $f_b^{univ} = \Omega_b/\Omega_m = 0.176$, such that the dark matter (gas) mass in the 100\,$h^{-1}$Mpc run is $4.06 (0.87) \times 10^{8}$\,$h^{-1} \mbox{M}_{\odot} $ and $6.34 (1.35) \times 10^{6}$\,$h^{-1} \mbox{M}_{\odot} $ for the 25\,$h^{-1}$Mpc box.  Baryonic particle masses are allowed to change during the simulation due to mass transfer from star to gas particles.  The comoving gravitational force softening was set to 1/25  of the initial mean interparticle spacing but is limited to a maximum physical scale of $2\, (0.5)$\,$h^{-1}$kpc for the $100\, (25)$\,$h^{-1}$Mpc boxes.  For a given box size the same initial conditions are used in each run.  This allows us to follow the same haloes with different implementations of the subgrid physics and provides a unique opportunity to systematically study the effects of the subgrid physics on haloes evolved within a cosmological framework.  

\subsection{Baryon physics}

In this work we have analysed a subset of five of the OWLS runs to explore the
effect of baryons on the orbits of dark matter haloes by studying simulations
with varying levels of feedback.  We have considered a dark matter only run; a
run which includes baryons and primordial element line cooling but no
feedback;  a weak stellar feedback run; a strong stellar feedback run and a
run which includes feedback from supermassive black holes in addition to weak
stellar feedback.  The
simulations used and their implemented subgrid physics are summarized in Table
\ref{table:simtypes}.  The baryon runs consider gas cooling and star formation
as well as feedback from stars and active galactic nuclei (AGN).  The implementation of each of these
processes is discussed briefly below.

\subsubsection{Cooling}

The cooling rates are computed element-by-element in the presence of the
cosmic microwave background and the \cite{bib:Haardt01} model for the evolving
ultraviolet/X-ray background radiation from quasars and galaxies.  Contributions from
hydrogen, helium, carbon, nitrogen, oxygen, neon, magnesium, silicon, sulphur,
calcium and iron are considered.  The contributions are interpolated as a
function of density, temperature and redshift from precomputed {\sc cloudy} tables
\citep{bib:Wiersma09}, assuming the gas to be optically thin and in
photo-ionisation equilibrium.  In the NOSN\_NOZCOOL run, cooling rates are calculated using primordial element abundances.  Cooling by both Bremsstrahlung emission and Compton cooling via interactions between the gas and cosmic microwave background is also taken into account.  Reionisation is modelled by `switching on' the \cite{bib:Haardt01} background at z = 9.  Collisional equilibrium is assumed before reionisation, and photo-ionisation after z = 9.
\subsubsection{Star formation and evolution}
Star formation is modelled by converting gas particles into collisionless stellar particles (representing a simple stellar population) according to the prescription of \cite{bib:Schaye08}.  A star formation density threshold of $n_H > 0.1$ cm$^{-3}$ is adopted; above this density an effective equation of state (P $\propto \rho^{\gamma_{\mbox{\scriptsize{eff}}}}$) is imposed, where $\gamma_{\mbox{\scriptsize eff}}$ is set to 4/3.  This acts to suppresses spurious fragmentation since neither the Jeans mass, nor the ratio of Jeans length to SPH smoothing length is density dependent.  Within the simulations, stars form at a rate dependent on their pressure.  This pressure-dependent rate is shown to reproduce the Kennicutt-Schmidt law (\citealt{bib:Kennicutt98}) $\dot{\Sigma}_* = A \left( {\Sigma_g}/{1 \mbox{M}_\odot \mbox{pc}^2} \right)^n $, where $\dot{\Sigma}_*$ is the rate of star formation per unit area per unit time and $\Sigma_g$ is the gas surface density, in \cite{bib:Schaye08}.  The simulations use a Chabrier initial mass function (\citealt{bib:Chabrier03}) with a star formation rate normalisation A of 1.515 $\times 10^{-4} \mbox{M}_\odot$ yr$^{-1}$ kpc$^{-2}$ and slope $n = 1.4$.  Stellar particles are assigned the metallicity of their parent gas particle, and their subsequent evolution is a function of this metallicity.

\begin{figure*}
\begin{center}
\begin{tabular}{c}
\includegraphics[width=18cm,height=15cm,angle=0,keepaspectratio]{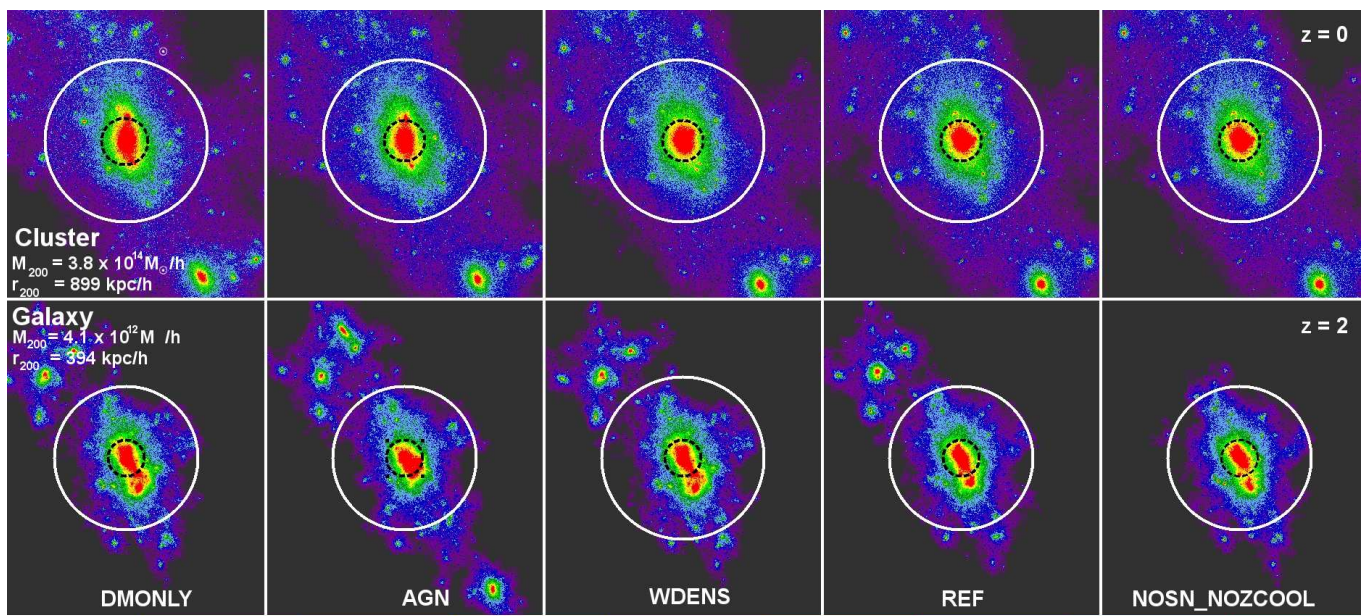}\\
\end{tabular} 
\end{center} 
\caption[Example of an OWLS halo.]{\label{imaging} An example of a halo
  extracted from each of the simulation runs considered (top: from the
  $100\,h^{-1}$Mpc runs at $z = 0$; bottom: from the $25\,h^{-1}$Mpc runs at
  $z = 2$); an increase in brightness corresponds to an increase in surface
  density.  In this figure the halo is extracted from (left to right) the
  dark matter only run (DMONLY); the stellar and AGN feedback (AGN) run; the
  strong stellar feedback (WDENS) run,  the weak stellar feedback (REF) run and
  the no feedback (NOSN\_NOZCOOL) run.  Top: a $z = 0$ cluster with $M_{200}$
  = 3.8$\times 10^{14}$ $h^{-1}$ M$_\odot$ and $r_{200}$ = 899 $h^{-1}$ kpc.
  Bottom: a $z = 2$ galaxy with $M_{200}$ = 4.1$\times 10^{12}$ $h^{-1}$
  M$_\odot$ and  $r_{200}$ = 394 $h^{-1}$ kpc.  In each panel $r_{200}$ is
  shown as a white circle, while the central region 0.25$r_{200}$ is marked
  by a dashed black circle.  It is evident that the baryons have a significant effect on the relaxed cluster.}
\end{figure*}

\subsubsection{Feedback}

As discussed by \cite{bib:Wiersma09}, the simulations follow the timed release
of mass and energy from massive stars [Type II supernovae (SNe) and stellar winds] and intermediate mass stars (Type 1a SNe and asymptotic giant branch stars. 
 
In the simulations we consider, energy is injected kinetically (stars `kick' nearby gas particles) using the prescription of \cite{bib:DallaVecchia08}. %
Kinetic energy is injected locally, and the winds are not hydrodynamically decoupled.  The efficiency of the feedback is characterised by the mass loading parameter, $\eta$, and the velocity added to the nearby gas particles, $v_w$.  The probability for a neighbouring particle $i$ to receive a `kick' of velocity $v_w$ from a new stellar particle $j$ is given by $\eta m_j / \sum_{i=1}^{N_{\rm ngb}} m_i$.  Typically each stellar particle `kicks' $\eta$ times its own mass and adds a randomly directed velocity, $v_w$, to each `kicked' gas particle.  

The REF simulations correspond to a `weak' feedback run, where $\eta = 2$ and $v_w = 600$\,km\,s$^{-1}$.  The WDENS run provides a more efficient form of feedback, where the mass loading depends on the local gas density in the following way: $v_w= 600$ km s$^{-1} \left( n_H/0.1 \mbox{cm}^{-3} \right) ^{1/6}$ and $\eta = 2 \left(v_w/600 \mbox{ km s}^{-1} \right)^{-2}$.  While the same amount of the SNe energy is injected into the surrounding gas particles, the distribution between mass loading and wind velocity results in a higher feedback efficiency.  Winds in the WDENS run are able to remove gas from higher mass haloes more efficiently than the REF model. 

The final feedback run that we consider includes feedback from AGN; this is, by far, the most efficient feedback model that we
consider.  The AGN run is implemented using the method of \cite{bib:Booth09}
which itself is based on that of \cite{bib:Springel05}.  %
Black holes grow both through merging and gas accretion (where the minimum of
the Eddington and Bondi-Hoyle-Lyttleton rate is assumed).  For star forming
gas the Bondi-Hoyle rate is multiplied by $(n_H/10^{-1}$ cm$^{-3})^2$ to
compensate for the lack of a cold, interstellar gas phase and the finite
resolution (\citealt{bib:Booth09}).  
The black hole is assumed to grow as $\dot{m}_{BH} = (1 -
\epsilon_r)\dot{m}_{accr}$ where $\epsilon_r$, the assumed radiative
efficiency, is 0.1.  Fifteen per cent of the radiated energy is assumed to be
coupled to the surrounding medium.  The simulation with AGN reproduces the $z
= 0$ observed relations between black holes and the mass and velocity
dispersion of their host galaxies (\citealt{bib:Booth09}) as well as the
observed optical and X-ray properties of the groups in which they reside
(\citealt{bib:McCarthy10}) and the steep drop-off in the cosmic star formation
rate below $z = 2$ (\citealt{bib:vandeVoort11}).

\subsection{Halo sample}

Haloes within the simulation are first identified using the friends-of-friends
(FOF) technique \citep{bib:Davis85} employing a linking length of $b = 0.2$
times the mean interparticle spacing.  The {\sc subfind} algorithm
(\citealt{bib:Springel01}; \citealt{bib:Dolag09}) is then used to separate the
{FOF} group into self bound structures.  The main halo itself is considered as
the main {\sc subfind} structure and substructures associated with the main halo are
recorded as subhaloes.  The final halo definition considered uses a slightly
modified version of the spherical overdensity (SO) algorithm described in
\cite{bib:Lacey94}.  A sphere is grown around the minimum potential position of a halo until a specified mean internal density is reached (for $r_{200}$ the overdensity is 200 times the critical density).  An SO halo consists of all particles within this sphere.  

\begin{figure*}
\begin{center}
\begin{tabular}{cc}
\includegraphics[width=7.cm,height=7.cm,angle=-90,keepaspectratio]{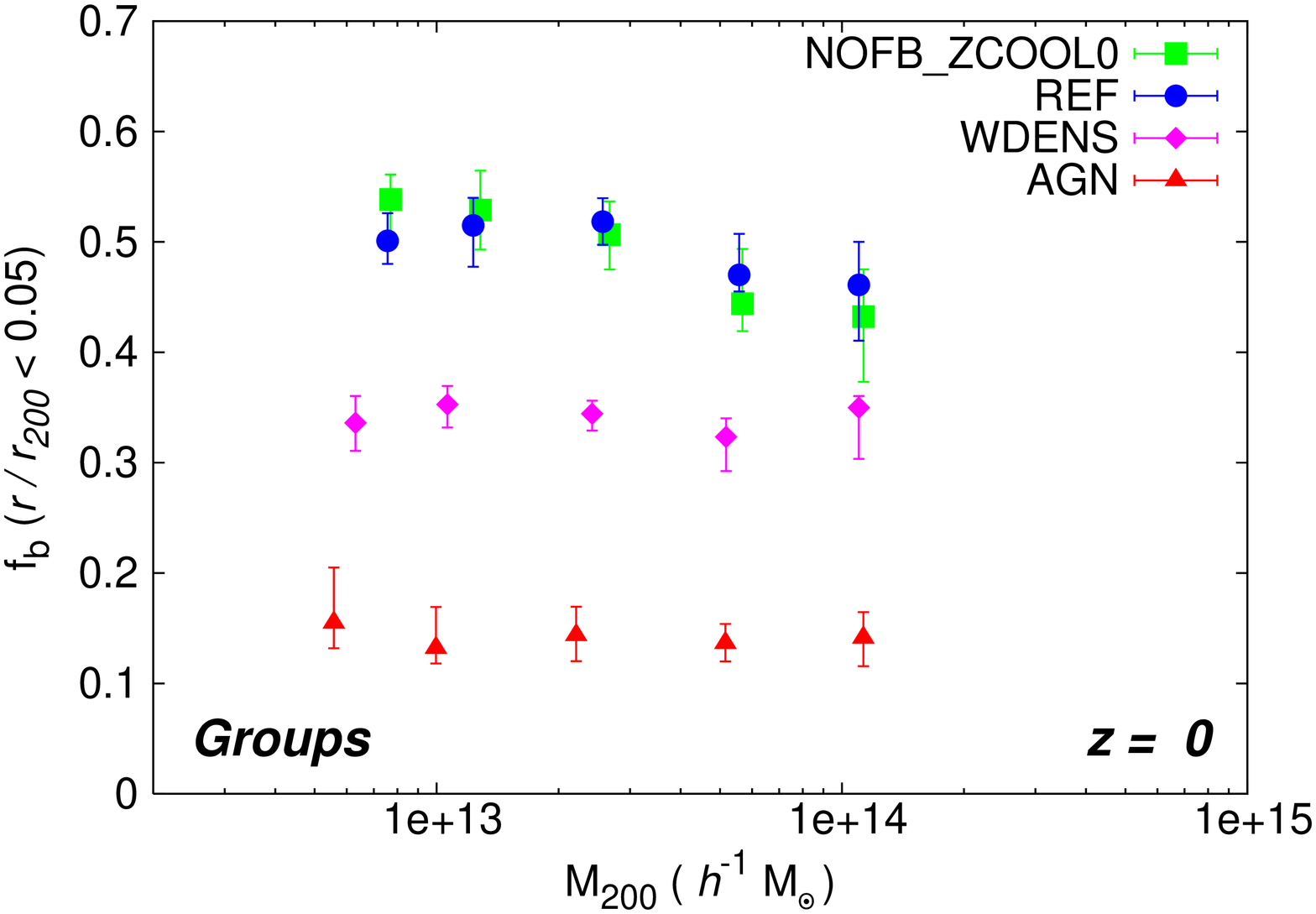} &
\includegraphics[width=7.cm,height=7.cm,angle=-90,keepaspectratio]{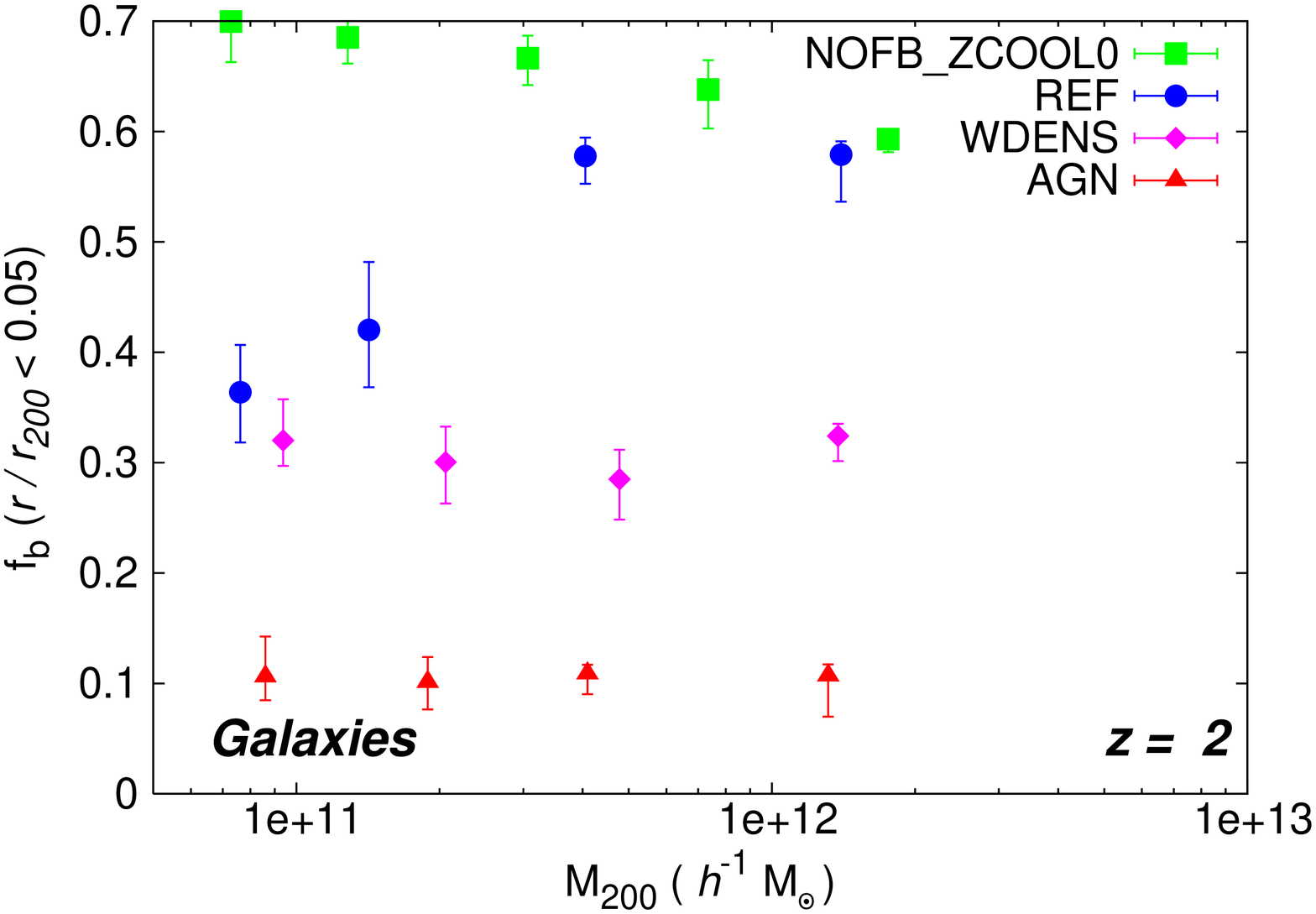}  
\end{tabular} 
\end{center} 
\caption[  The baryon fraction in each of the simulation
  runs.]{\label{baryonfraction}  The baryon fraction within 5 per cent of
  $r_{200}$ versus $M_{200}$ of the halo in each of the simulation runs.
  Error bars represent the quartile scatter.  The left-hand panel corresponds to $z =
  0$, and the right-hand panel to $z = 2$.  The strength of the different feedback models
  is clearly visible.
 
} 
\end{figure*}

An example of a single halo extracted from each of the five $100\, (25) \,h^{-1}$Mpc
simulations at $z = 0 \, (2)$ is shown in the top (bottom) panel of Fig.
\ref{imaging}, where an increase in brightness corresponds to an increase in
surface mass density.  The top row shows the most massive FOF halo identified
at $z = 0$, a cluster-sized object with $M_{200}$ (mass within $r_{200})$ of
3.8$\times 10^{14}$\,$h^{-1}$ M$_\odot$ and $r_{200}$ of 899\,$h^{-1}$\,kpc.
The bottom row shows the most massive FOF halo at $z = 2$, a galaxy-sized
object with $M_{200}$ of 4.1$\times 10^{12}$\,$h^{-1}$\,M$_\odot$ and
$r_{200}$ of 394\,$h^{-1}$\,kpc.  The panels are ordered so that the
efficiency of galaxy formation (as measured by the central baryon fractions,
see Fig. \ref{baryonfraction} and \citealt{bib:Duffy10}) of the simulations increases from left to
right' showing the dark matter only run (DMONLY), the stellar and AGN
feedback (AGN) run, the strong stellar feedback (WDENS) run, the weak stellar
feedback (REF) run and the no feedback (NOSN\_NOZCOOL) run, respectively. In
each panel $r_{200}$ is shown as a solid white circle, while the central
region 0.25$r_{200}$ is marked by the dashed black circle.  It is clear from
the top panel that the baryons act to make the central regions of the relaxed
cluster more spherical.  It is interesting to note the remarkable similarity
between the images of the cluster in the DMONLY and the AGN run (two top left-most images).  The galaxy at $z = 2$ is less relaxed and the effects of the baryons are not obvious.  

We have selected the 50 most massive FOF haloes from each of the five
different physics runs (discussed above) for this analysis.  Haloes are
selected from the 100 (25) \,$h^{-1}$Mpc box at $z = 0$ (2).  At $z = 0$
the mean dark matter halo mass of this sample is $6 \times 10^{13} \, h^{-1}$ M$_\odot$,
while at $z = 2$  the mean dark matter halo mass is $7 \times
10^{11} \, h^{-1}$ M$_\odot$.  These values approximately correspond to group
and galaxy-scale haloes, respectively, so for the sake of clarity we will
refer to them as our `group' and `galaxy' samples.  For each halo we use particles associated with
the main {\sc subfind} group to describe the smooth potential of the system in
order to compute the orbital content of the haloes.  These groups contain
between $2 \times 10^4$ and $6 \times 10^5$ dark matter particles at $z = 0$
and between $2 \times 10^4$ and $5 \times 10^5$ dark matter particles at $z = 2$. This corresponds to dark matter
masses of between  $1 \times 10^{13}$ and $3 \times 10^{14} \, h^{-1}$ M$_\odot$
at $z = 0$ and between $2 \times 10^{11}$ and $4 \times 10^{12}$\,$h^{-1}$ M$_\odot$ at $z = 2$.  When discussing the radial dependence of the orbital content, we scale our results by the SO definition of $r_{200}$.  

For each halo we consider its mass, dynamical state, spin, concentration,
velocity anisotropy parameter, halo shape and baryon fraction.  Halo
properties are computed using the particles belonging to the main {\sc subfind} group.  We explore effect of these properties on the orbital content of our sample of haloes.  A brief description of how these quantities are computed is given below.  
 
The dynamical state of the halo (whether it is considered to be relaxed or
not) is measured as the displacement of the centre of mass from the minimum
potential position as a fraction of $r_{200}$.  If this fraction is less than
7 per cent we consider the halo to be relaxed (\citealt{bib:Neto07}).  Of the 50 most massive haloes at $z = 0$ (2), 27 (20) are found to be relaxed.

The spin parameter of each halo is defined as in \cite{bib:Bullock01} as:
\begin{equation*}
\lambda' = \frac{J}{\sqrt{2} M \, V \, R},
\end{equation*}
where $J$ is the angular momentum within a sphere of radius $R$ containing mass $M$.  The halo circular velocity $V$ is defined at a radius $R$ as $V^2=GM/R$.  This spin parameter reduces to the standard spin parameter (\citealt{bib:Peebles69}) when measured at the virial radius of a truncated singular isothermal halo. 

We use the halo concentrations (defined as $r_{200}/r_s$, where $r_s$ is the characteristic scalelength of the NFW profile) obtained by \cite{bib:Duffy10}.

 The velocity anisotropy parameter $\beta$ measures the proportion of radial to tangential orbits and is given by
\begin{equation}
  \beta = 1 - 0.5 \frac{\sigma^2_t}{\sigma_r^2},
\end{equation}
where $\sigma_t$ is the tangential velocity dispersion, and $\sigma_r$ the
radial velocity dispersion.  A value of $\beta = 0$ corresponds to isotropic
orbits while a value of $\beta = 1$ corresponds to purely radial orbits. 

In order to characterize the halo shape, we use the definition of inertia tensor given in \cite{bib:Bailin05} as
\begin{equation*}
I_{ij} = \sum_k{\frac{r_{k,i} r_{k,j}}{r_k^2}}.
\end{equation*}
The inertia tensor is diagonalised and the eigenvalues and eigenvectors are
computed.  The values of $a, b, c$ are defined to be the square roots of the
eigenvalues (where $a \geq b \geq c$).  The shape parameters are defined as
follows: $s=c/a$ is used as a measure of halo sphericity (where $s = 1$ for a
spherical halo)  and $T = (a^2-b^2)/(a^2 - c^2)$ as a measure of the
triaxiality of the halo [where $T = 1\,(0)$ for an prolate (oblate) halo]. Computation of the inertia tensor in a spherical region biases the shapes towards higher sphericity; this is corrected for (as in \citealt{bib:Bailin05}) by adopting the empirically motivated modified axis ratios $\left(c/a\right)_{\mbox{\tiny{true}}} \equiv (c/a)_{\mbox{\tiny{measured}}}^{\sqrt{3}}$ and  $\left(b/a\right)_{\mbox{\tiny{true}}} \equiv (b/a)_{\mbox{\tiny{measured}}}^{\sqrt{3}}$.

Finally, we consider the central baryon fraction of the halo (the baryon to
total mass fraction within 0.05 $r_{200}$, see also \citealt{bib:Duffy10}).
The central baryon fraction versus $M_{200}$ of the OWLS haloes are shown in
Fig. \ref{baryonfraction} where the haloes are divided into 5 mass bins, equally spaced in log$(M_{200})$, error bars represent the quartile scatter.
This figure illustrates how the central baryonic mass concentration is
affected by the strength of the different feedback models.  The left-hand figure
corresponds to $z = 0$ haloes, while the right-hand figure corresponds to haloes at
$z = 2$.  As expected, the runs with weak or no feedback have a much higher
central baryon concentration than the stronger feedback runs.  The AGN run
clearly has a significantly lower central baryonic concentration than any of
the other baryon runs considered here.  Haloes from the AGN run therefore
appear most similar to the haloes in the DMONLY (for example, see the two
left-most panels in Fig. \ref{imaging}).  Over the range of halo mass probed,
the central baryon fraction does not appear to vary significantly as a
function of the halo mass at $z = 0$.  At $z = 2$, neither of the strong
feedback runs are mass dependent, but in the no feedback run low-mass haloes
have higher central concentrations of baryons than their high-mass counterparts, while the opposite is true for the weak feedback run. 

\section{Orbital Content Computation}

\label{orbitsmethod}
We aim to identify the orbital content of cosmological haloes, to explore how the baryonic process affects the orbital content of dark matter haloes and to link the orbital content of haloes with their observable and intrinsic properties.  To this end the orbits of dark matter particles, stellar particles and subhaloes are integrated (using a Runge-Kutta-Fehlberg routine) within the smooth gravitational potentials of the OWLS haloes.  These orbits are then classified using the spectral classification routine of \cite{bib:Carpintero98} into box, tube and irregular orbits allowing for a quantitative comparison between different simulation runs.  Below, we summarise the key steps of this procedure.  

\subsection{Calculating the potential}
\label{potential}
The first step in determining the orbital content of a halo is to estimate the
gravitational potential of the halo.  There are a number of techniques that
can be used to do this.  As galaxies are regarded as collisionless systems, an
estimate of the smooth mean gravitational field of the system which minimises
the effects of discrete particle representations on the halo potential is
particularly useful.  One such approach is the self consistent field (SCF)
method.  This method is used to obtain an estimate of the mean gravitational
field by expanding the density and potential into a set of basis functions and
using the forces derived from this expansion to integrate the equations of
motion of the particles.  If the first few terms of the basis are sufficient
to provide a good representation of the system, then higher order terms may be
neglected, minimizing the effects of discreteness.  In this work (as in
\citealt{bib:Jesseit05, bib:Hoffman10, bib:Lowing11}) we have used the SCF basis
functions to reconstruct the potential of the haloes. Density and potential are given as 
\begin{align}
\rho(r,\theta,\phi) &= \displaystyle\sum_{n,l,m} \mbox{A}_{nlm}\, \rho_{nl} \, Y_{lm}(\theta,\phi), \\
\Phi(r,\theta,\phi) &= \displaystyle\sum_{n,l,m} \mbox{B}_{nlm} \, \Phi_{nl} \, Y_{lm}(\theta,\phi),
\end{align}
where $n$ denotes the radial expansion terms and $l$ and $m$ the angular
terms.  There are two commonly used basis functions: those suggested by
\cite{bib:CluttonBrock73} and  by \cite{bib:HernquistOstriker92}.   The basis
set used here is constructed from the latter (the code was generously provided by the authors) so that the lowest order terms
represent the Hernquist profile \citep{bib:Hernquist90}.  We find that the choice of basis set does not affect the reconstruction of the halo potentials significantly for the radial region we explore here (see Appendix A for further discussion).  The Hernquist model density-potential pair is given by  
\begin{align}
\rho\left(r\right) &= \frac{M}{2\pi}\frac{a}{r}\frac{1}{\left(r+a\right)^3},\\
\phi\left(r\right) &= - \frac{G M}{r+a},   
\end{align}
where $M$ is the total mass and $a$ is a scale-length that is related to the half mass radius $r_{1/2}$ as follows:
\begin{equation}
a=\left(\sqrt{2}-1\right) r_{1/2}.
\label{scalelength}
\end{equation} 
Twelve radial terms and six angular terms are used as this has been found to be sufficient to reproduce the potential to within a few percent of the $N$-body potential (discussed in Appendix A).  

It is important to choose a reasonable scale length in the Hernquist profile for the potential reconstruction.  To optimize the potential
reconstruction, particles are divided into two components: a diffuse component
consisting of dark matter and hot gas ($T > 10^5$ K) and a compact component
consisting of stars, cold gas and black holes.  The scale length is determined
separately for each component based on its half-mass radius using equation (\ref{scalelength}), and the
corresponding potentials for these components are computed.  These potentials are then
summed to give the resulting potential of the system as a whole.

\subsection{Computing the orbits}
\label{orbits}

We consider the orbits of dark matter particles, stellar particles and
subhaloes.  In determining the orbital content described by the particle
distribution, a subsample of 500 particles is selected from each halo, and the
orbits of these particles are followed within the underlying potential of the
halo (as estimated using the SCF basis functions).   One hundred particles are
chosen, at random, from five radial bins equally spaced in log$(r)$, with the
outermost bin edges defined to be at $0.048, 0.072, 0.109, 0.166$ and $0.251
\times r_{200}$.  In this way we focus on the central region of the haloes
where baryons are expected to dominate (see discussion on convergence testing
in Appendix A).  We assigned particles to a bin according to their `initial'
position.  We have explored binning by energy and found similar results.  Particles in the innermost region are integrated for 100 Gyr; this time interval is then increased with radius.  Orbits are also computed for all subhaloes with masses greater than $10^{10}$\,$h^{-1}$M$_\odot$.  Subhaloes are integrated for 1000 Gyr.  The motion of each particle/subhalo is integrated assuming that the potential remains static.  A static potential is adequate for the purposes of this paper as we are interested only in characterizing the orbital content of a halo at a given point in time; we do not consider the evolution of this quantity.    
Also, the figure rotation of these haloes is assumed to be slow (\citealt{bib:Bailin04}; \citealt{bib:Bryan07}) and would probably have a negligible effect on the quantities calculated here.  A full investigation of the figure rotation of these haloes is deferred to future work.  

\subsection{Classifying the orbit}
\label{classifications}

\begin{figure*}
\begin{center}
\begin{tabular}{ccc}
\includegraphics[width=7cm,height=7cm,angle=-90,keepaspectratio]{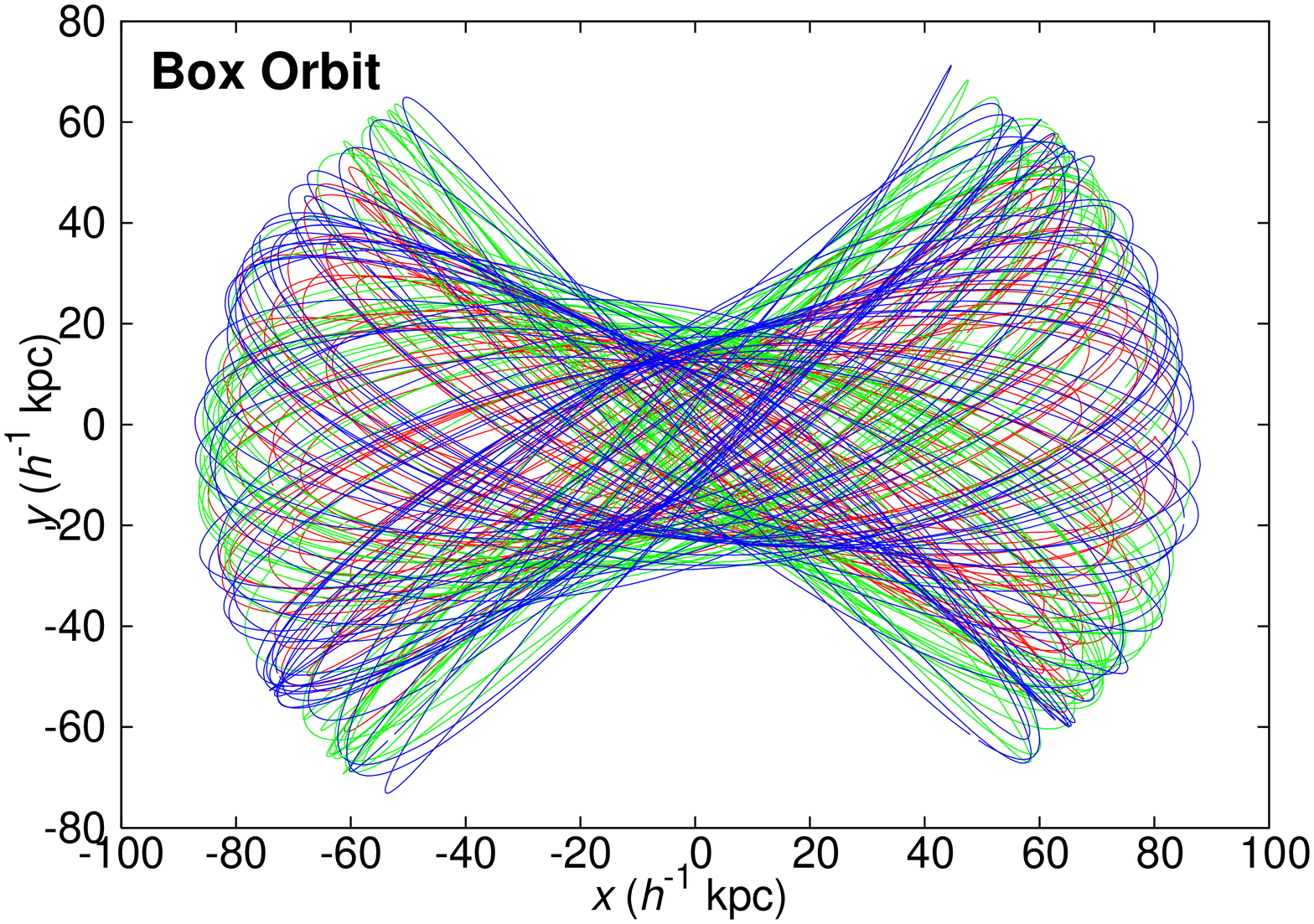} &
\includegraphics[width=7cm,height=7cm,angle=-90,keepaspectratio]{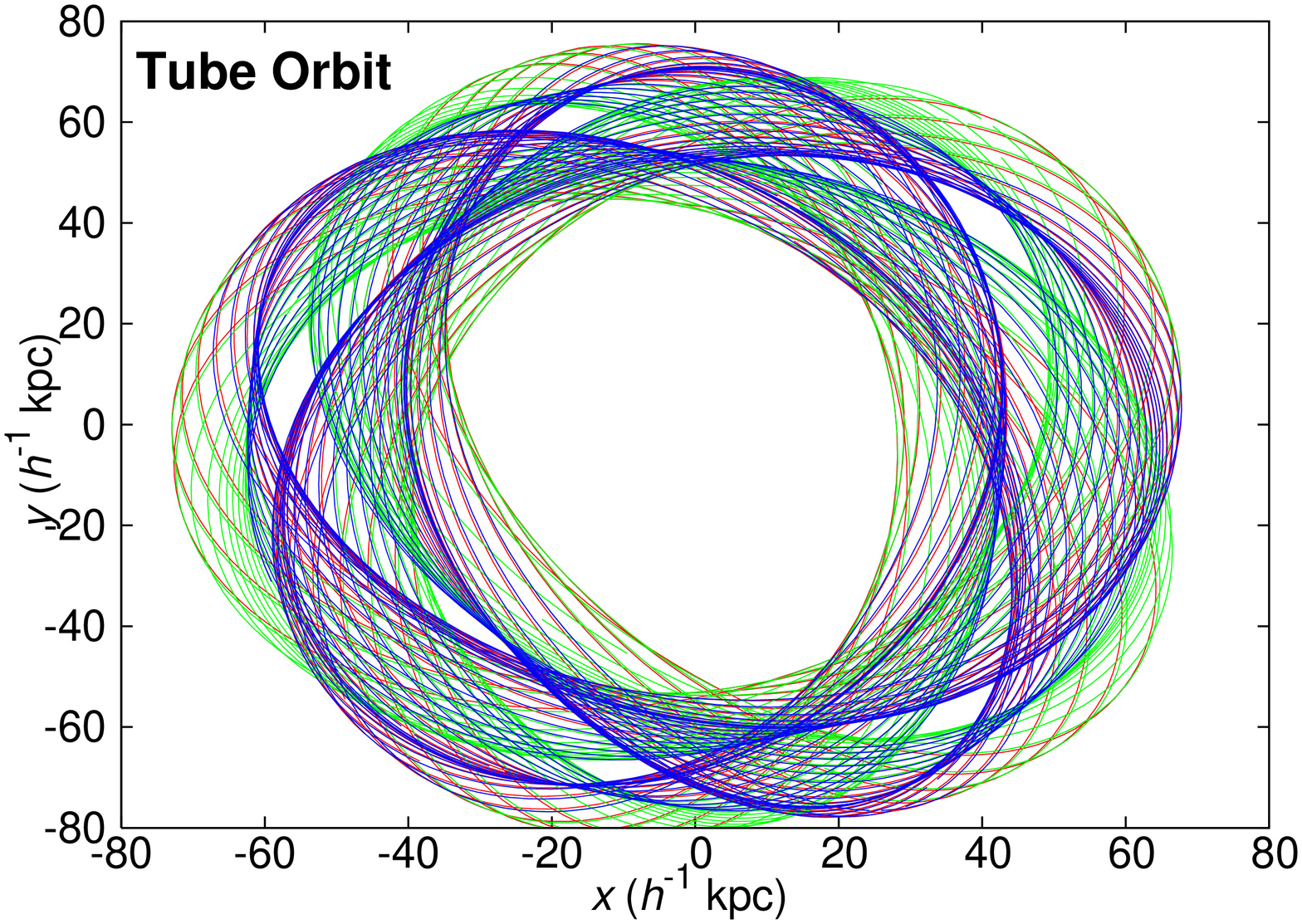} \\ 

\includegraphics[width=7cm,height=7cm,angle=-90,keepaspectratio]{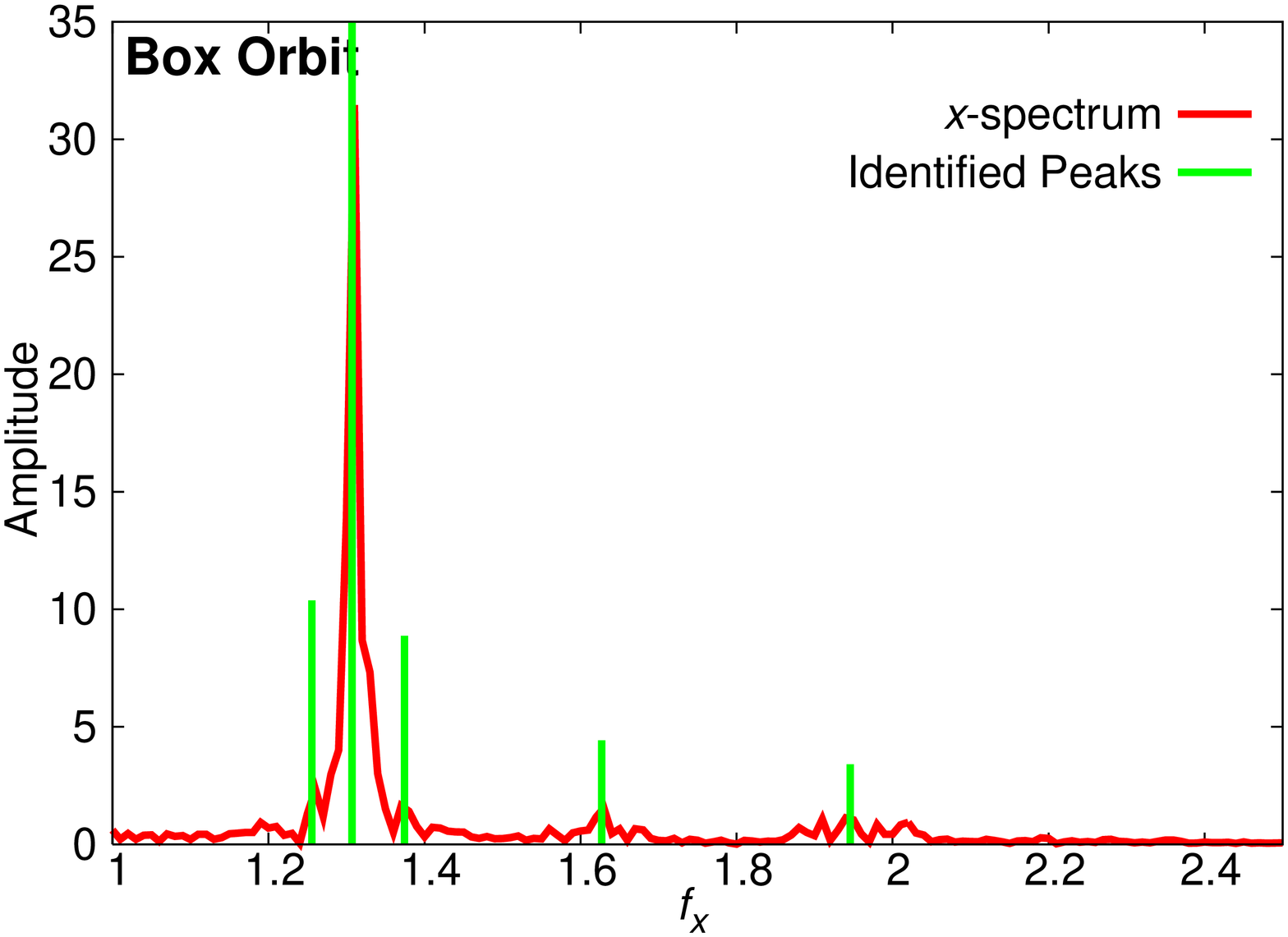}   &
\includegraphics[width=7cm,height=7cm,angle=-90,keepaspectratio]{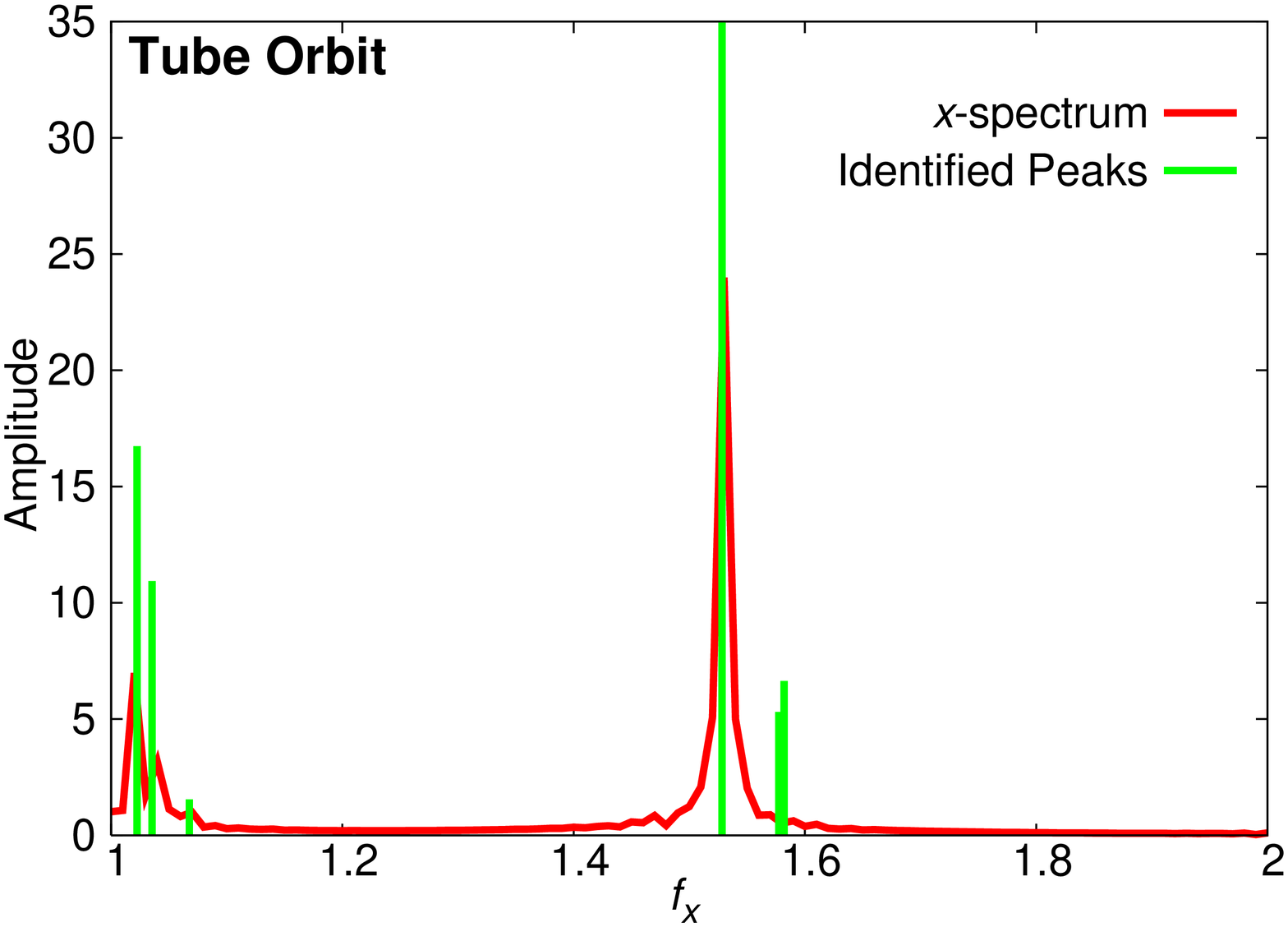}   \\

\includegraphics[width=7cm,height=7cm,angle=-90,keepaspectratio]{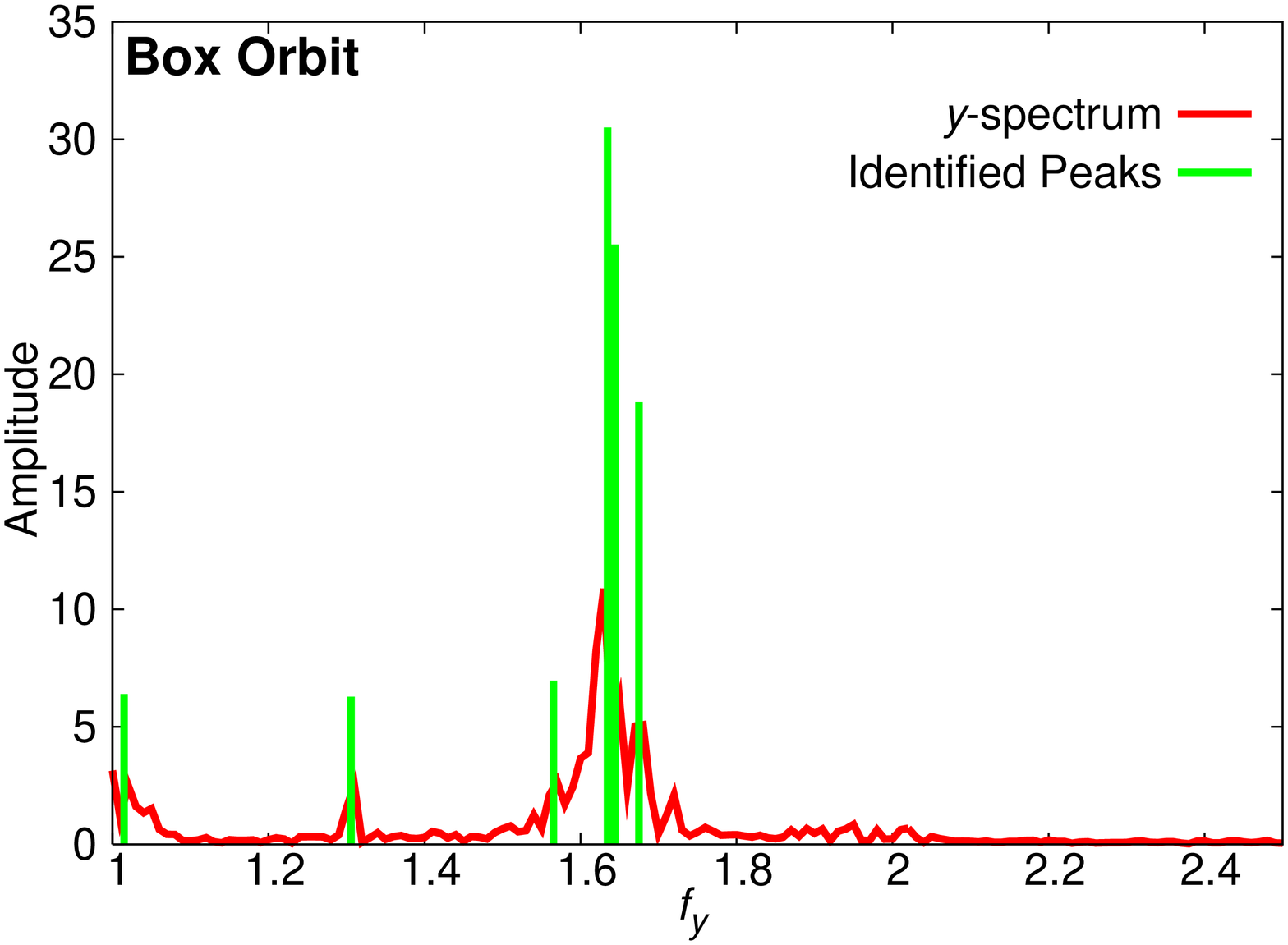} &
\includegraphics[width=7cm,height=7cm,angle=-90,keepaspectratio]{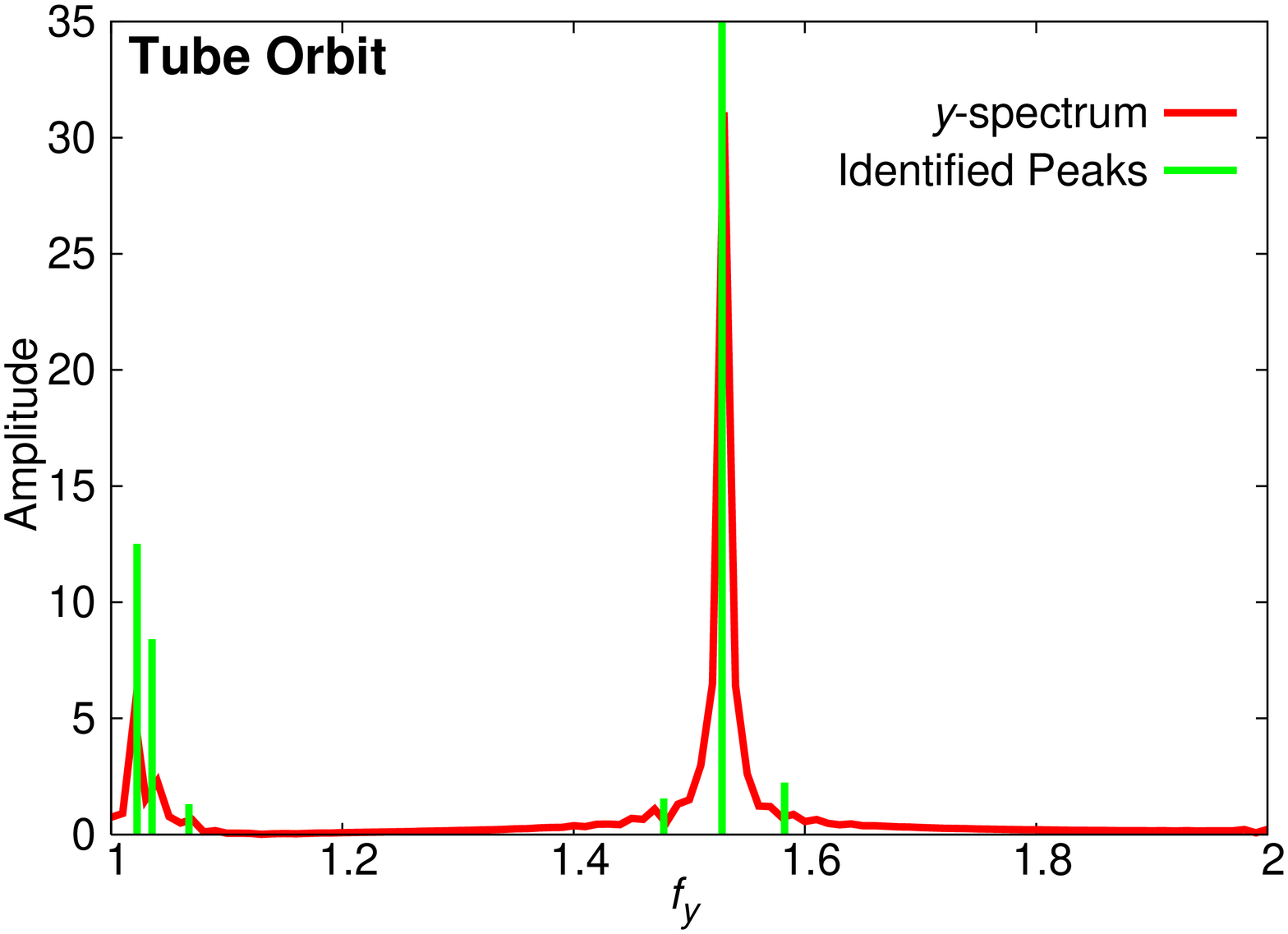}   \\
\end{tabular} 
\end{center} 
\caption[Examples of orbital types extracted from the DM
  simulations.]{\label{egorbits}Examples of the orbital types extracted from
  the dark matter only simulations.  The left (right) columns show an example
  of a box ($z$-tube) orbit. The $x$-$y$ projection of the orbit is shown in
  the top panel; the Fourier spectra of the $x$- and $y$-motion are shown in
  the middle and bottom panels, respectively.  Peaks identified by the routine
  are marked as green vertical lines.  Both of these orbits have three base
  frequencies.  While the box orbit has no resonances, one can clearly see that
  the dominant peaks in the $x$- and $y$-spectra of the tube orbit show a 1:1
  resonance ($f_x/f_y = 1$) - this is the only resonance found.  Colour indicates the time evolution of the orbit, from red to green to blue indicates progression with time.}
\end{figure*}

To classify the orbits obtained, the spectral classification routine of \cite{bib:Carpintero98}, hereafter CA98, is used.  This is a fully automated classification routine based on the Fourier spectra of the motions of the particles.  For a full description of the technique the reader is referred to their paper, but the method behind this routine is discussed briefly here.  

CA98 uses the result that, once a frequency spectrum of an orbit is decomposed into its fundamental frequencies, the relationship between these frequencies can be used to classify the orbit in a 3D potential into the major orbital families: box, major($x$)-axis and minor($z$)-axis tubes where orbits are orientated such that the major axis corresponds to the $x$-axis and the minor axis to the $z$-axis.  Since regular orbits are quasi-periodic, the Fourier spectra of the time series of each coordinate will consist of discrete peaks (this is not the case for irregular orbits).  The Fourier transform of the time series of each coordinate is performed and the dominant peak frequency determined. 
For each pair of coordinates ($x$-$y$, $y$-$z$ and $x$-$z$) these frequencies are compared, searching for linear combinations (resonances).  If the peak frequency in each direction of motion $i$ is represented by $\omega_i$, then a resonance is defined as 
\begin{equation}
l\omega_x + m\omega_y + n\omega_z = 0,
\end{equation}
for non-trivial combinations of the integers $n$, $l$ and $m$.   If all
dominant frequencies are a multiple of a single unit frequency, then there is
one base frequency.  If there is no resonance, then all dominant frequencies are irrationally related.  Once the dominant frequencies have been compared, the spectra are searched for additional base frequencies.  The number of base frequencies specifies whether an orbit is regular (open, closed or thin) or irregular, while the number of resonances specifies the orbital family as box or $x$-tube, $y$-tube or $z$-tube.   Only particles/subhaloes that have undergone at least 40 orbits are classified; this ensures that they have clearly defined spectra.  

A 3D orbit with 4 or more base frequencies is classified as irregular; if it has 3 (or fewer) base frequencies it is classified as regular.  
The base frequencies of a box orbit are incommensurable; this is the only class which does not exhibit resonance between the dominant frequencies.  The orbit is classified as a $z$-tube if the $x$- and $y$-spectra show a 1:1 resonance, that is $l=1$, $m=1$ and $n$ is arbitrary.  If $y$ and $z$ show a 1:1 resonance ($m=1$ and $n=1$), then the orbit is classified as an $x$-tube.  As orbits around the intermediate axis are unstable, it is only in rare cases that $y$-tubes are identified.  These show resonances between the $x$- and $z$-base frequencies.  A summary of the orbit classifications (taken from \citealt{bib:Carpintero98}) is given in Table \ref{orbitclass}.  

Examples of the orbital types extracted from the dark matter only simulations
are shown in Fig. \ref{egorbits}.  The left (right) panels show an example
of a box ($z$-tube) orbit.   The $x$-$y$ projection of the orbit is shown in
the top row while the Fourier spectra of the $x$- and $y$-motion are shown in
the middle and bottom panels, respectively.  Peaks identified by the routine
are marked as vertical lines.  Both of these orbits have three base
  frequencies.  While the box orbit has no resonances one can clearly see that
  the dominant peaks in the $x$- and $y$-spectra of the tube orbit show a 1:1
  resonance ($f_x/f_y = 1$) - this is the only resonance found.

The CA98 algorithm has been tested rigorously using a number of analytic
potentials.  As it is fully-automated, it allows for the classification of
large numbers of orbits.  It also distinguishes more orbital classes than
classifications based on the sign of a component of the orbits' angular
momentum.  For comparison, the orbits considered here have also been
classified using the spin classification technique (\citealt{bib:Barnes92}).  While the fraction of box orbits is in general higher than that obtained using the method of CA98, the same general trends are found using both classification schemes.

\begin{table*}
\caption[Orbit classifications.]{\label{orbitclass}Classifications of orbits (as in \citealt{bib:Carpintero98}).} 
\centering
\begin{tabular}{|c|l||l|l|l|l|}
\hline
&  &\multicolumn{4}{|c|}{Number of base frequencies} \\
&& 1 & 2 & 3 & $ > 4$\\ 
\hline
Number & 0 & axial & 2-D box & 3-D box & \\
\ \\
of & 1 & closed $0:m:n$ box & thin $\pi:m:n$ box & open $\pi:m:n$ box & \\
resonances & & closed $0:1:1$ tube &  thin $\pi:1:1$ tube& open  $\pi:1:1$ tube & Irregular\\
\ \\

& 3& closed $l:m:n$ box & thin $l:m:n$ box & open $l:m:n$ box &\\
& & closed $l:1:1$ tube &  thin $l:1:1$ tube & open $l:1:1$ tube &   \\ 
\hline
\end{tabular} 
\end{table*}

\section{Results}
\label{orbitsresults}

In this section the results of the spectral analysis of the orbital content of the OWLS haloes are presented.  Particular emphasis is placed on the fraction of box orbits, as these orbits are known to be important in conveying information from the central region to the outskirts of the halo and are thought to be responsible for supporting the triaxial shape of haloes.  We quantify the fraction of different orbital types and show how the orbital content is affected by the addition of baryons and feedback processes.  We then show how the orbits of dark matter particles are influenced by halo properties (such as concentration, shape, spin and central baryon fraction).  Finally, we compare the orbits of dark matter particles to those of stellar particles and subhaloes.

\subsection{Orbits of dark matter particles}

\begin{figure*}
\begin{center}
\begin{tabular}{cc}

\includegraphics[width=7cm,height=7cm,angle=-90,keepaspectratio]{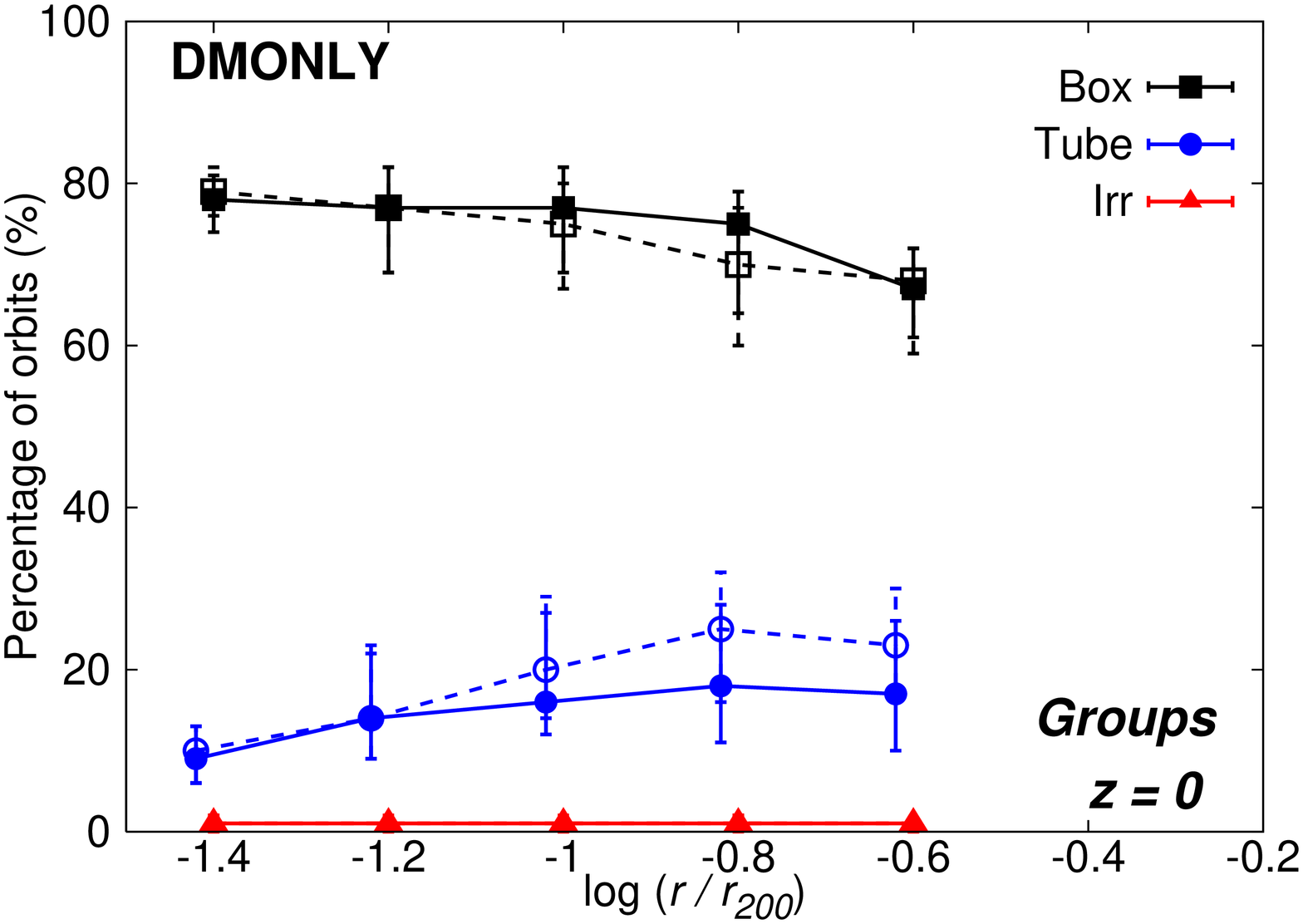} &
\includegraphics[width=7cm,height=7cm,angle=-90,keepaspectratio]{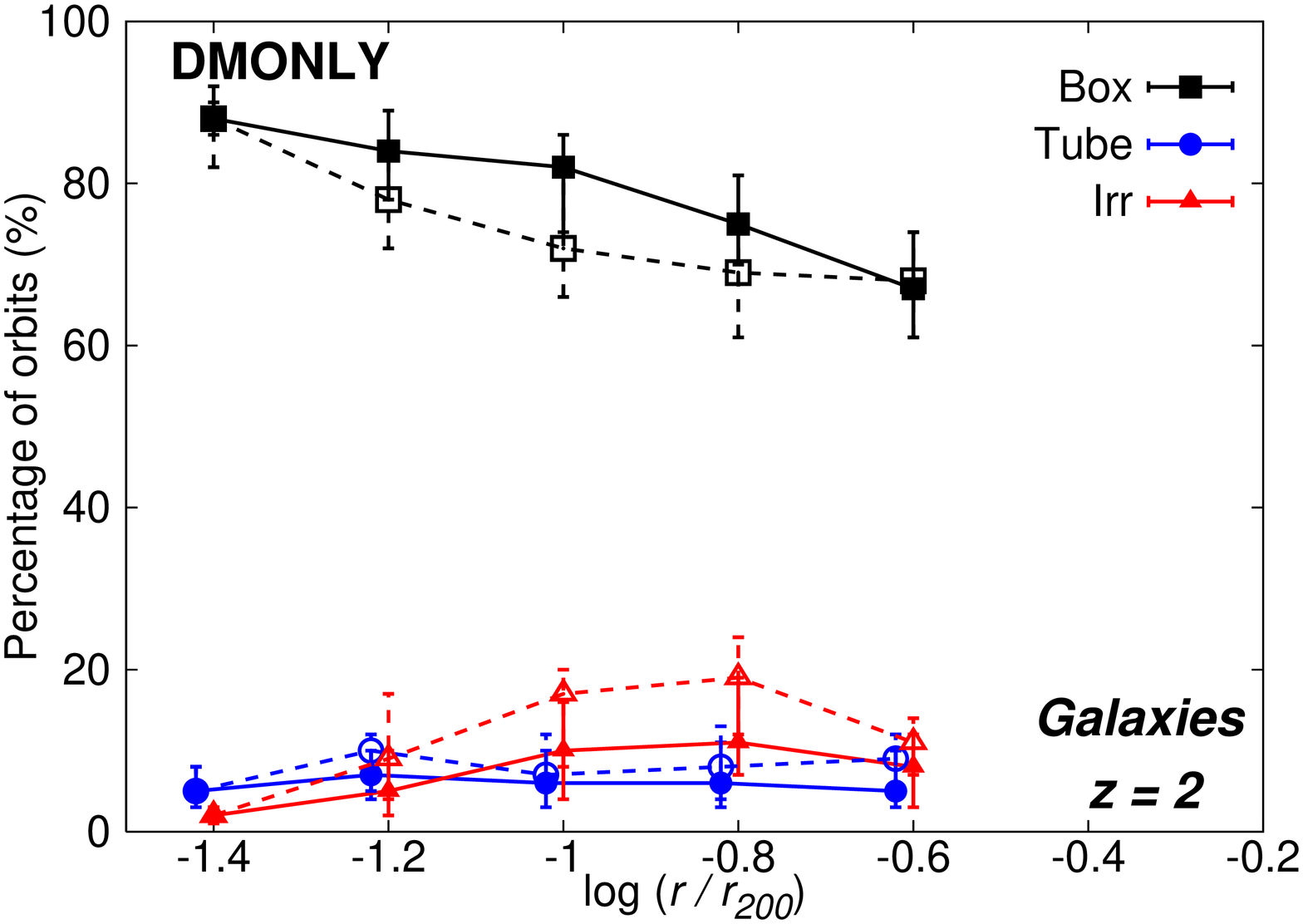} \\
\end{tabular} 
\end{center}
\caption[ The radial distribution of orbital content of DM
  haloes.]{\label{dmorbits} The orbital content of the dark matter only
  simulation.  We show the median percentage of dark matter particles on box
  (black squares), tube (blue circles) and irregular orbits (red triangles)
  estimated over the 50 most massive haloes (solid lines) and for the
  relaxed subsample (dashed lines).  Error bars show the quartile halo-to-halo
  scatter.  The left-hand panel shows the orbital content of the haloes at $z = 0$
  where the mean dark matter halo mass is $6 \times
  10^{13}$\,$h^{-1}$M$_\odot$.  The orbital content at $z = 2$ is shown
  in the right-hand panel; this sample has a mean dark matter halo mass of $7 \times 10^{11}$\,$h^{-1}$M$_\odot$.  }
\end{figure*}

We begin by considering the orbital content of dark matter particles in each of the five simulation runs.  We focus on the central region of the haloes (within 25 per cent of $r_{200}$) where baryonic physics is likely to play a large role.  These results are presented as the percentage of each type of orbit as a function of radius in Fig. \ref{dmorbits}.  For clarity we show only the fraction of box (black squares), loop (blue circles) and irregular (red triangles) orbits.  The number of orbits that are not classified ($<10$ per cent) can be determined by subtracting the sum of box, tube and irregular from 100 per cent.    

\subsubsection{Dark matter only simulations}

\begin{figure*}
\begin{center}
\begin{tabular}{cc}

\includegraphics[width=7cm,height=7cm,angle=-90,keepaspectratio]{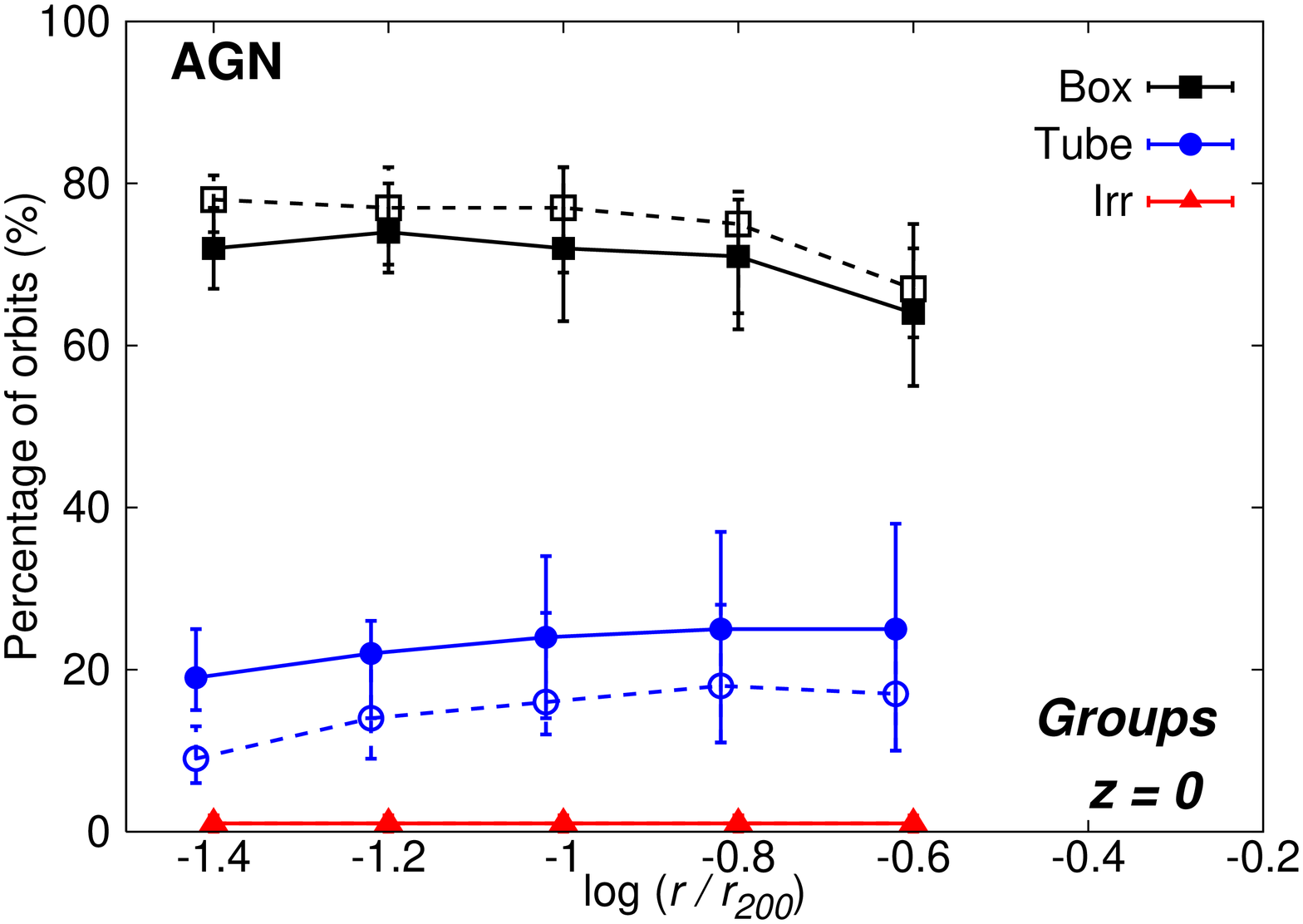} &
\includegraphics[width=7cm,height=7cm,angle=-90,keepaspectratio]{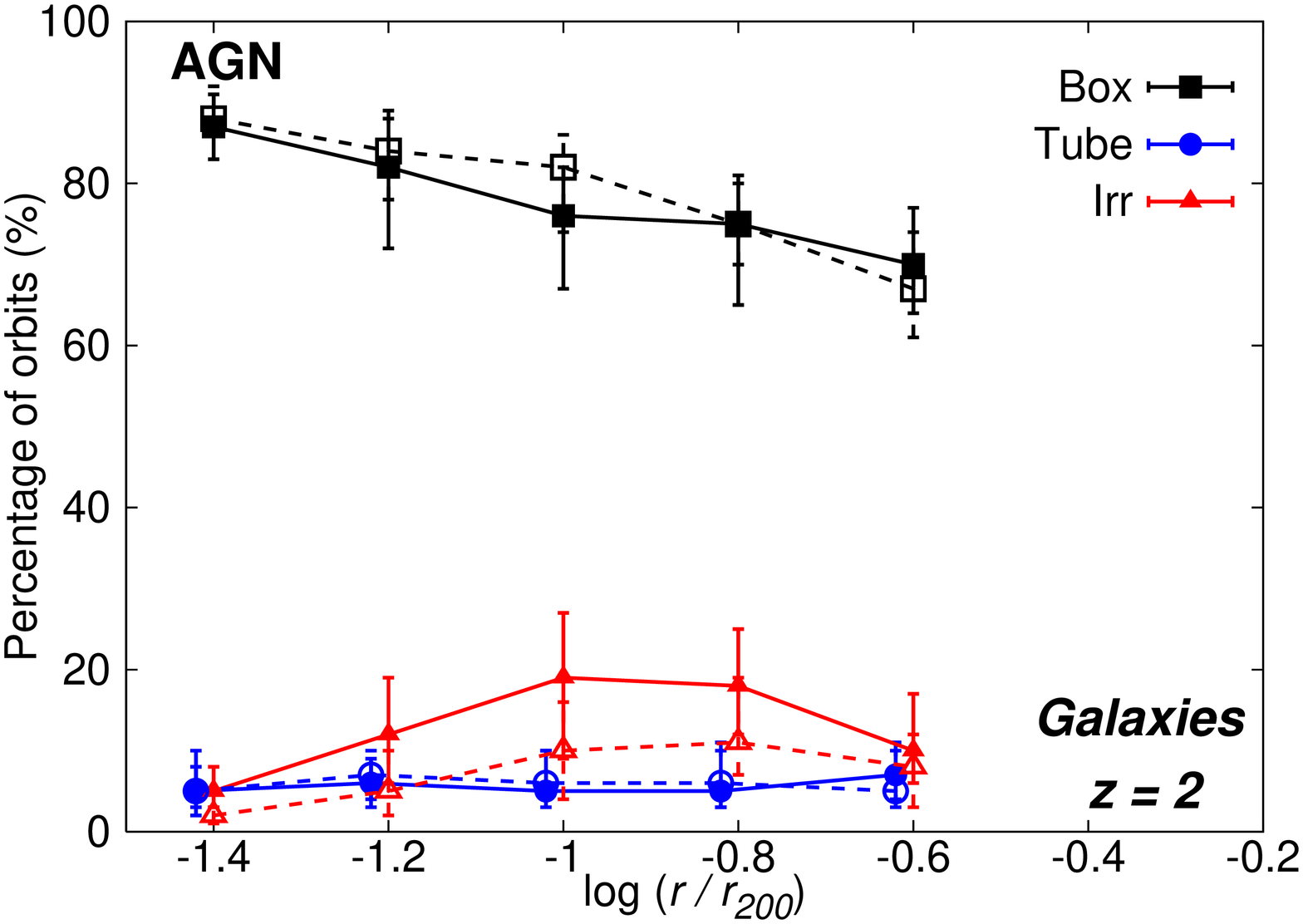} \\

\includegraphics[width=7cm,height=7cm,angle=-90,keepaspectratio]{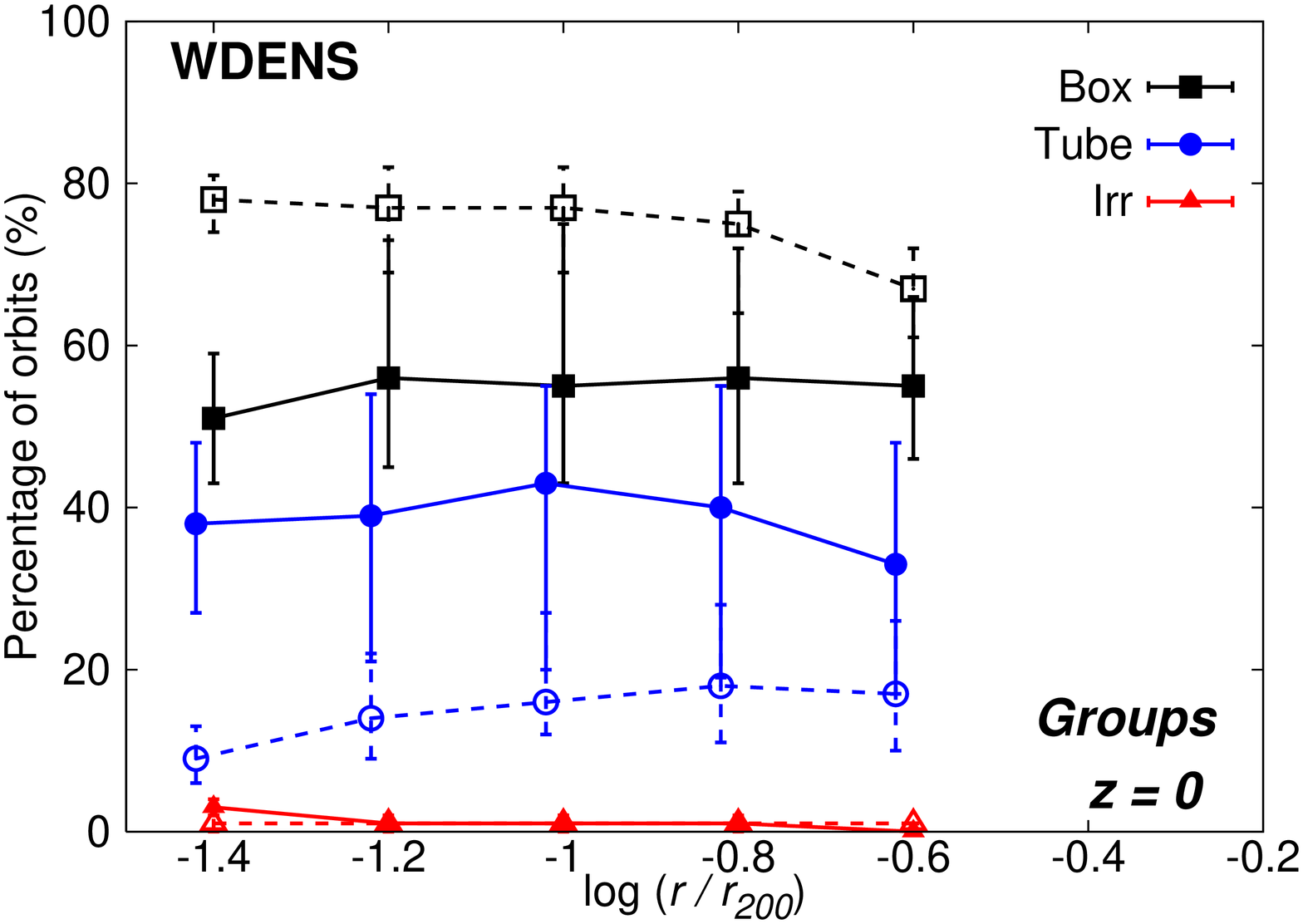} &
\includegraphics[width=7cm,height=7cm,angle=-90,keepaspectratio]{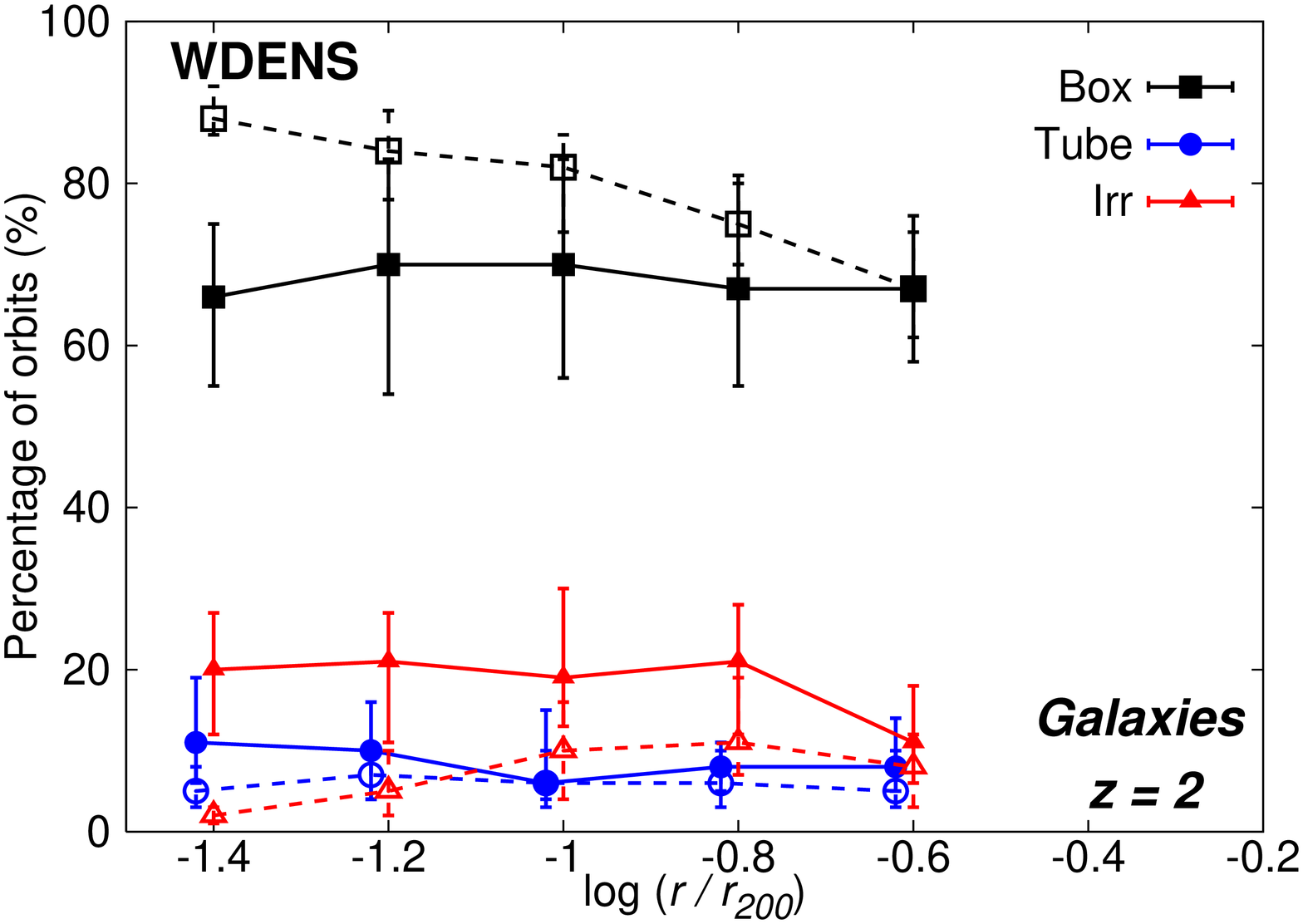} \\

\includegraphics[width=7cm,height=7cm,angle=-90,keepaspectratio]{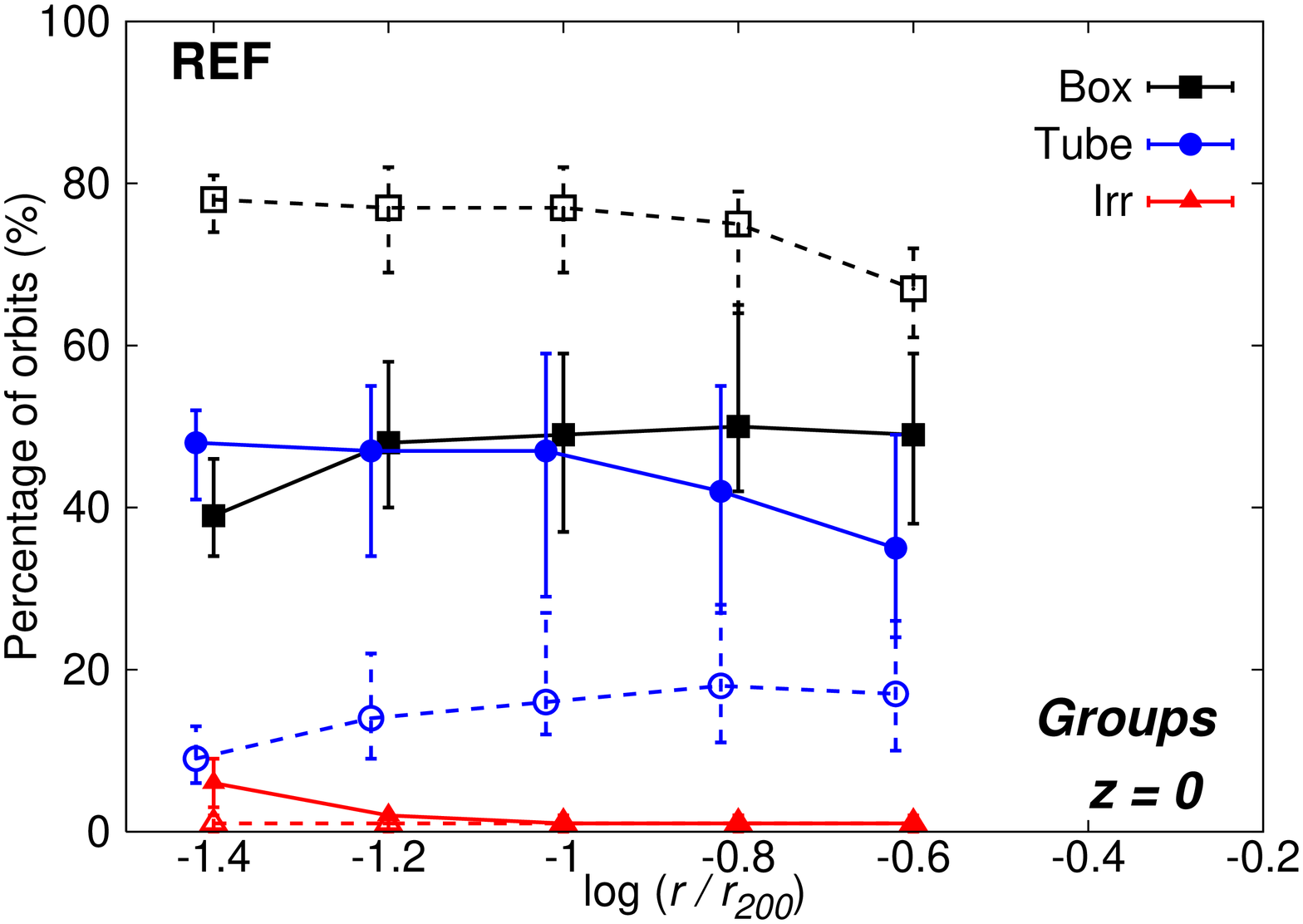} &
\includegraphics[width=7cm,height=7cm,angle=-90,keepaspectratio]{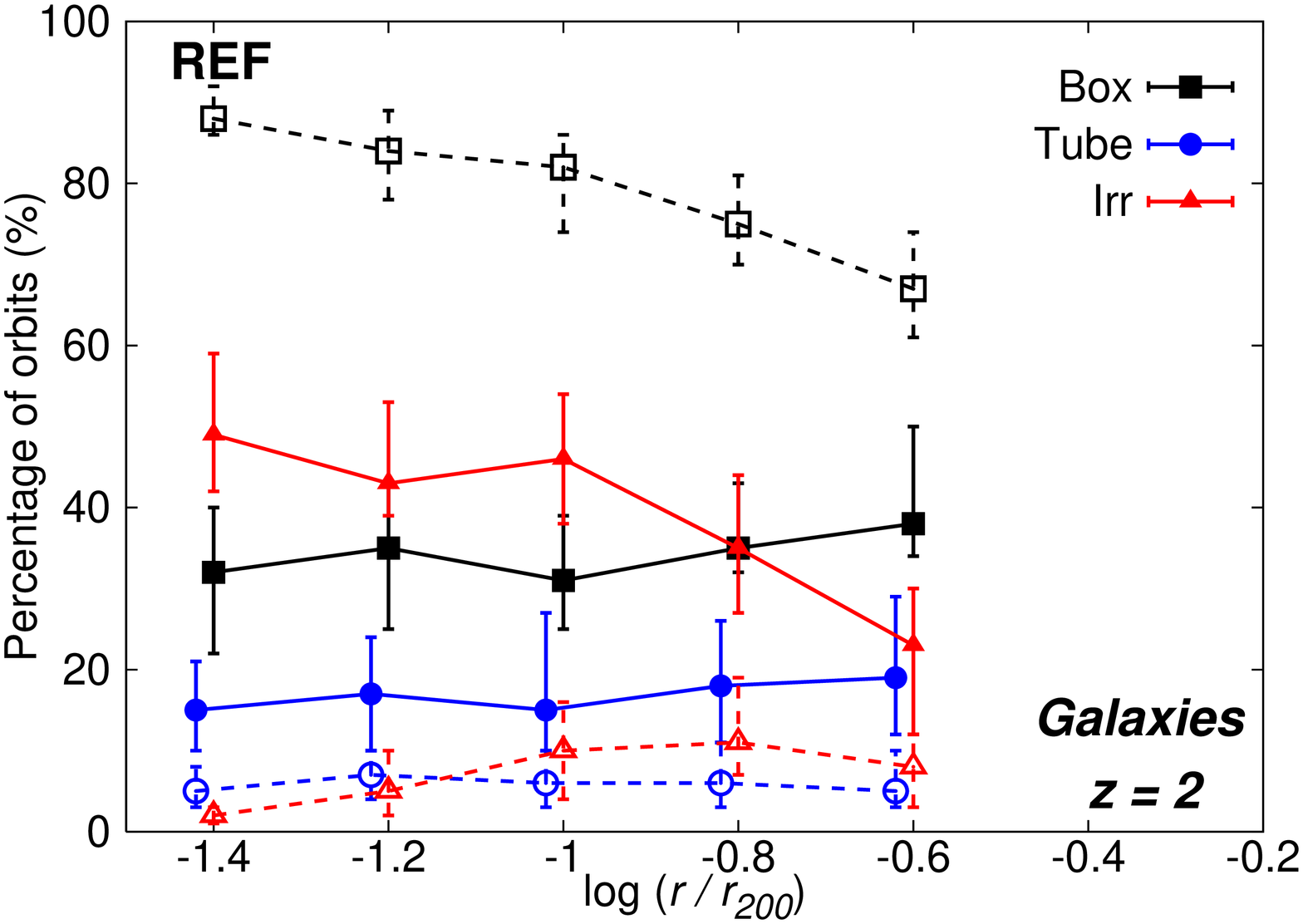} \\

\includegraphics[width=7cm,height=7cm,angle=-90,keepaspectratio]{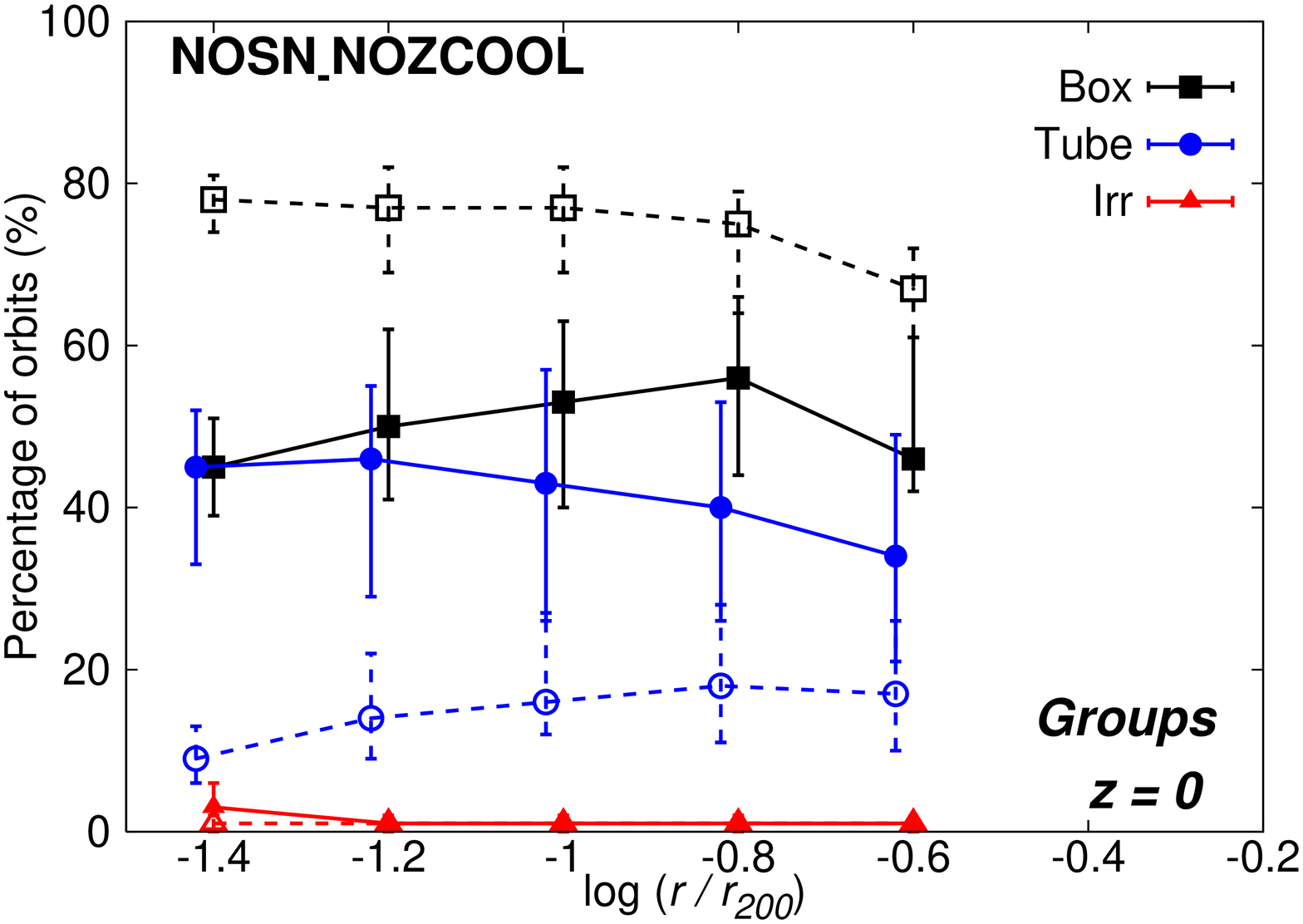} &
\includegraphics[width=7cm,height=7cm,angle=-90,keepaspectratio]{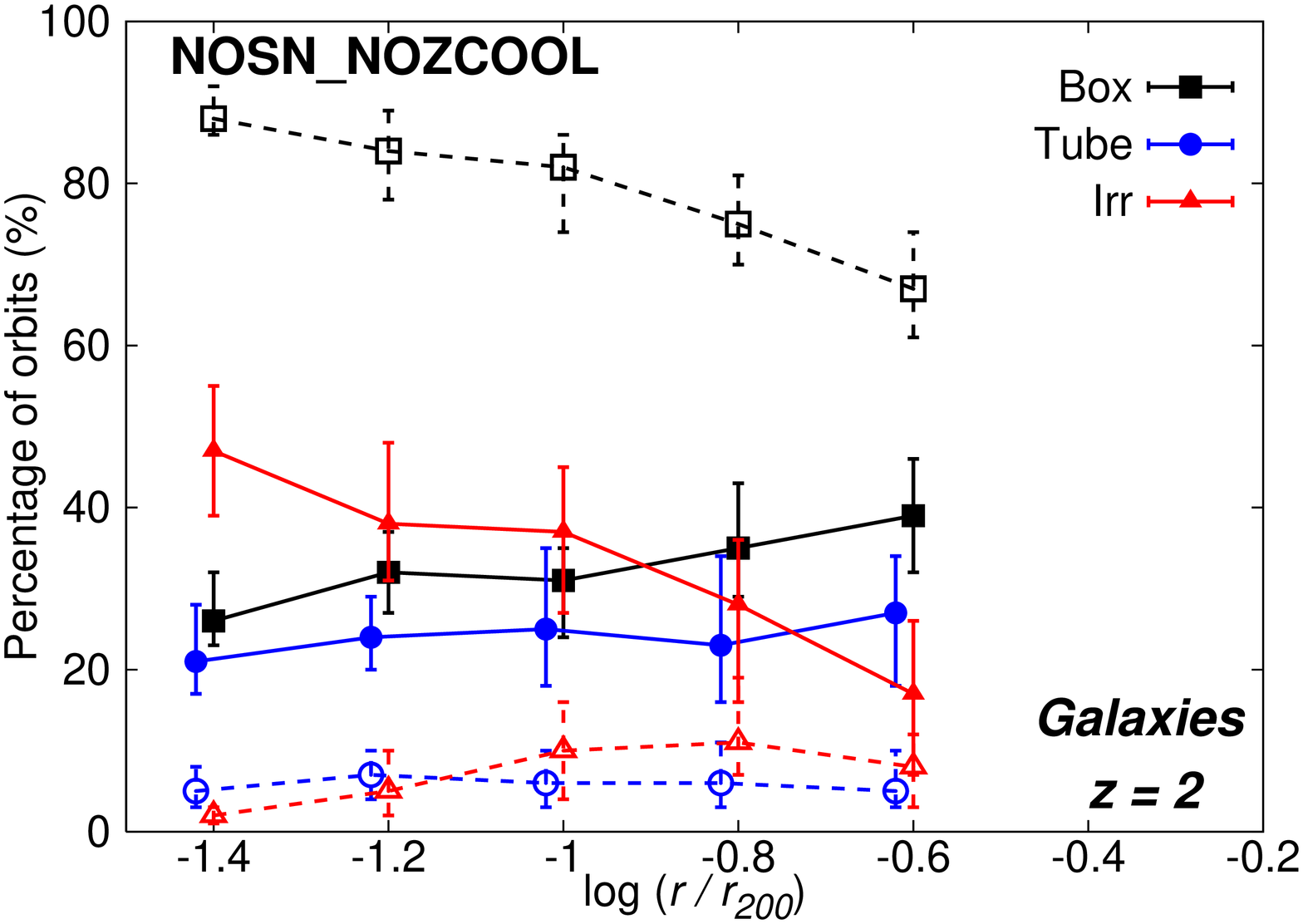} \\

\end{tabular} 
\end{center}

\caption[ The radial distribution of orbital content of DM
  haloes.]{\label{fborbits} The median percentage of dark matter particles on
  box (black squares), tube (blue circles) and irregular orbits (red
  triangles) estimated over the 50 most massive haloes in each simulation
  (solid lines).  Error bars show the quartile halo-to-halo scatter. Dashed
  lines show the orbital content of the 50 haloes in the dark matter only
  run, for comparison.  The left-hand panels show the orbital content of the haloes
  at $z = 0$, where the mean dark matter halo mass is $6 \times
  10^{13}$\,$h^{-1}$M$_\odot$.  The orbital content at $z = 2$ is shown in the right-hand
  panels (mean dark matter halo mass of $7 \times 10^{11}$\,$h^{-1}$M$_\odot$).  }

\end{figure*}
  
The orbital content of the DMONLY haloes is shown in Fig. \ref{dmorbits}.
In the left-hand panel we show the orbital content of the $z = 0$ haloes (mean halo
mass of $6 \times 10^{13}$\,$h^{-1}$M$_\odot$); haloes at $z = 2$ (mean halo
mass of $7 \times 10^{11}$\,$h^{-1}$M$_\odot$) are shown in the right-hand panel.  Symbols show the median fraction of orbits of a particular type averaged over all haloes and error bars represent the quartile halo-to-halo scatter.  In these two panels, we compare the complete sample of 50 haloes (solid lines) to the subset of these haloes that are found to be relaxed (dashed).   The orbital content of the relaxed sample does not appear to differ significantly from that of the whole sample so all 50 haloes are used for the rest of the analysis.  

At both redshifts, the haloes are dominated by box orbits out to
0.25$r_{200}$.  The dominance of box orbits is unsurprising; these orbits are
required to support the triaxial haloes characteristic of dark matter
simulations.  The fraction of tube orbits in the galaxy sample at $z = 2$ is
considerably lower than that found in the $z = 0$ group sample, while the
fraction of box orbits is slightly higher in the $z = 2$ galaxy sample.  We
find that the fraction of irregular orbits in the $z = 2$ galaxy sample is
much higher than that seen in the $z = 0$ group sample, as expected when major mergers dominate the formation process or the mass accretion is rapid (\citealt{bib:Zhao09}).

There is a weak trend for the fraction of box orbits to decrease with
increasing radius; this is accompanied by an increase in the fraction of tube
orbits.  Resonant box orbits (defined in Table \ref{orbitclass}, considering integer values up to $n,m=12$) account for approximately half of the box orbits shown here and are also found to decrease with increasing radius.  The fraction of $y$-tubes, irregular and non-classified orbits is negligible.  The fractions of both $x$- and $z$-tubes increase with radius.  While $x$-tubes dominate the tube contribution at small radii, the fraction of $z$-tubes becomes increasingly important at larger radii (not shown).

\subsubsection{Baryon simulations and the effect of feedback}

The central baryonic mass concentration is significantly affected by the
strength of the different feedback models, as shown in
Fig. \ref{baryonfraction}.  In this section the impact of this central
concentration on the orbital content of the haloes is discussed. The impact of
baryons on the orbital content of haloes can be seen in Fig. \ref{fborbits}
which is presented in the same way as Fig. \ref{dmorbits}.  The $z = 0$ (2)
sample is shown in the left (right) column.  In these plots the solid lines
show the fraction of orbits in the run with baryons while the dashed lines show the orbital content of the dark matter only simulation for comparison.  

All of the baryon runs are found to have a smaller fraction of box orbits (at
all radii out to 0.25$r_{200}$) than the dark matter only haloes, but this
decrease is most noticeable in the very central regions where baryonic
condensation is most significant.  The central concentration of baryons acts to
transform the box orbits into tube orbits.  This is a result of the decrease in the
elongations of the orbits in response to the central mass (\citealt{bib:Dubinski94}).  While the orbital content of
haloes extracted from the AGN run is remarkably similar to the dark matter
only haloes, the efficient cooling in the weak feedback and no feedback runs
shows the most significant reduction in the fraction of box orbits in the
central region.  These results are not unexpected.  The AGN feedback expels
most of the baryonic component from the central regions (as is evident in
Fig. \ref{baryonfraction}).  The runs with no or weak feedback have a much
higher central baryon concentration and hence fewer box orbits than the
stronger feedback runs.  These results are in accord with expectations that
increased galaxy formation efficiency/central baryon fraction lowers the fraction of
box orbits. 

The right column of Fig. \ref{fborbits} shows the orbital content of the
50 most massive haloes at $z = 2$,  where the mean dark matter halo mass is $7 \times
  10^{11}$\,$h^{-1}$M$_\odot$.  Again we note that there is a
significant decrease (increase) in the fraction of tube (irregular) orbits
compared to the $z = 0$ sample.  This is particularly apparent  in the weak
feedback (REF) and no feedback (NOSN\_NOZCOOL) simulations. These runs appear
very similar in the most central regions; perhaps unsurprisingly in that they
appear to share very similar baryon fractions at $z = 2$ (see Fig. \ref{baryonfraction}).  

\subsubsection{Orbital content versus halo properties}

\begin{figure*}
\begin{center}
\begin{tabular}{cc}
\includegraphics[width=7.cm,height=7.cm,angle=-90,keepaspectratio]{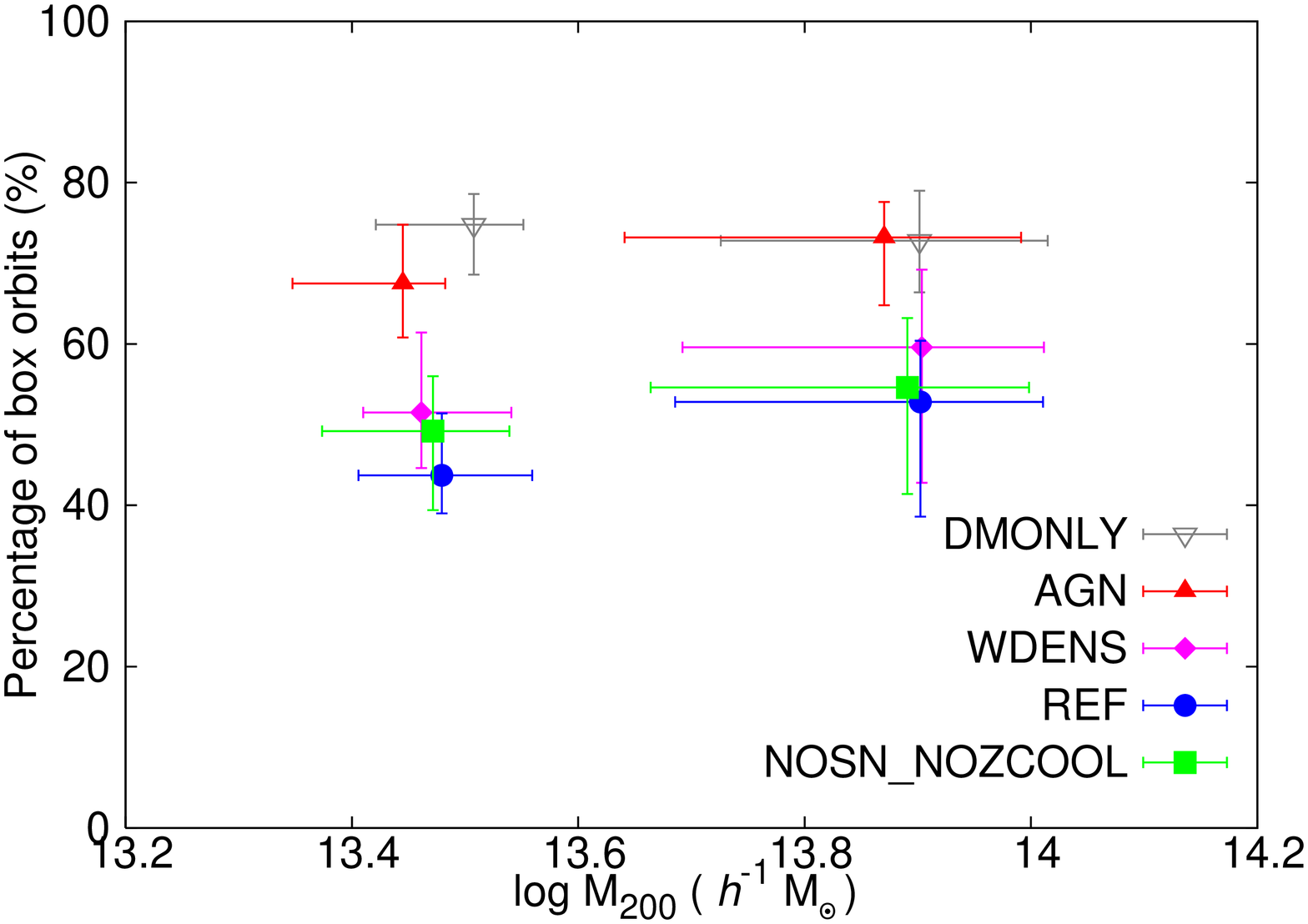} &
\includegraphics[width=7.cm,height=7.cm,angle=-90,keepaspectratio]{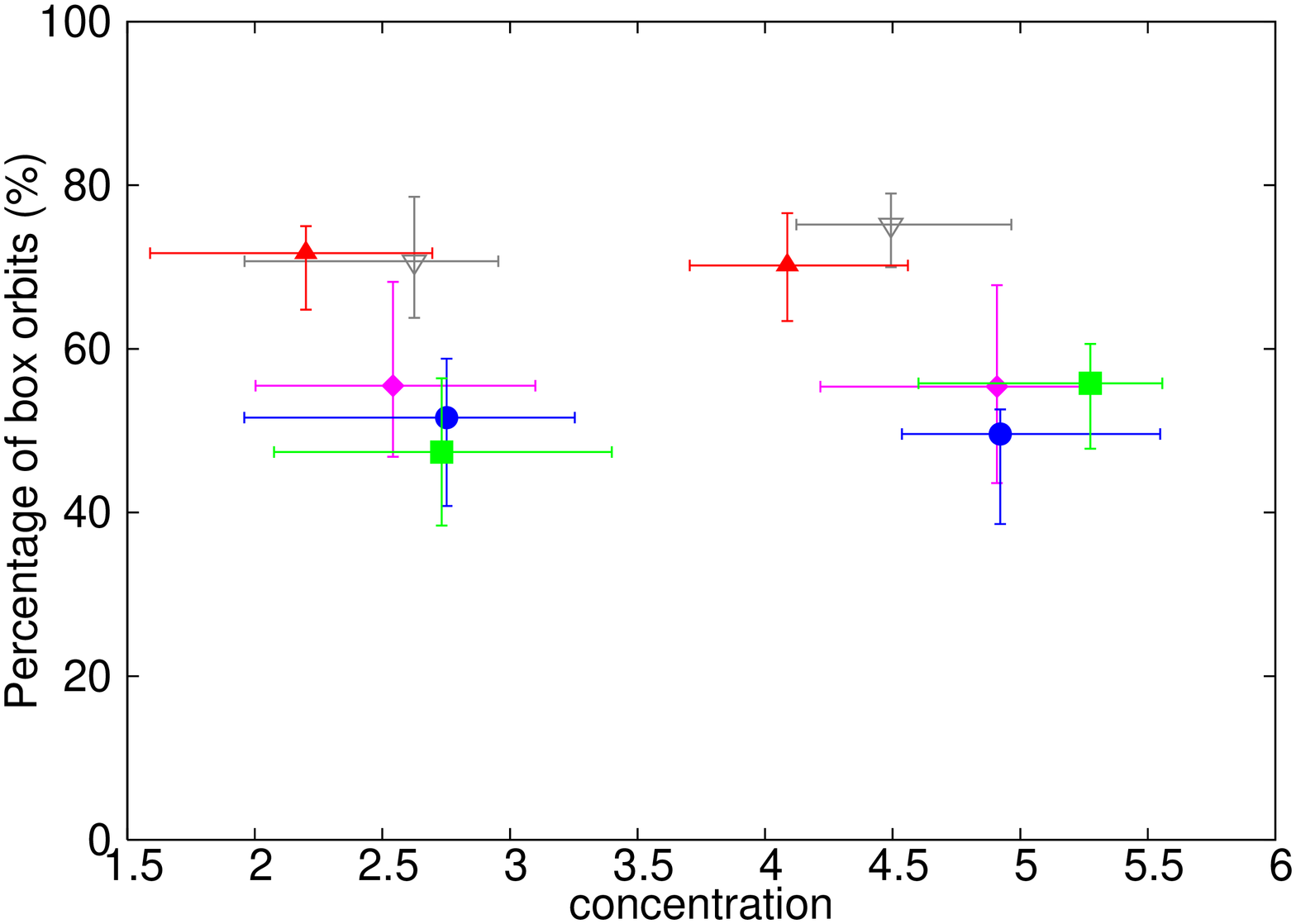}\\
\includegraphics[width=7.cm,height=7.cm,angle=-90,keepaspectratio]{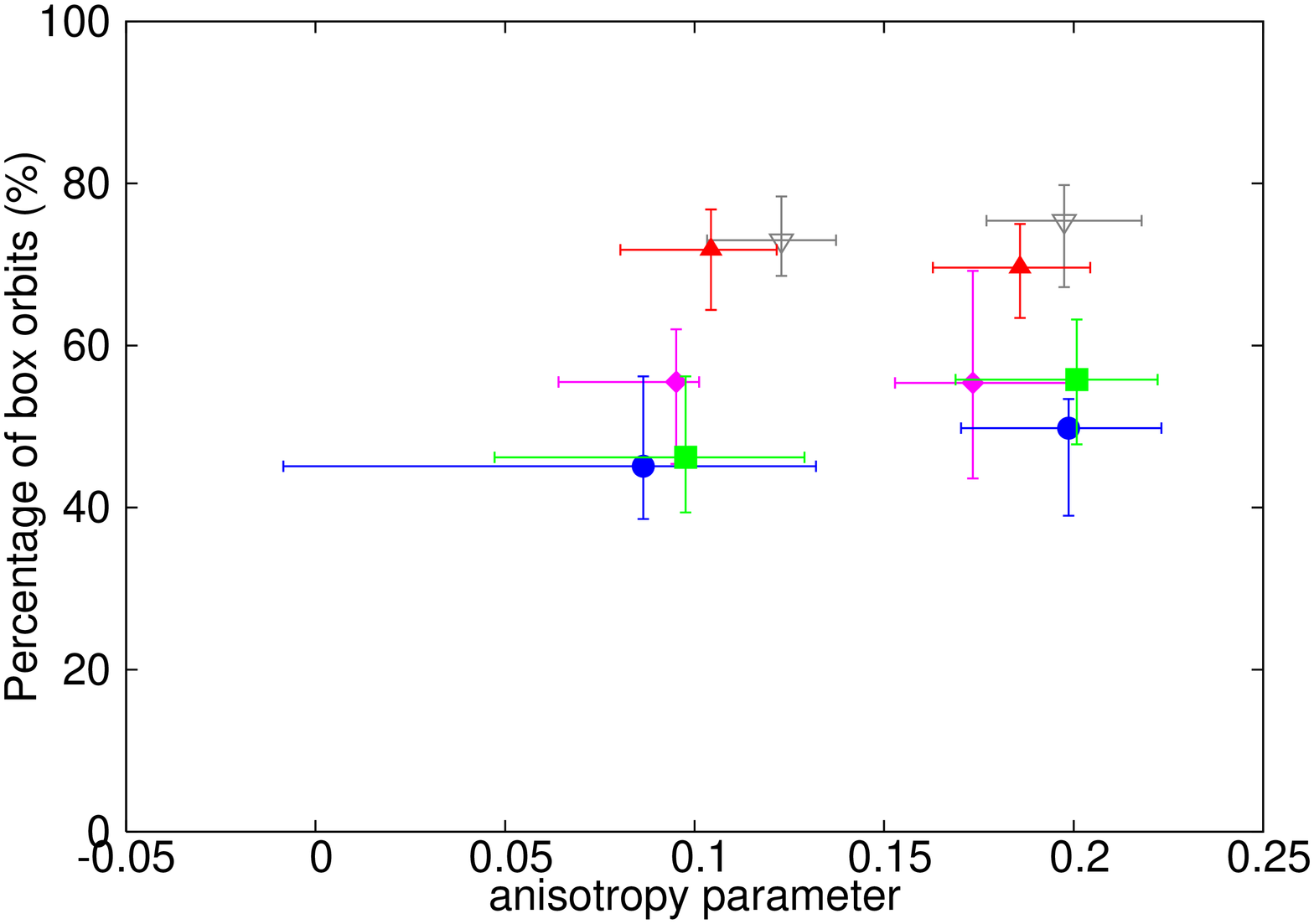} &
\includegraphics[width=7.cm,height=7.cm,angle=-90,keepaspectratio]{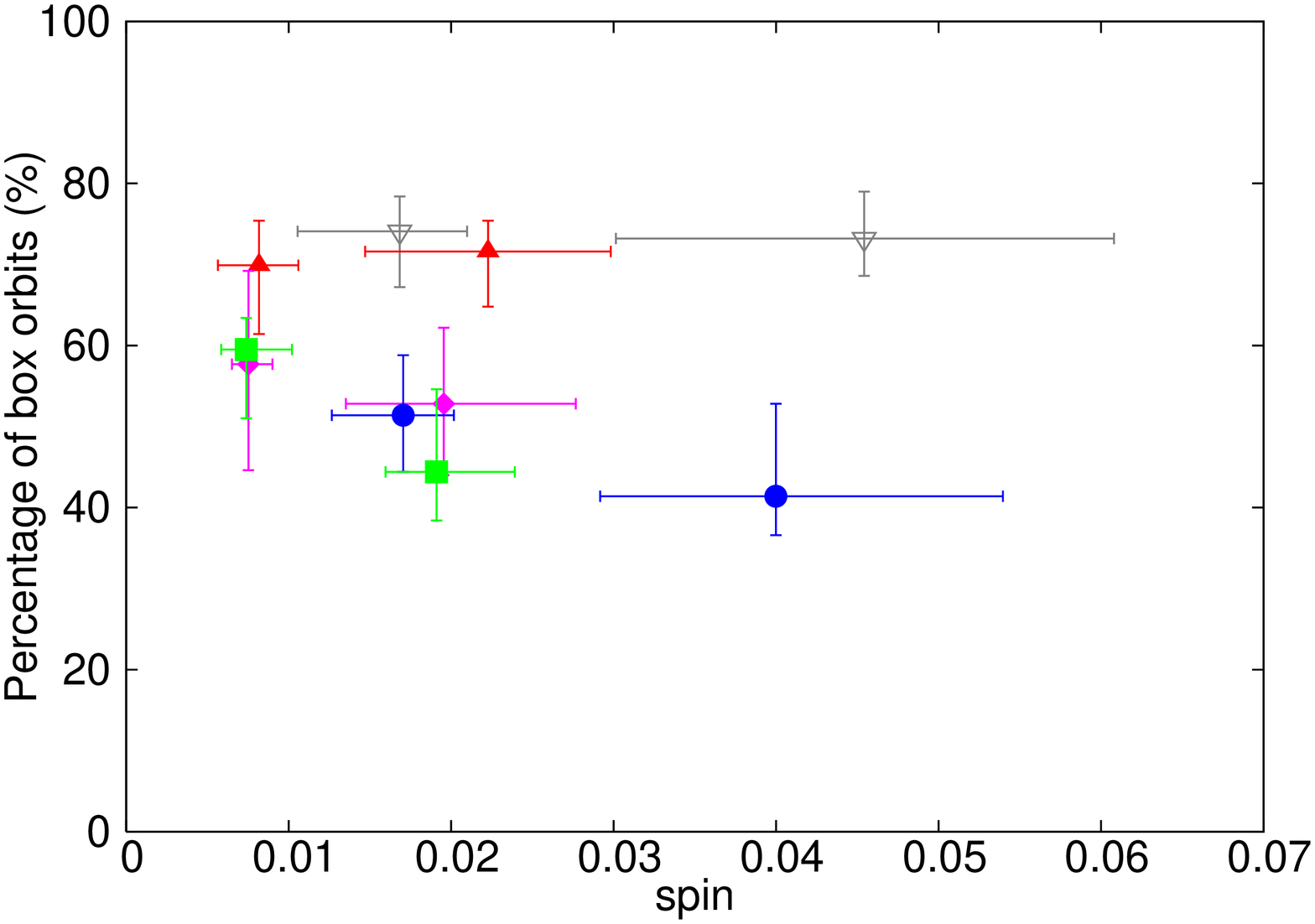}\\
\includegraphics[width=7.cm,height=7.cm,angle=-90,keepaspectratio]{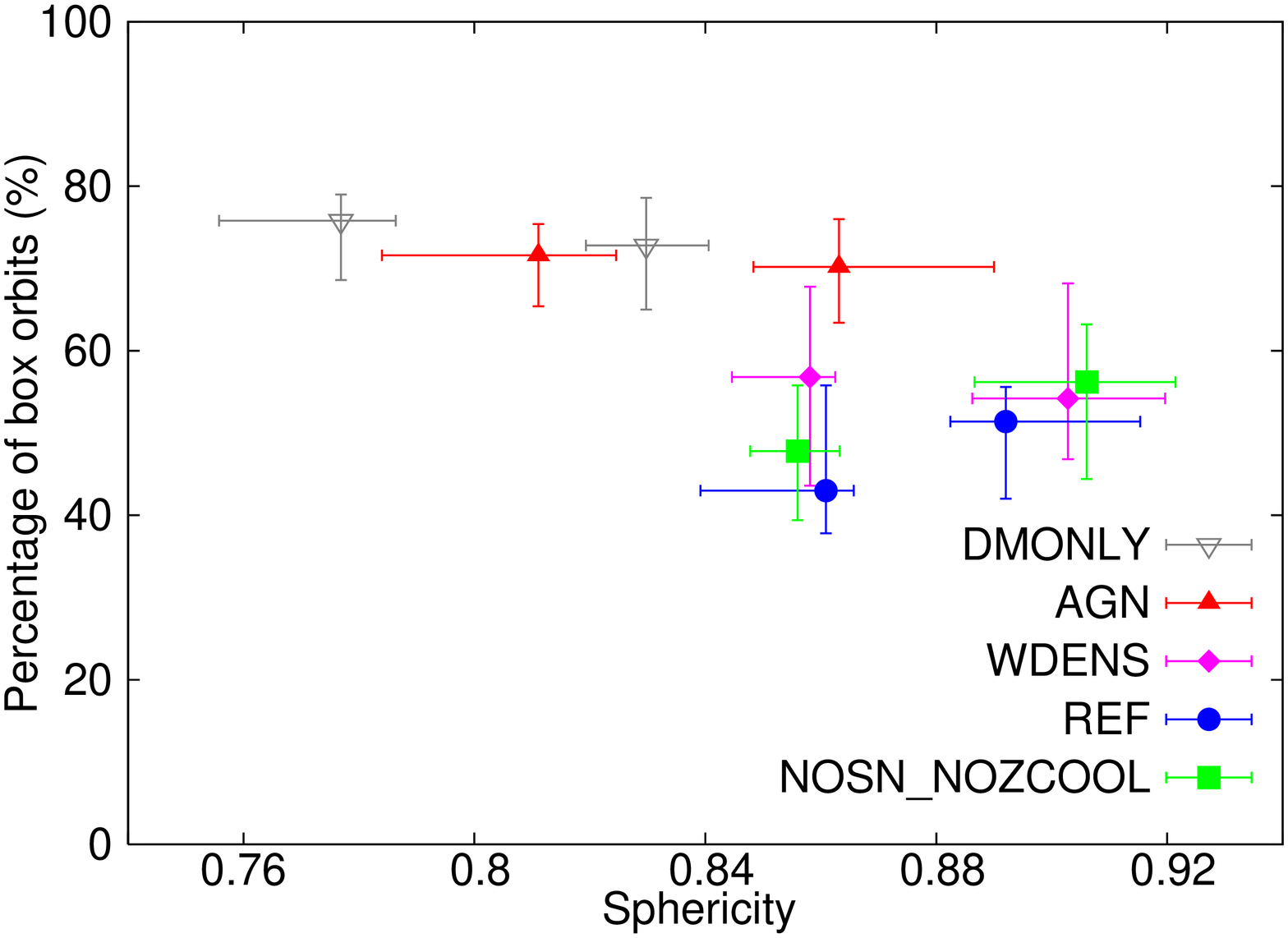} &
\includegraphics[width=7.cm,height=7.cm,angle=-90,keepaspectratio]{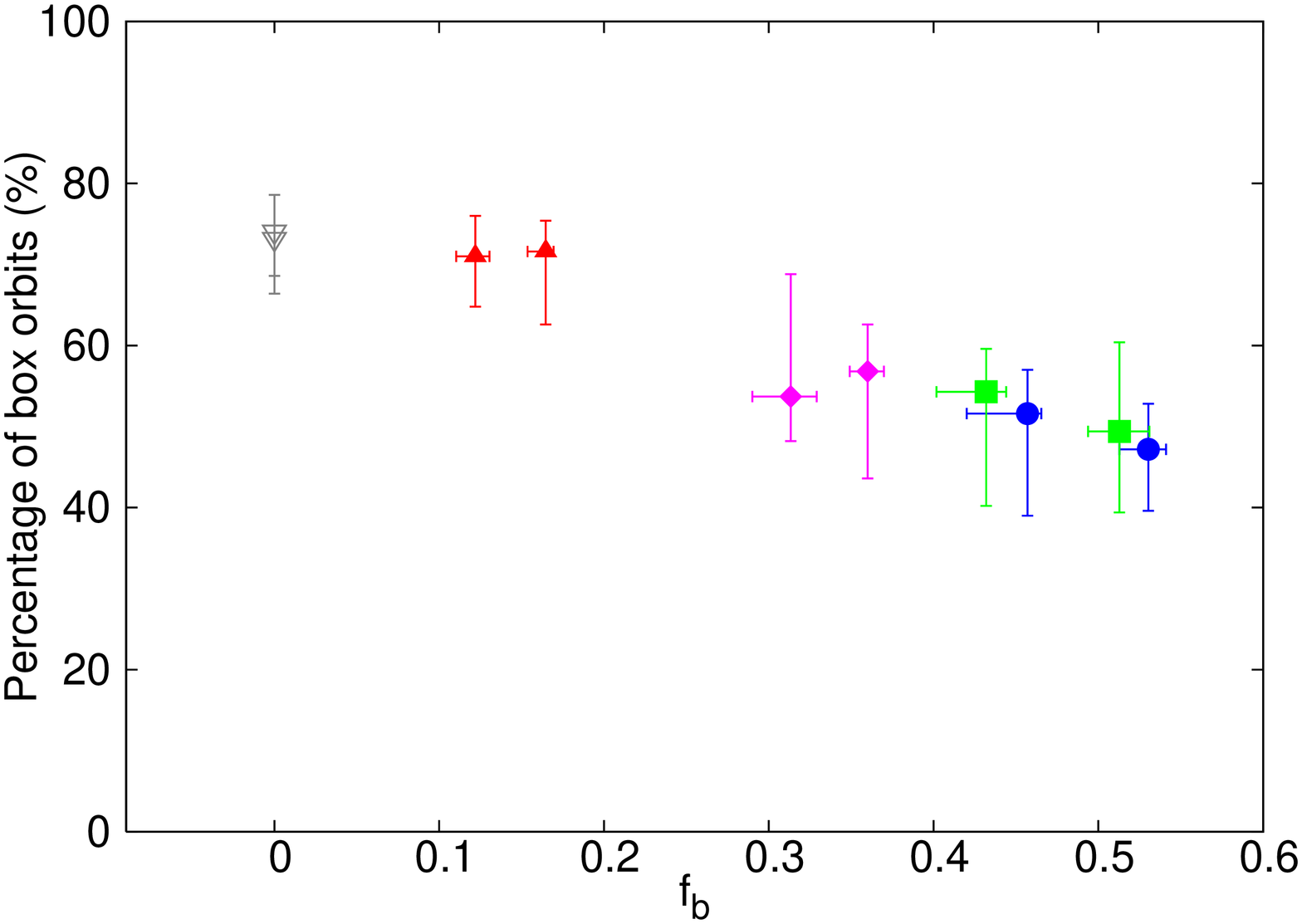}\\
\end{tabular} 
\end{center}
\caption[]{\label{hp2} The effect of basic halo properties on the fraction of
  box orbits in the central region (within 0.25$r_{200}$) of group-sized haloes at $z =
  0$.  From top-left to bottom-right: halo mass, concentration, velocity
  anisotropy, spin, sphericity and central baryon fraction ($f_b$ within 0.05
  $r_{200}$) are considered.  Halo properties are computed within
  0.25$r_{200}$. The haloes from each run are divided into two subsets to
  emphasize the effect of a given halo property within a simulation run.  The
  first subset contains the 25 haloes with the highest value of the specified
  property and the second the 25 haloes with the lowest value of the same
  property.  Error bars show the quartile halo-to-halo scatter for the subset of haloes.}
\end{figure*}

In this section the dependence of the orbital content of the haloes extracted
from the cosmological simulations on several key halo parameters is
considered. The effect of the halo mass, concentration, velocity anisotropy,
spin, sphericity and central baryon fraction on the percentage of box orbits
can be seen in Fig. \ref{hp2}.  As the radial dependence of the orbital
content is found to be weak, we focus here on the orbital content averaged
over the inner region of the haloes (within $r < 0.25r_{200}$).

For each of the halo properties considered we divide the halo sample (from each run) into two subsets.  The  first containing the 25 haloes with the
highest value of the specified property and the second the 25 haloes with the
lowest value of the same property.

Fig. \ref{hp2} emphasizes the impact of baryons on the fraction of box
orbits and illustrates their effect on several other halo properties.  From
top-left to bottom-right, the panels show halo mass, concentration, velocity
anisotropy, spin, sphericity and central baryon fraction, respectively.  In
each panel one can clearly see that the fraction of box orbits is inversely
proportional to the galaxy formation efficiency of the simulation.  From the top-right
panel it is clear that strong (weak) feedback runs result in haloes that are
less (more) concentrated than the dark matter only case (as in
\citealt{bib:Duffy10}).   
It is also clear that efficient cooling results in more spherical haloes (bottom-left panel).  

From Fig. \ref{hp2} we can see that the percentage of box orbits is not sensitive
to the halo mass, concentration, anisotropy or spin parameter (although there
is a tentative trend for haloes with high spin parameters to have fewer box
orbits in the weak/no feedback runs) over the range of parameters considered here.  Trends are apparent when we consider halo shape and central baryon fraction.  It is clear from the bottom-left panel of Fig. \ref{hp2} that an increase in sphericity corresponds to a decrease in the percentage of box orbits.  It is also clear from the bottom-right panel that the central baryon fraction has a significant effect on the orbital content; an increase in central baryon fraction corresponds directly to a decrease in the fraction of box orbits.

\subsection{Orbits of stellar particles}

\begin{figure*}
\begin{center}
\begin{tabular}{cc}
\includegraphics[width=7cm,height=7cm,angle=-90,keepaspectratio]{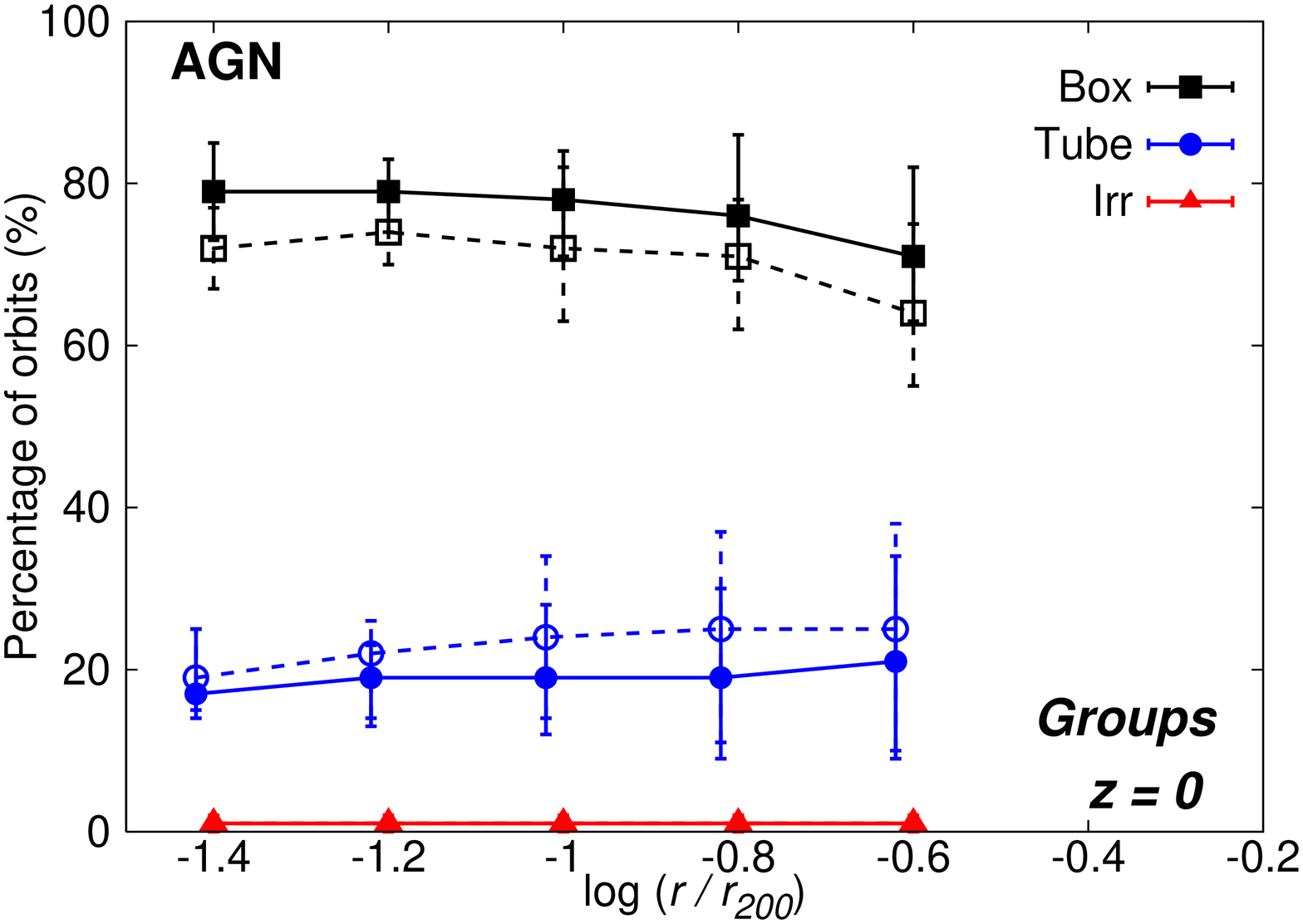} &
\includegraphics[width=7cm,height=7cm,angle=-90,keepaspectratio]{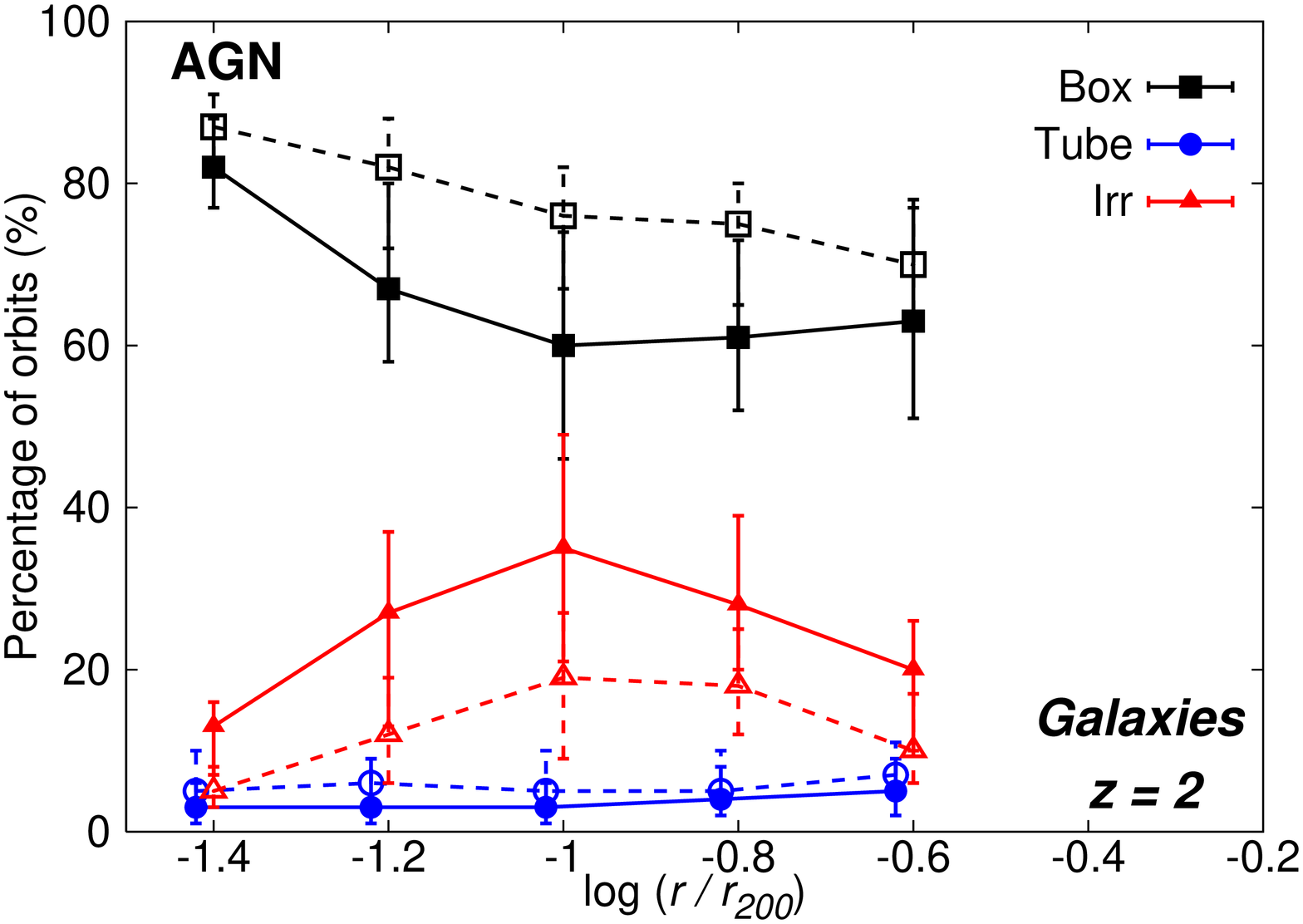} \\

\includegraphics[width=7cm,height=7cm,angle=-90,keepaspectratio]{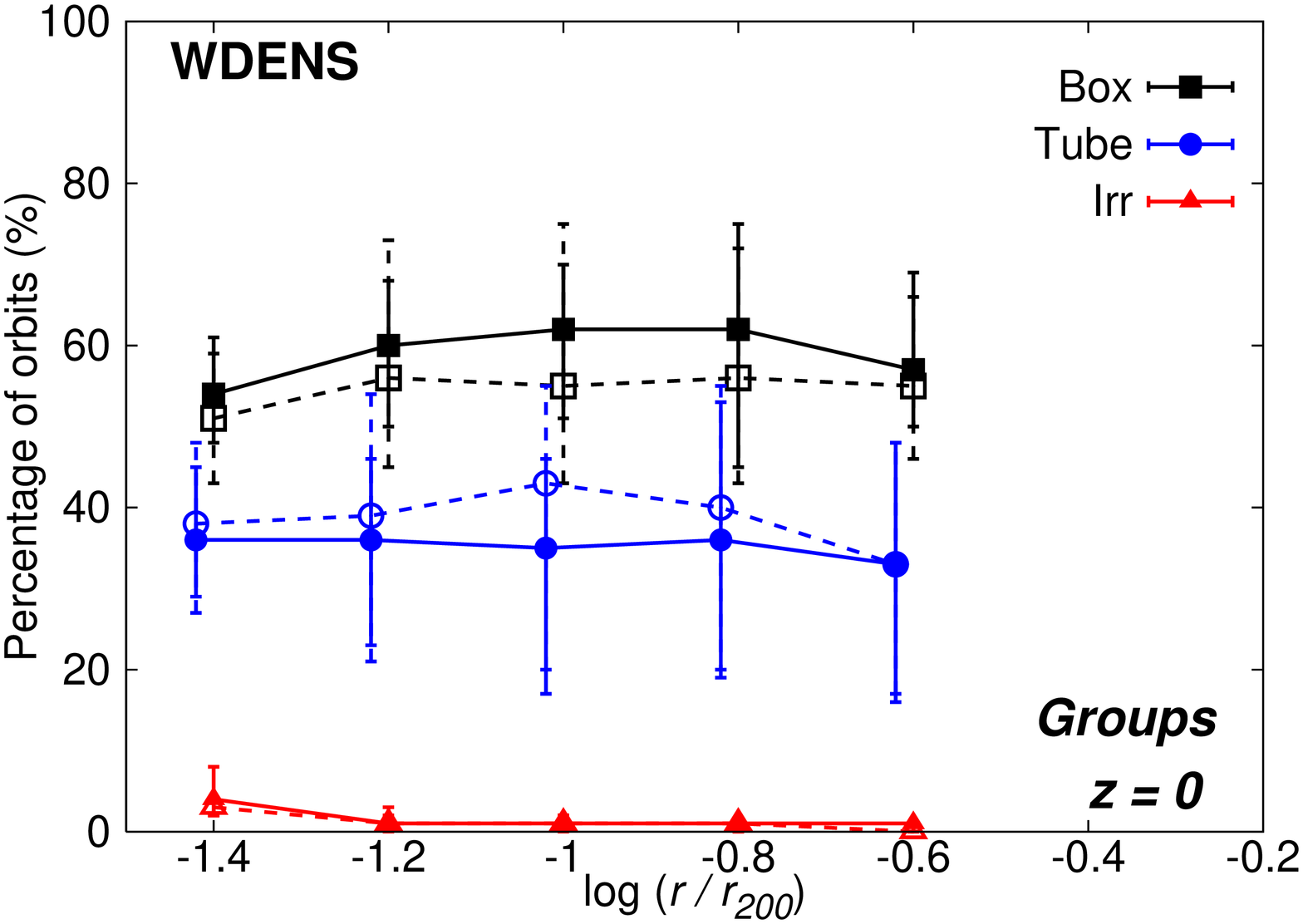} &
\includegraphics[width=7cm,height=7cm,angle=-90,keepaspectratio]{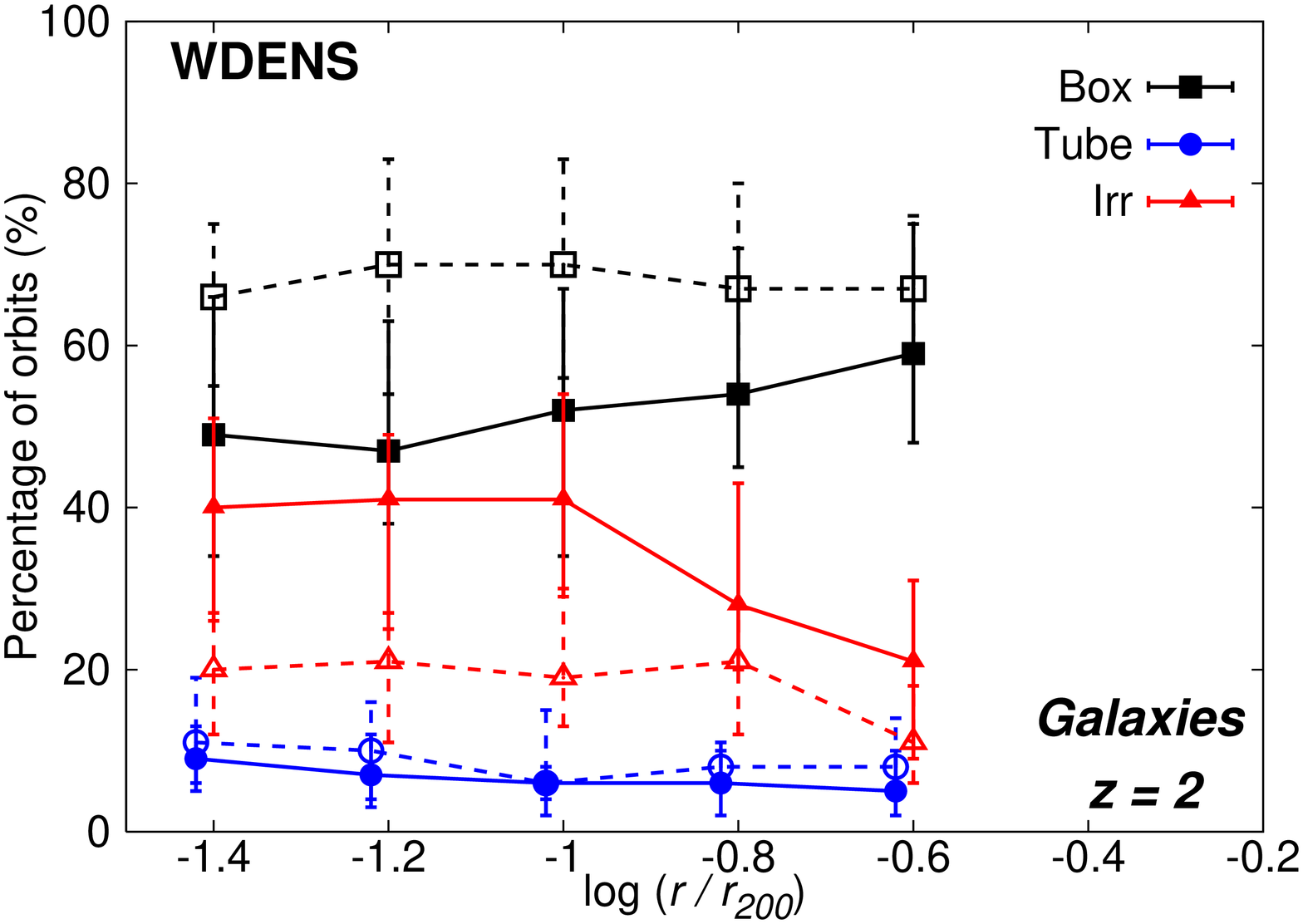} \\

\includegraphics[width=7cm,height=7cm,angle=-90,keepaspectratio]{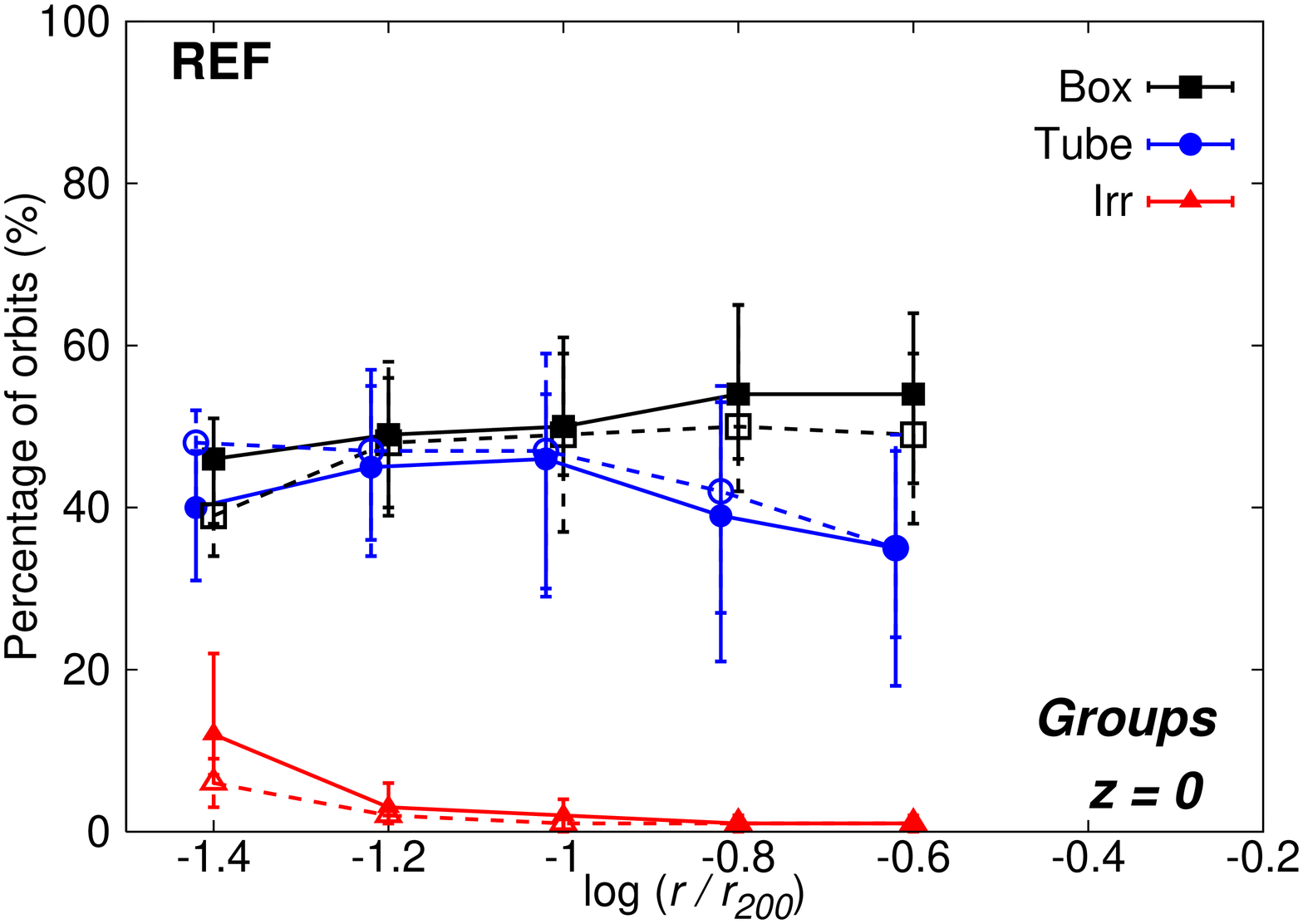} &
\includegraphics[width=7cm,height=7cm,angle=-90,keepaspectratio]{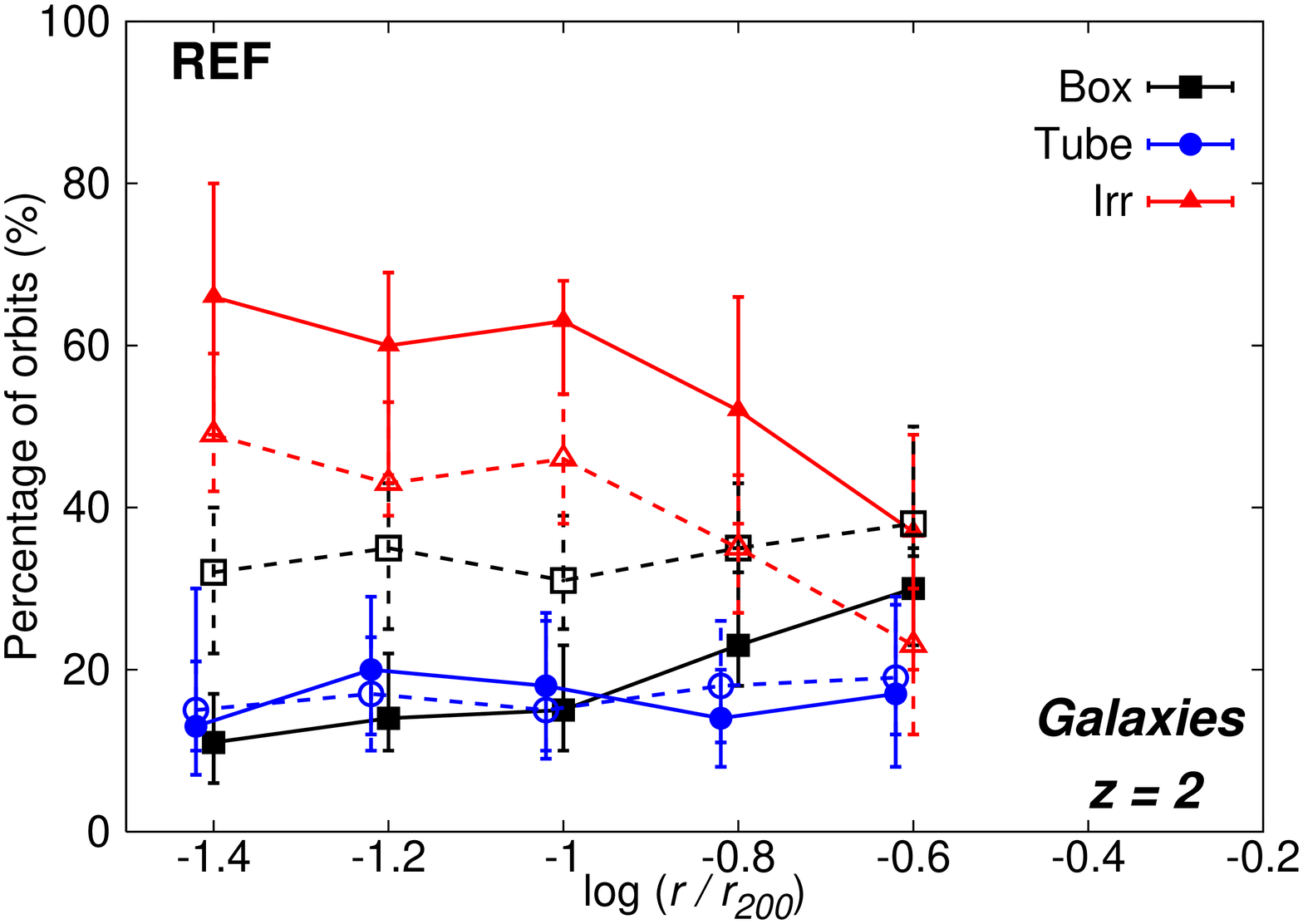} \\

\includegraphics[width=7cm,height=7cm,angle=-90,keepaspectratio]{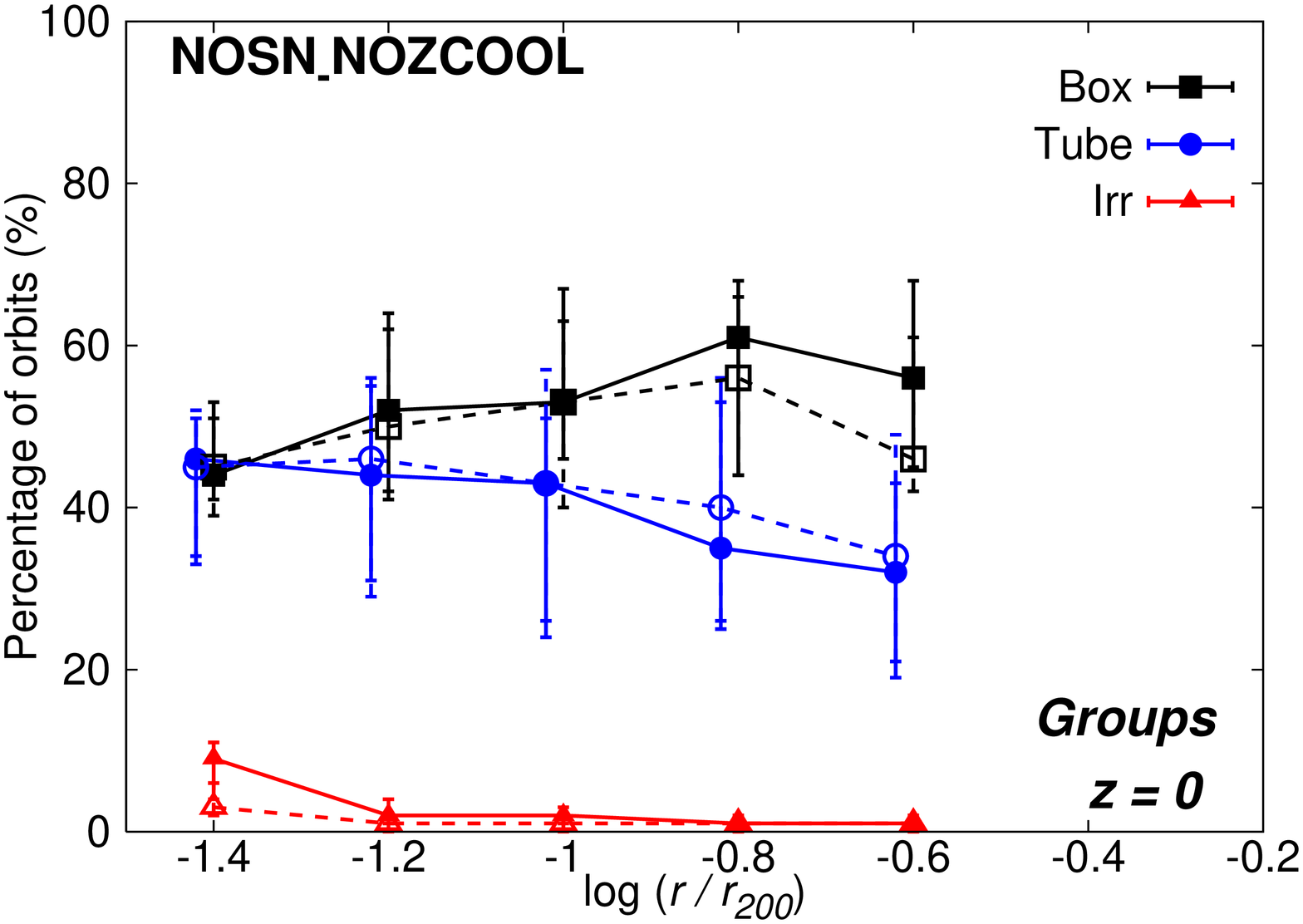} &
\includegraphics[width=7cm,height=7cm,angle=-90,keepaspectratio]{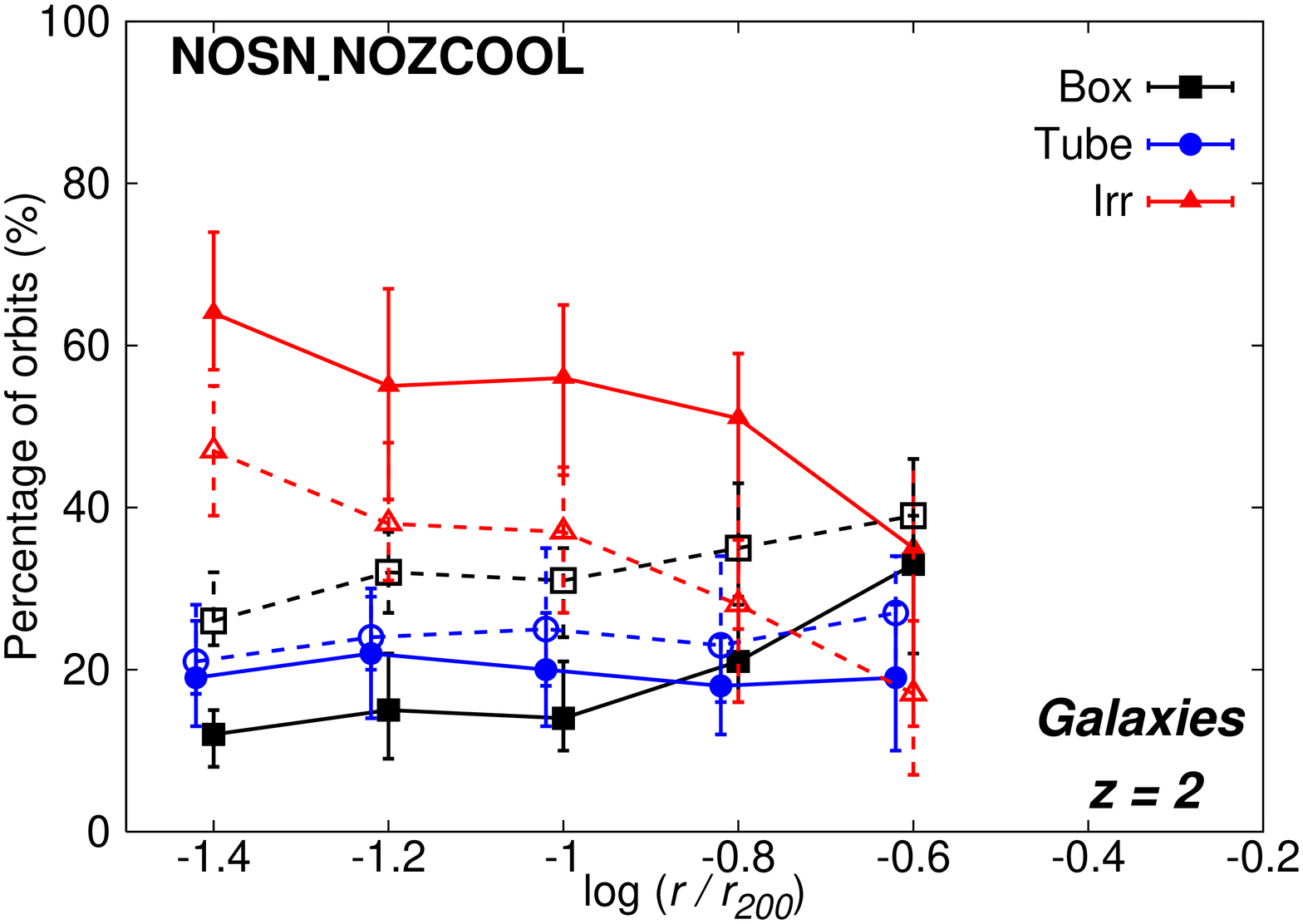} \\

\end{tabular} 
\end{center}
 
\caption[The orbital content of stellar particles at $z = 0$.]{\label{StellarOrbitsz0} 
The median percentage of stellar particles on box (black squares), tube (blue circles) and irregular orbits (red triangles) estimated over the 50 most massive haloes in each simulation (solid lines).  Error bars show the quartile halo-to-halo scatter.  For comparison, in all panels, dashed lines show the percentage of dark matter particles (from the same simulation) on each type of orbit.
The left-hand panels show the orbital content of the haloes at $z = 0$, where the mean dark matter halo mass is $6 \times
  10^{13}$\,$h^{-1}$M$_\odot$.  The
orbital content at $z = 2$ is shown in the right-hand panels (mean dark matter halo mass of $7 \times 10^{11}$\,$h^{-1}$M$_\odot$). 
}
\end{figure*}

Next we compare the orbits of dark matter particles to those of the stellar particles.  We note, however, that since the potential and initial conditions are drawn from the main {\sc subfind} halo, distinct subhaloes and satellites are not included and the stellar population considered here is associated with the central galaxy and the diffuse intra-halo component.  

We show the orbital classifications of stellar particles in Fig.
\ref{StellarOrbitsz0}, presented in the same way as in Fig. \ref{dmorbits}.
Symbols show the median fraction of orbits of a particular type averaged over
all haloes and error bars represent the quartile halo-to-halo scatter. The
dashed curves show the orbits of the dark matter particles taken from the same
simulation, for comparison. In the left (right) column we show the orbital
content of the $z = 0$ (2) haloes.  As in Fig. \ref{dmorbits}, the different
feedback implementations are compared, with the galaxy formation efficiency of the simulations increasing from top to bottom. 

The orbital content described by the stellar particles at $z = 0$ is
remarkably similar to that drawn from the orbits of dark matter particles.
While the dark matter and stellar particles are selected from the same radius,
one might expect a different trend due to the formation history of the stellar
particles.  A full analysis of the history of the stellar particles (such as
when they were stripped from parent subhaloes) could prove insightful.  The
results presented here seem to indicate that they were either stripped a long time ago and have forgotten their dynamical history, or subhaloes bringing in stars are not biased significantly with respect to the main distribution - the velocity bias is weak (e.g. \citealt{bib:Springel01}). %

At $z = 2$ the same trends seen in the orbital content of the dark matter
particles are visible in the orbital content of the stellar particles.
However, the fraction of stellar particles on irregular orbits is
significantly enhanced compared to that for dark matter particles, coming at the expense of the box orbits.  The fraction of dark matter and stellar particles on tube orbits is remarkably similar.

\begin{figure}
\begin{center}
\begin{tabular}{c}

\includegraphics[width=7cm,height=7cm,angle=-90,keepaspectratio]{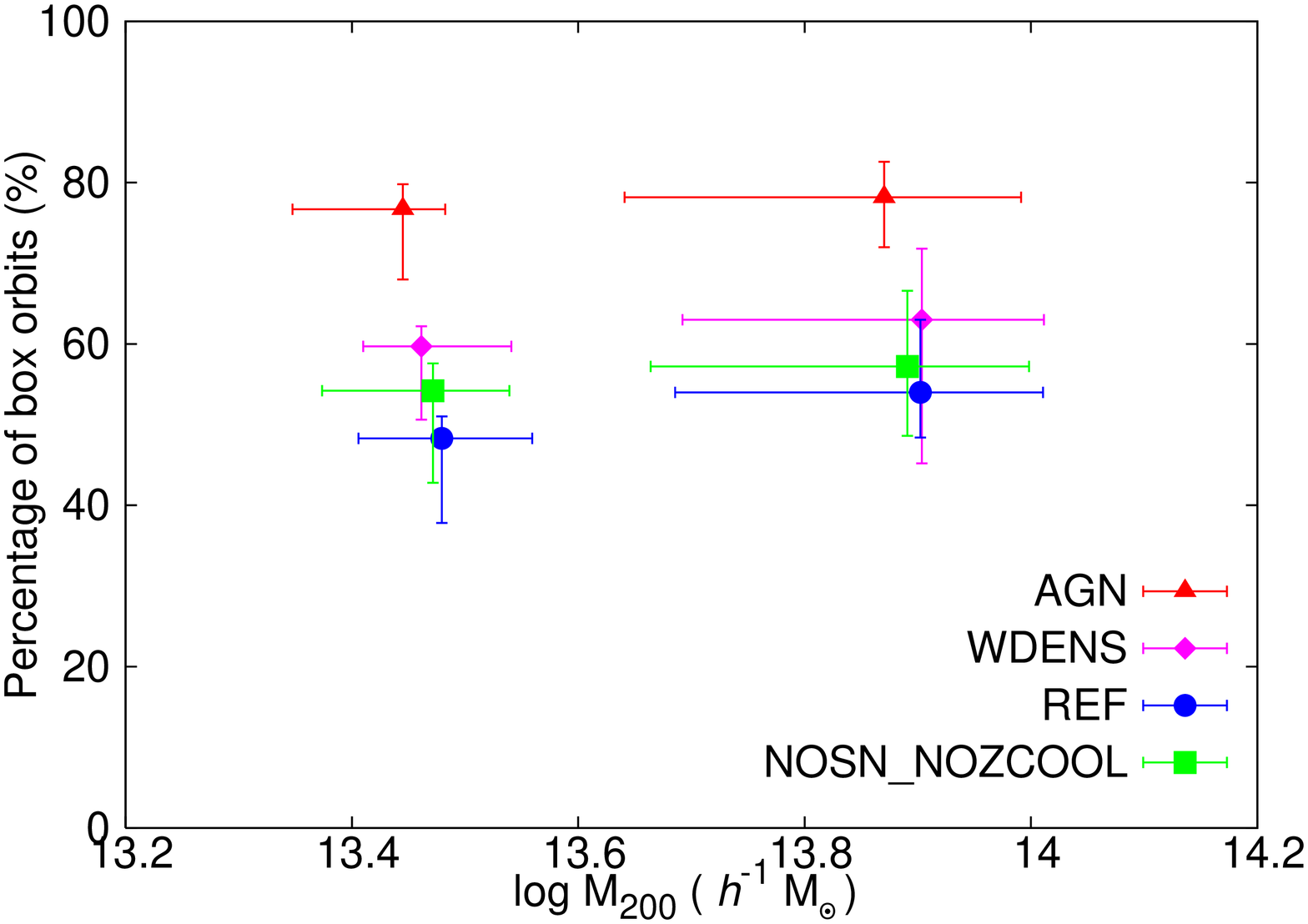} \\ 
\includegraphics[width=7cm,height=7cm,angle=-90,keepaspectratio]{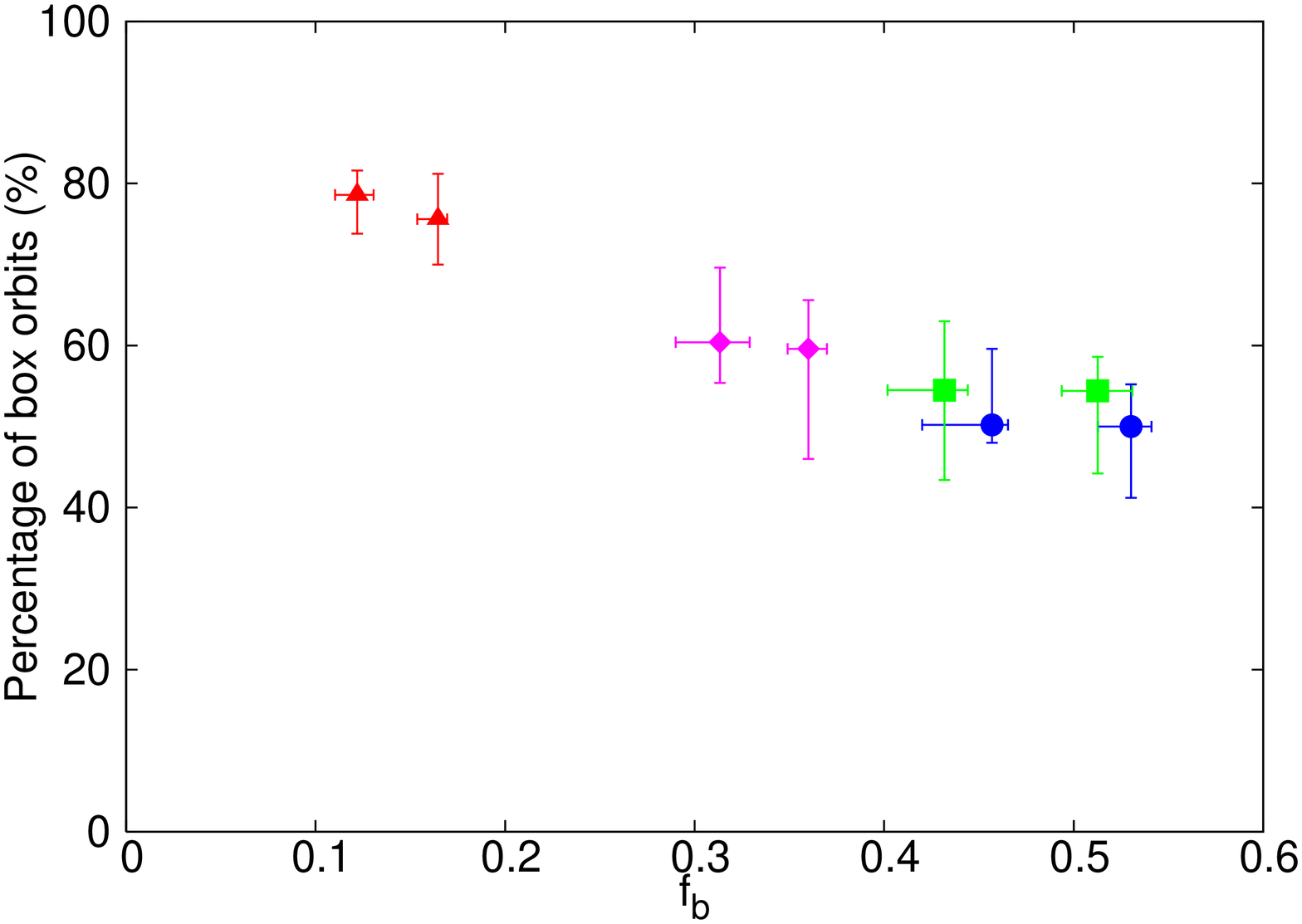}

\end{tabular} 
\end{center}
\caption[Stellar particle box orbits as a function of $f_b(r/r_{200} < 0.05)$.]{\label{starsbf} The fraction of stellar particles on box orbits for the different simulation runs at $z = 0$.  Top: fraction of box orbits as a function of the halo mass $M_{200}$. Bottom:  fraction of box orbits as a function of the central baryon fraction (within 0.05 $r_{200}$).  Error bars show the quartile halo-to-halo scatter.}

\end{figure}

As a final comparison between the orbits of dark matter and stellar particles
we show, in Fig.  \ref{starsbf}, the dependence of the fraction of stellar
particles on box orbits (within $0.25r_{200}$) on the halo mass and central
baryon fraction $f_b$ (within 0.05 $r_{200}$).  This is directly comparable to
the top-left and bottom-right panels of Fig. \ref{hp2}.  Once again we see a
clear indication of the impact that baryons have on reducing the fraction of box orbits.

\subsection{Orbits of subhaloes}

Finally, we consider the orbits of the subhaloes associated with each {\sc
  subfind} main halo at $z = 0$. We consider all subhaloes with masses greater 
than $10^{10}$\,$h^{-1}$M$_\odot$, there are not enough objects to restrict
the sample to the central ($r<0.25r_{200}$) region.  The subhaloes chosen in this way trace a region much further out than discussed previously (mean $r/r_{200} = 0.6$ compared to a mean value of $r/r_{200} = 0.12$ for the dark matter and stellar particles) and are likely to probe a region less strongly affected by the presence of baryons.  Subhaloes on box orbits are also likely to be strongly affected by tidal disruption.

The initial position of each subhalo was
taken to be an average over the 10 most bound particles, and the velocity of the subhalo is assumed to be that of the most bound particle.  The orbits of these subhaloes were integrated, within the gravitational potential of the main halo, for 1000 Gyr.  After this $\sim$$85 - 90$ per cent of subhaloes from the baryon runs are classified;  however, only $\sim$70 per cent of the dark matter subhaloes had undergone more than 40 orbits. 

\begin{figure}
\begin{center}
\begin{tabular}{cc }

\includegraphics[width=7cm,height=7cm,angle=-90,keepaspectratio]{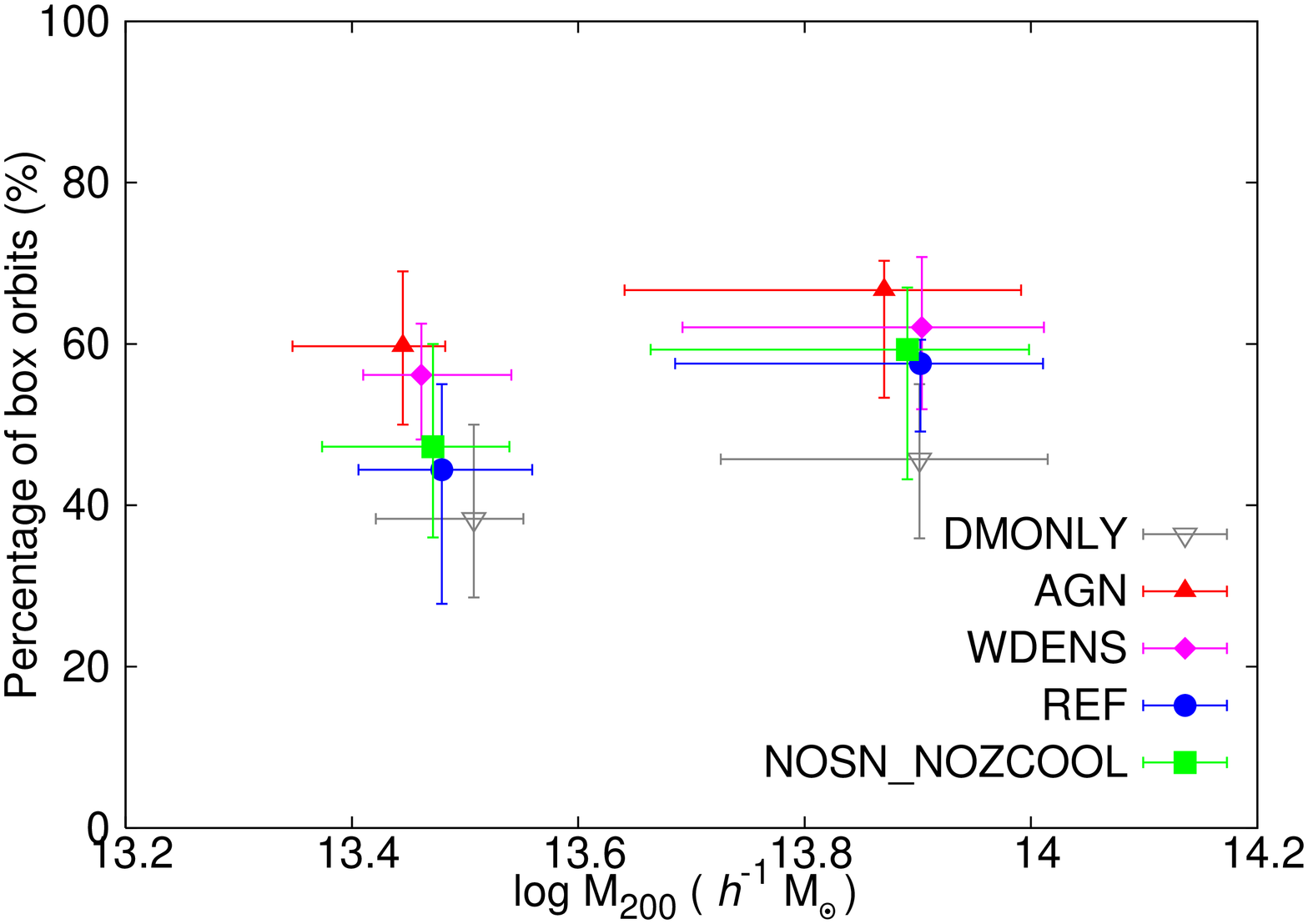} \\
\includegraphics[width=7cm,height=7cm,angle=-90,keepaspectratio]{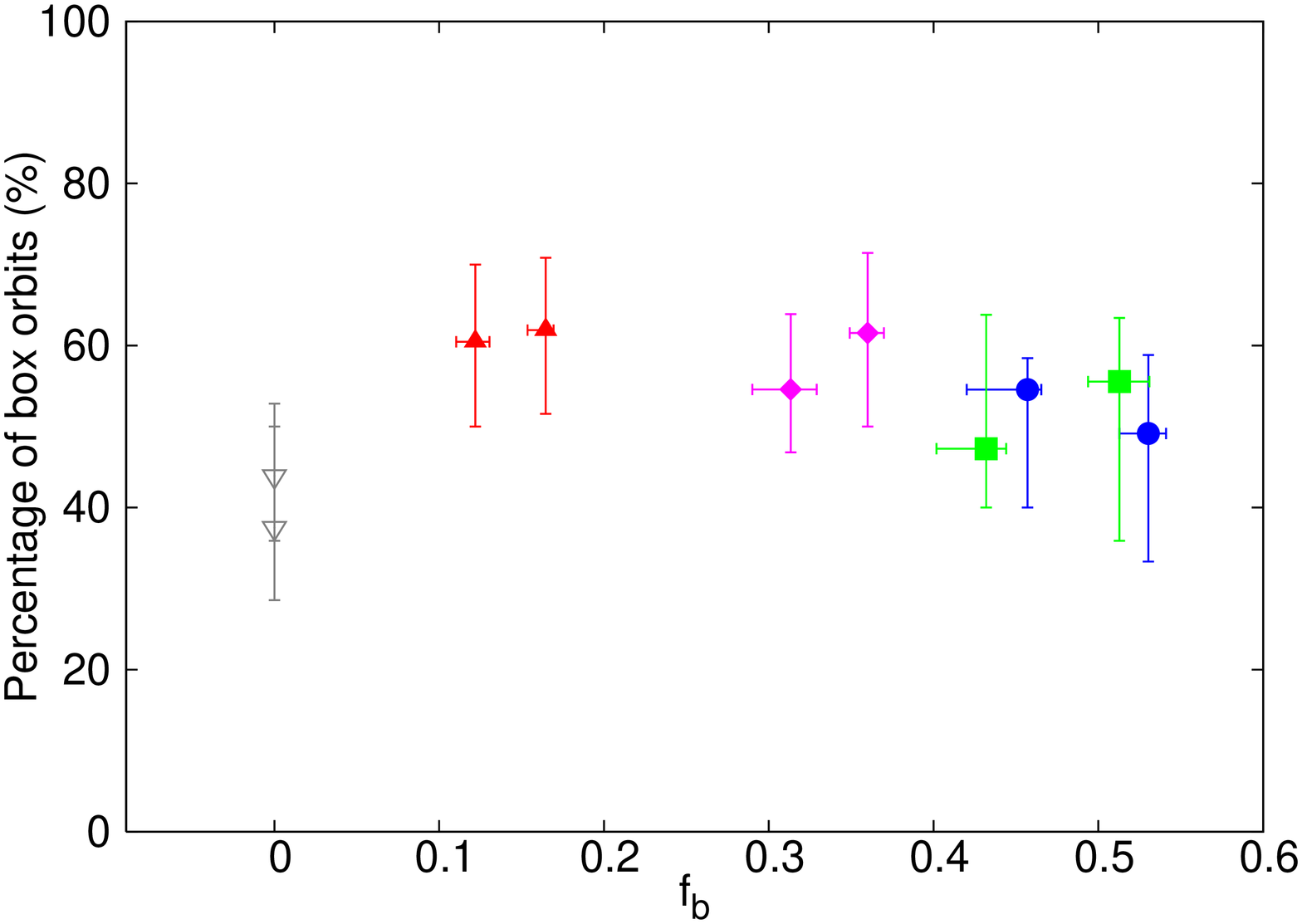} &

\end{tabular} 
\end{center}
\caption[Subhaloes on box orbits as a function of $M_{200}$.]{\label{SUBBox}
  The fraction of subhaloes on box orbits for the different simulation runs at
  $z = 0$.  Top: fraction of box orbits as a function of the halo mass
  $M_{200}$. Bottom:  fraction of box orbits as a function of the central
  baryon fraction (within 0.05 $r_{200}$).  Error bars show the quartile halo-to-halo scatter.}

\end{figure}

The fraction of subhaloes on box orbits as a function of the halo mass
$M_{200}$ and central baryon fraction (within 0.05 $r_{200}$) is shown in
Fig. \ref{SUBBox}.  All of the baryon runs indicate a similar fraction of
box orbits; this fraction is higher than that seen in the dark matter only
simulation only because of the fraction of subhaloes that remain unclassified
in this simulation.  If we consider the fraction of classified subhaloes on
box orbits, then the fraction in the dark matter only simulation increases to $\sim$60 per cent, in agreement with that seen in the AGN run.

Comparing the orbital content of the subhaloes to that of the dark matter
particles (Fig. \ref{dmorbits} and Fig. \ref{fborbits}), we find that the orbits
of the subhaloes are in broad agreement with those seen in the outermost
radial bins of the particle distributions.  Perhaps suggesting that subhaloes
bringing stars into the main galaxy are not biased significantly with respect to the main distribution and explaining why
the orbits of the diffuse intra-halo stellar component are so similar to those
of the dark matter particles.

\section{Summary and Discussion}
\label{orbitsconclusions}

The orbital content of a large sample of haloes extracted from
state-of-the-art high-resolution cosmological hydrodynamical simulations from
the OWLS project is explored in order to identify potential signatures of the
formation process.  We focus on the central regions
(0.25$r_{200}$) of haloes with virial masses $\sim \, 6 \times
10^{13} ( \sim \, 7\times 10^{11})$\,$h^{-1}$M$_\odot$ at $z = 0$ $(2)$ and study
how the orbital content of these haloes is affected in the presence of baryons.

Haloes in dark matter only simulations are dominated by box orbits out to 0.25$r_{200}$.  This is not surprising as box orbits are known to be required to support the triaxial haloes characteristic of dark matter only haloes and are thought to dominate in systems that have undergone major mergers.  The fraction of box orbits is found to decrease with increasing distance from the halo centre; this is mirrored by an increase in the fraction of tube orbits. While at $z = 0$ very few of the orbits were classified as irregular, at $z = 2$ irregular orbits are more common. 

While spherical haloes tend to have fewer box orbits, the orbital content of
the haloes does not appear to be strongly dependent on halo properties such as
mass, concentration, velocity anisotropy, spin and dynamical state for the
range of parameters considered here.  It is, however, strongly dependent on the central baryon fraction. 

By comparing simulations run with no feedback, with stellar feedback and with feedback from AGN, the fraction of box orbits in the central region is found to decrease when baryonic physics is included.  Baryons are able to cool and condense to the centre of the halo, and this central concentration tends to transform box orbits into tube orbits.  Increasing the strength of the feedback implementation is found to reduce the central concentration of baryons and increase the fraction of box orbits.  The orbital content of the strongest feedback run (AGN) is very similar to that seen in the dark matter only case.
  
We then compared the orbital content described by the dark matter particles to
that of the stellar distribution and the subhaloes.  The orbital content
described by the stellar particles (within the central galaxy and the diffuse
intra-halo light) is found to be remarkably similar to that drawn from the
orbits of dark matter particles.  Typically $\sim$\,50 -- 60 per cent of the
subhaloes are found to be on box orbits regardless of the baryonic physics
implemented.  This fraction is in broad agreement with that found in the
outermost radial orbits of the dark matter particles.  The subhaloes we have analysed probe a more extended region of the halo where the effects of baryons do not appear to be as significant.

While the results presented here highlight the importance of the baryons on
the orbital content of haloes, we are limited by the resolution of the
simulations. Ideally this analysis would be extended to a study of the
innermost regions of galaxies.  The stellar half-mass radius of the haloes
considered here is well below the innermost radial bin that we are able to
consider.  By studying high-resolution haloes that have been resimulated from
cosmological conditions, we would be better placed to make direct comparisons
with observations.  The {\it Gaia} satellite, soon to be launched, will
provide us with a kinematic census of our Galaxy and place strong constraints
on galaxy formation models.  In order to fully exploit such observational
datasets, a comprehensive comparison with simulations is essential.  This has
motivated much work on the topic including a recent paper by
\cite{bib:Valluri11} who find, in agreement with our study, that orbital analysis can provide constraints on the underlying potential and may prove to be a useful probe of the formation history of the system.

Analysis of the sort presented here may also prove useful to modelling
approaches such as Schwarzschild's method (\citealt{bib:Schwarzschild79}),
Made-to-Measure techniques (\citealt{bib:Syer96}; \citealt{bib:deLorenzi07};
\citealt{bib:Dehnen09}; \citealt{bib:Long10}) and torus modelling (\citealt{bib:McMillan08}).  These methods attempt to
reproduce observables using a superposition of the orbital density
distributions.  While these techniques are, in many cases, able to reproduce a number of observables, the chosen orbital distribution may not be unique.  It is possible that different combinations of orbits with distinctly different shapes may produce the same triaxial density distribution \citep{bib:deZeeuw91}.  Studying the orbital content of simulated merger remnants provides unique insight into the types of orbital distributions expected to be present in galaxies with specific properties, and may help to provide initial conditions, or additional constraints, for these methods.  

With better future data (particularly those from integral field units), a
comparison of orbit classifications from dynamical models and cosmological
simulations may provide insights into baryonic processes in galaxy formation.

\section*{Acknowledgements}
We thank, in particular, Alan Duffy, Dandan Xu and Mareike Haberichter for
useful discussions and invaluable assistance, Lars Hernquist for the generous
provision of the Self Consistent Field potential reconstruction code and
Daniel Carpintero and Luis Aguilar for the generous provision of the Orbital
Classification routine, and Volker Springel for the use of {\sc subfind} and
{\sc gadget}. We would also like to thank the referee, Adrian Jenkins, for constructive comments.   The simulations presented here were run on Stella,
the LOFAR Blue Gene/L system in Groningen, on the Cosmology Machine at the
Institute for Computational Cosmology in Durham (which is part of the DiRAC Facility jointly funded by STFC, the Large Facilities Capital Fund of BIS, and Durham University) as part of the Virgo
Consortium research programme, and on Darwin in Cambridge.
 This work was
sponsored by National Computing Facilities Foundation (NCF) for the use of
supercomputer facilities, with financial support from the Netherlands
Organization for Scientific Research (NWO). This work was supported by an NWO
VIDI grant and by the Marie Curie Initial Training Network CosmoComp
(PITN-GA-2009-238356).  SEB acknowledges the support provided by the EU
Framework 6 Marie Curie Early Stage Training Programme under contract number
MEST-CT-2005-19669 `ESTRELA'.  SM and STK were supported by the Science and Technology Facilities Council (STFC) through grant 
ST/G002592/1.   SM also thanks the Chinese Academy of Sciences for financial support.

\bibliographystyle{mn2e}
\bibliography{ms}

\begin{appendix}

\section{Numerical Issues}
\label{orbitsextras}

In this section the convergence radius $r_{conv}$ and resolution effects are considered in order to show that the results presented in this paper are well converged.   The choices of halo definition and basis sets are also discussed. 
 
\subsection{Convergence radius}
\begin{figure}
\begin{center}
\begin{tabular}{c}

\includegraphics[width=7.cm,height=7cm,angle=-90,keepaspectratio]{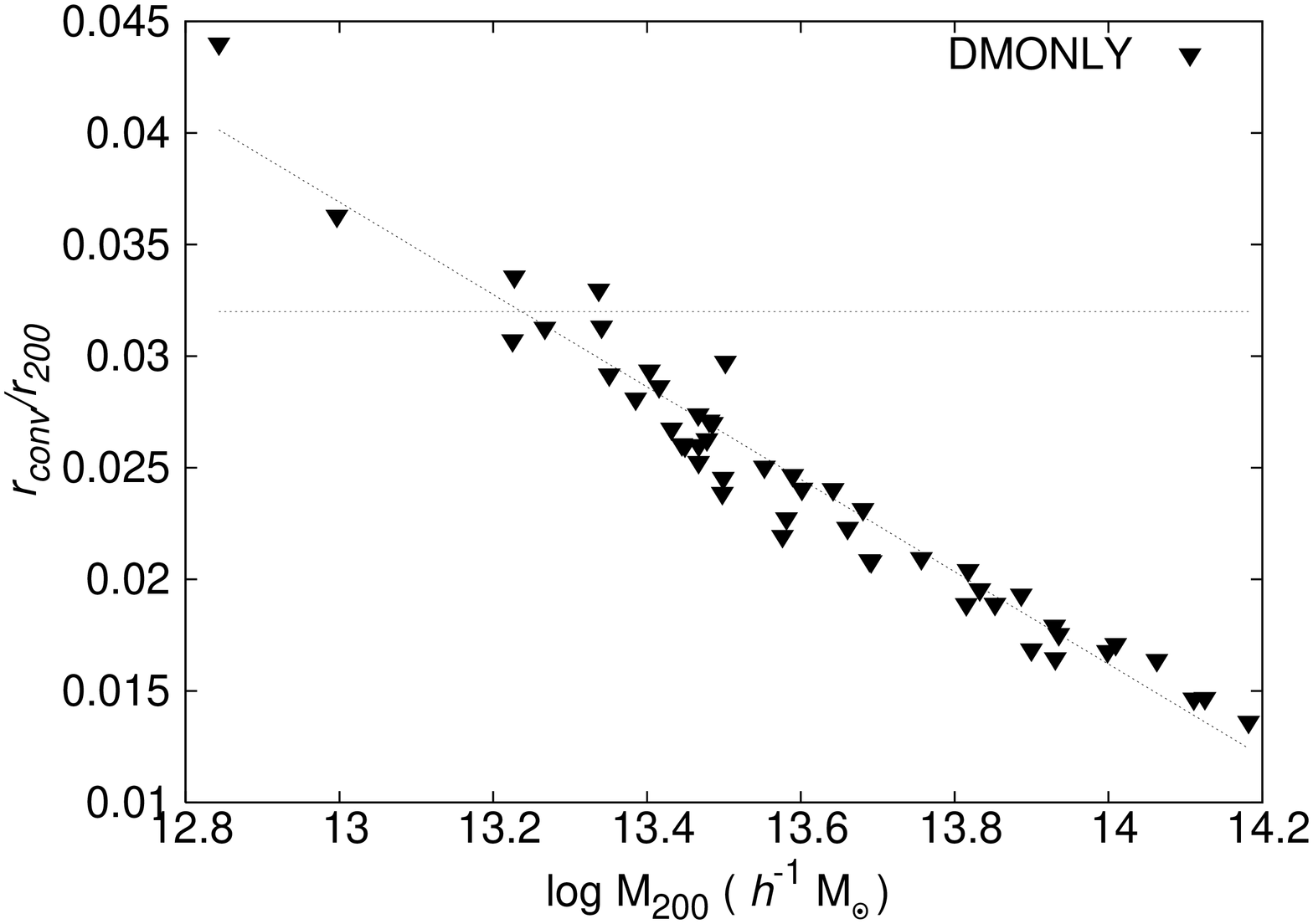} \\
\includegraphics[width=7.cm,height=7cm,angle=-90,keepaspectratio]{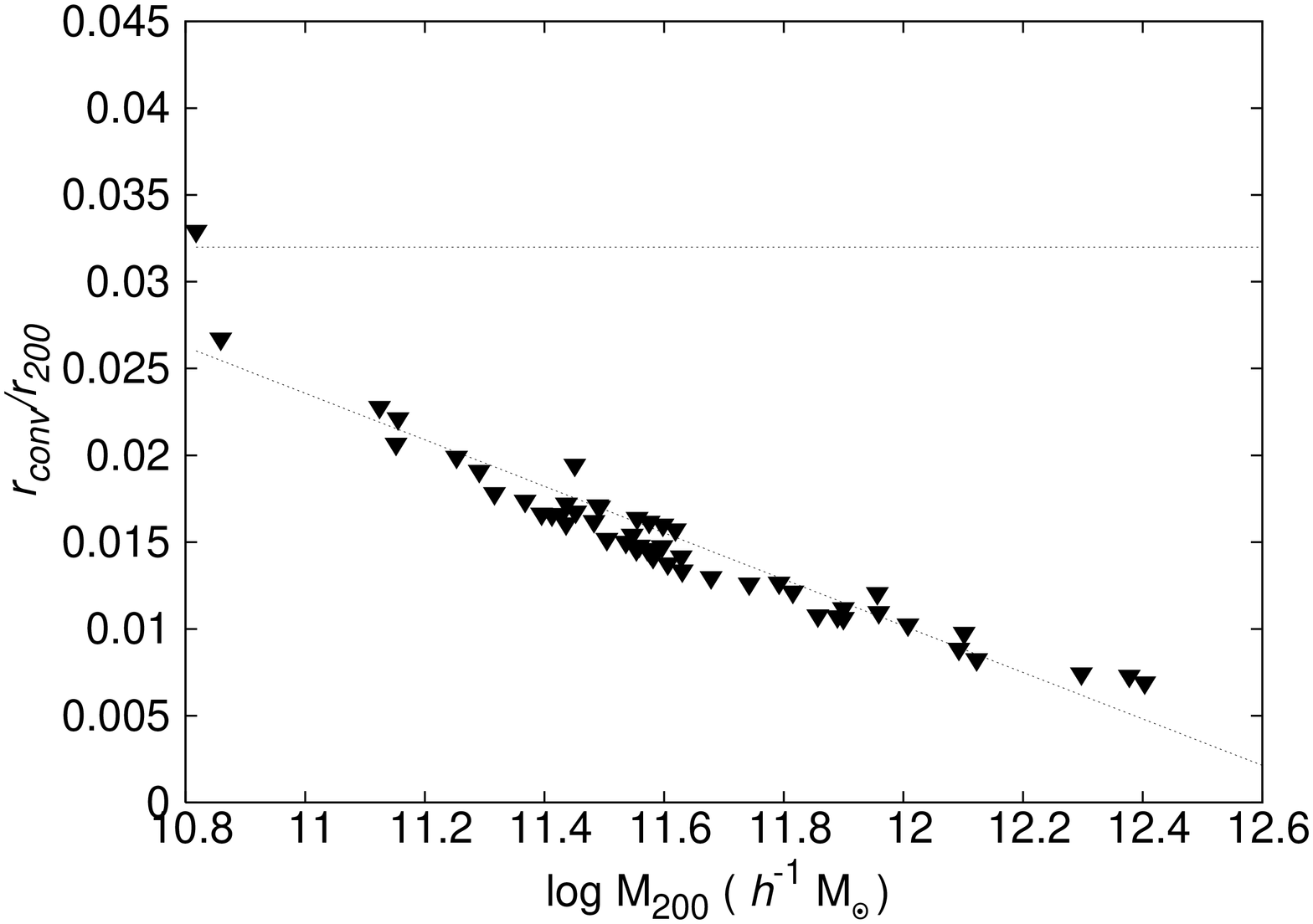}

\end{tabular} 
\end{center}
 
\caption[Convergence radii.]{\label{convr}Convergence radius $r_{conv}$ for the dark matter only simulation at  $z = 0$ (top) and $z = 2$ (bottom).  The horizontal line shows the innermost bin considered in this analysis.  Orbits of particles beyond the convergence radius are studied.}
\end{figure}

\cite{bib:Power03} showed that numerical convergence in the inner regions of
dark matter haloes was achieved outside of the convergence radius, $r_{conv}$,
defined to ensure that the two-body dynamical relaxation time within this
radius is comparable to the age of the Universe.  The convergence radius
depends on halo size and on the resolution of the simulation; it sets a minimum resolved length scale for the analysis.  Fig. \ref{convr} shows an estimate of the convergence radius as a fraction of $M_{200}$ for the haloes used in this analysis.  The horizontal line shows the innermost radial bin we consider; this has been chosen to ensure that our results are converged.

\begin{figure}
\begin{center}
\begin{tabular}{c}

\includegraphics[width=7.cm,height=7cm,angle=-90,keepaspectratio]{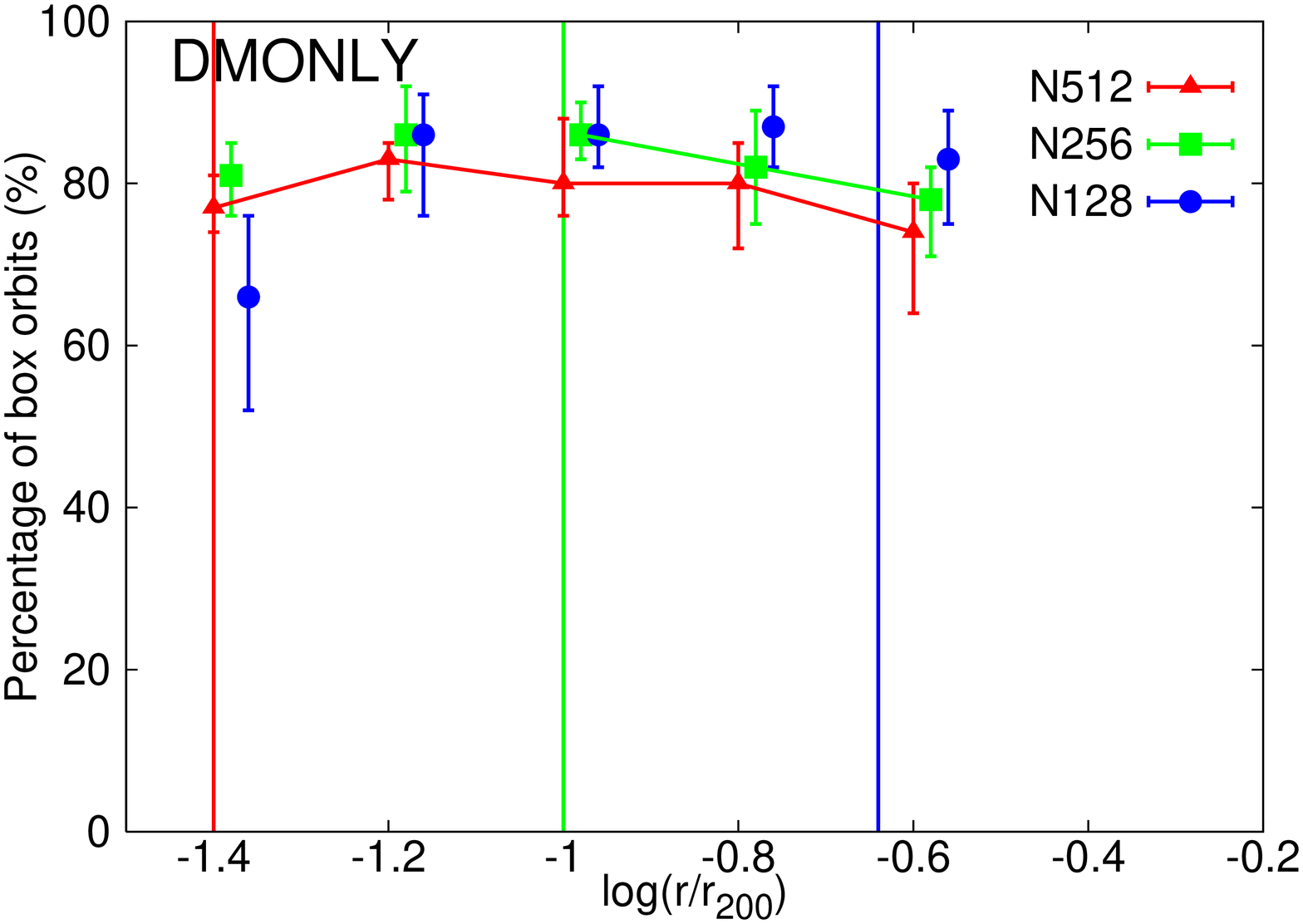} \\
\includegraphics[width=7.cm,height=7cm,angle=-90,keepaspectratio]{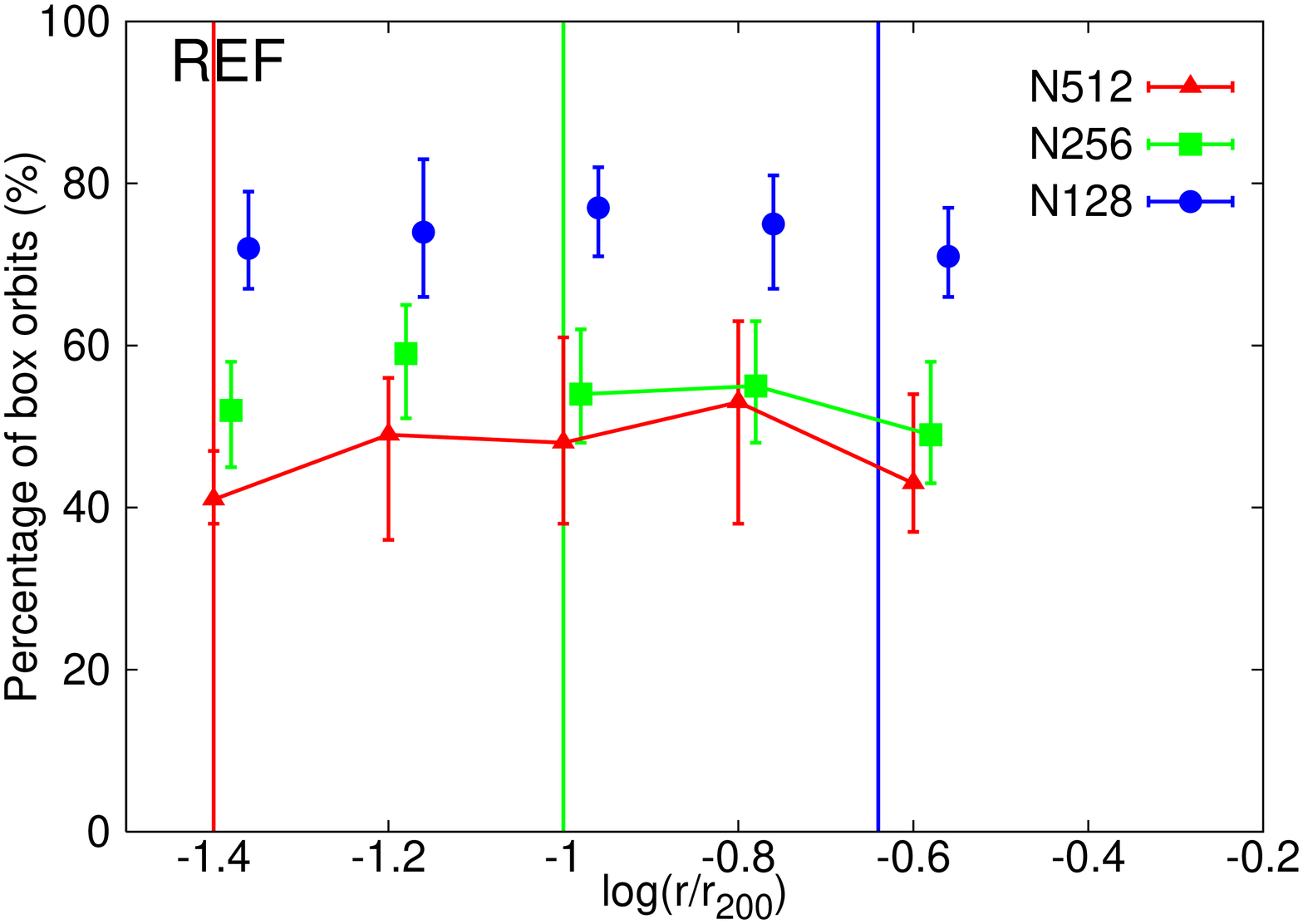}

\end{tabular} 
\end{center}
 
\caption[Effect of resolution on box orbit fraction.]{\label{resltn} The
  effect of resolution on $z = 0$ results.  Results from the $128^3$
  simulation are shown in blue, the $256^3$ simulation in green and the
  $512^3$ simulation used for this analysis in red.  The fraction of box
  orbits found in the dark matter only run is shown in the top panel.  The
  fraction of box orbits for dark matter particles found in the weak stellar feedback
  run (REF) is shown in the bottom panel.  Error bars show the quartile halo-to-halo scatter. The convergence radius for each simulation is shown as a vertical line.}
\end{figure}

\subsection{Resolution effects}
\label{resolution}
In order to quantify the effects of resolution on the orbital content, the
$512^3$-particle run from the DMONLY simulations (with a maximum softening
length of 2 $h^{-1}$ kpc)  is compared with the corresponding lower resolution
runs (containing $256^3$ and $128^3$ particles and with maximum softening
lengths of 4 and 8 $h^{-1}$ kpc, respectively).  In Fig. \ref{resltn} the
fraction of box orbits found in the $128^3$ simulation is shown in blue, the
$256^3$ simulation in green and the $512^3$ simulation used for this analysis
in red.  The fraction of box orbits found in the dark matter only run is shown
in the top plot while the fraction of box orbits found in the weak feedback
run (REF) is shown on the bottom.  Only relaxed haloes that are matched
between the different resolution runs are considered.  Vertical lines show the
convergence radii for each simulation; bins that are considered to be
converged are connected by solid lines to aid in the comparison.

\subsection{Effect of halo definition}
\label{appb}

\begin{figure}
\begin{center}
\begin{tabular}{c}

\includegraphics[width=7.cm,height=7.cm,angle=-90,keepaspectratio]{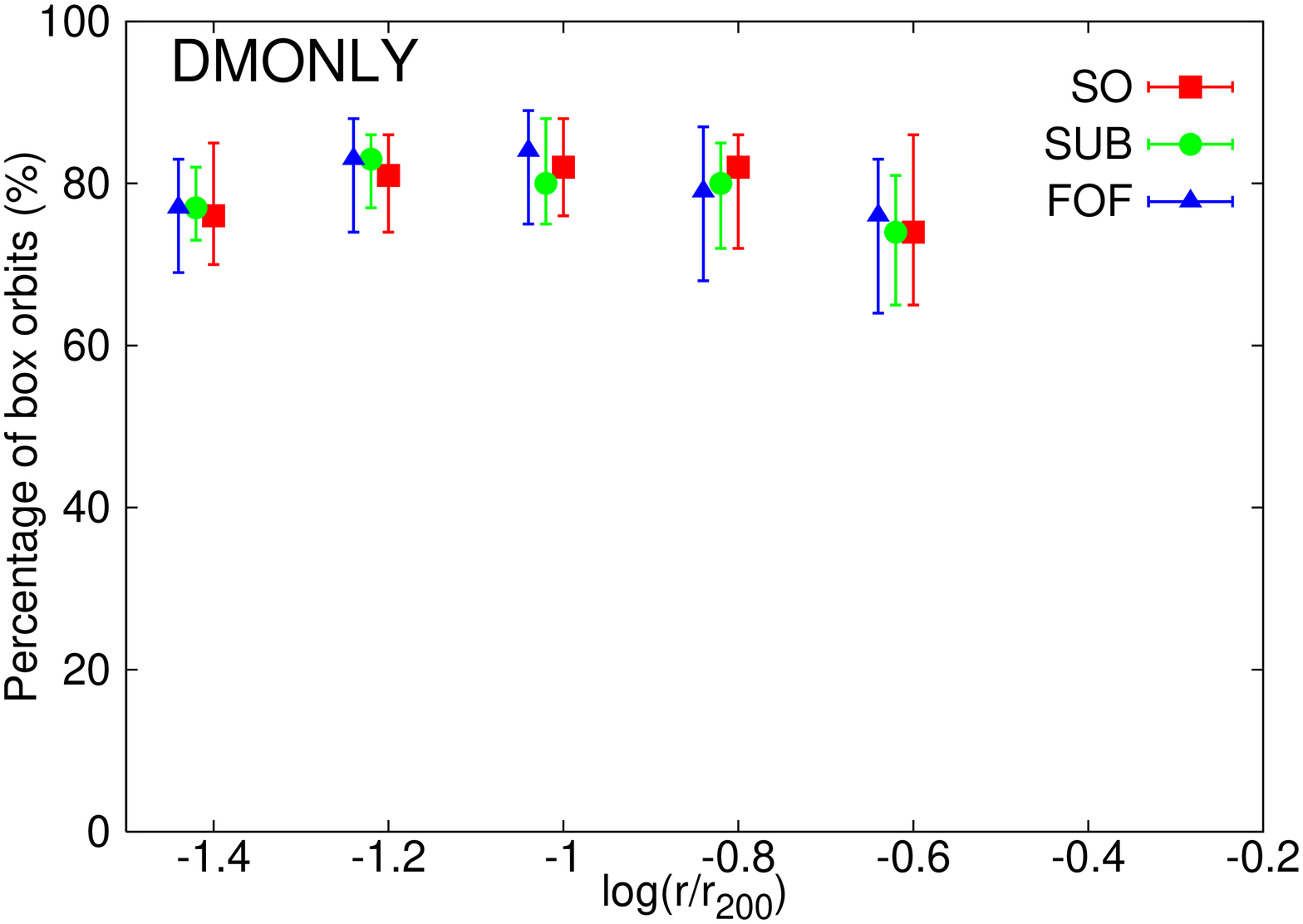}  \\
\includegraphics[width=7.cm,height=7.cm,angle=-90,keepaspectratio]{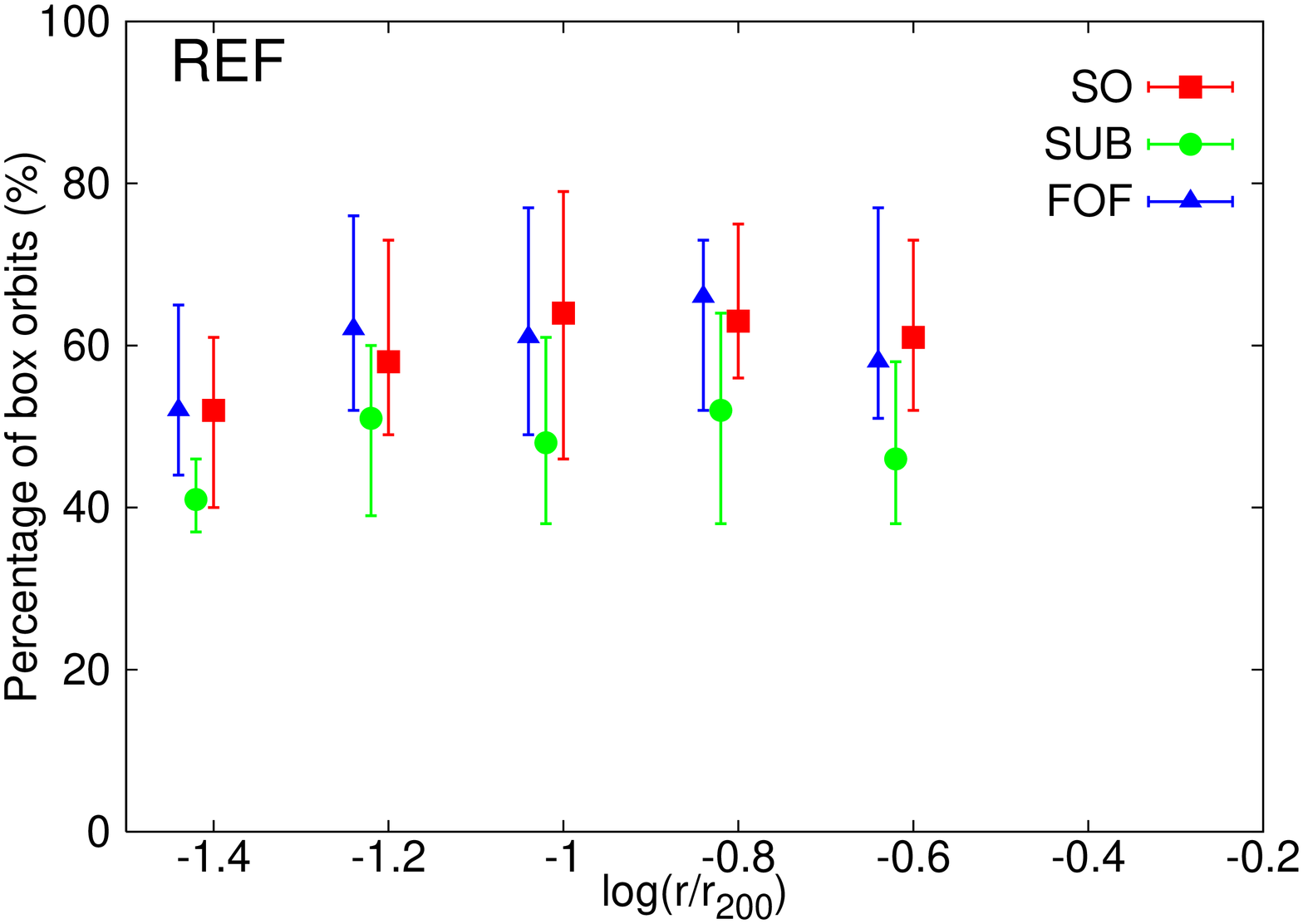}

\end{tabular} 
\end{center}
 
\caption[The effect of halo definition on the orbital content of DM
  haloes.]{\label{halodefn} The effect of halo definition on the orbital
  content of haloes, using the $512^3$-particle simulations at $z = 0$.  The
  top plot shows the fraction of box orbits in the dark matter only
  simulations, while the bottom plot shows the fraction of box orbits in the
  weak feedback run (REF).  Results obtained if haloes are defined using the
  FOF algorithm are shown in red, those using the main {\sc subfind}
  halo are shown in green and those obtained using SO are shown in blue.  Error bars show the quartile halo-to-halo scatter.  The orbital content of a halo is not particularly sensitive to the halo definition. }
\end{figure}

To explore the effect of the halo definition on the results presented here,
the three common definitions of a group -- FOF, main subhalo as
identified by {\sc subfind} and the SO approach -- were used.  The orbital
content of the haloes is not significantly affected by the halo definition, as
shown in Fig. \ref{halodefn}.  In the dark matter only simulations the
fraction of box orbits does not depend on the choice of groupfinder.  In the
weak feedback run (REF) the {\sc subfind} main haloes show a slightly lower
fraction of box orbits due to the effect that removing subhaloes has on the
halo potential.  This difference is not sufficient to account for the trends
discussed in this paper.   The main subhalo is therefore used throughout this
analysis.  This has the advantage of providing a smooth potential, unperturbed by substructure. 

\subsection{Choice of basis set and expansion coefficients}
\begin{figure}
\begin{center}
\begin{tabular}{c}

\includegraphics[width=7.cm,height=7.cm,angle=-90,keepaspectratio]{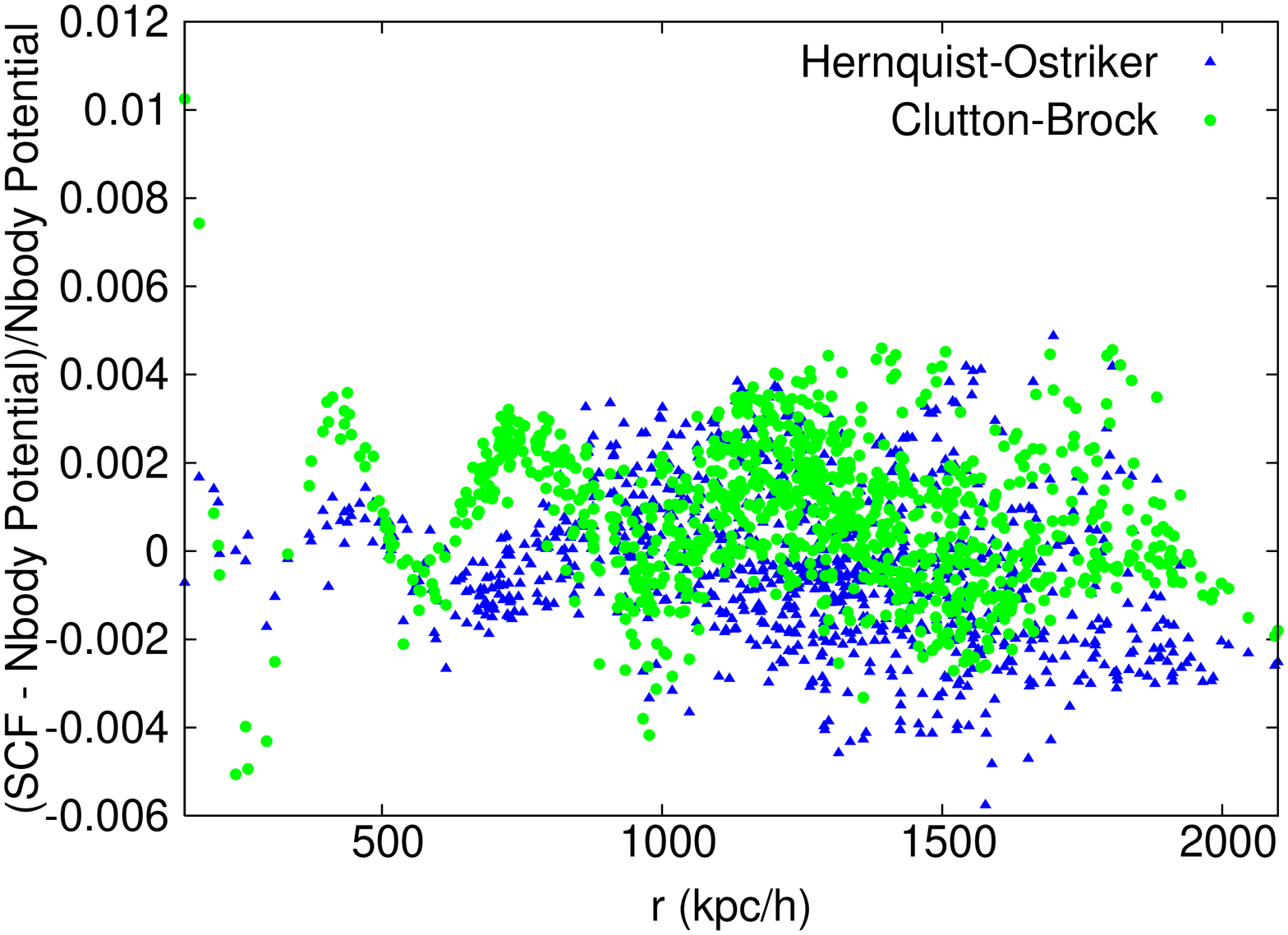} \\

\end{tabular} 
\end{center}

\caption[Comparison between SCF and $N$-body potential.]{\label{cbcomp}
  Difference in potential as computed by the SCF basis functions (using $n=12$, $l=6$) and the Direct-Summation approach for the most massive weak feedback (REF) halo.  For the green points the potential has been computed using the basis set of Hernquist-Ostriker, while for blue triangles the potential is calculated using the Clutton-Brock basis set.}
\end{figure}

The basis set used in this analysis is constructed so that the lowest order
terms represent the Hernquist profile (\citealt{bib:Hernquist90}).  Twelve
radial terms and six angular terms are used as this has been found to be
sufficient to reproduce the potential to within a few percent of the $N$-body
potential (illustrated in Fig \ref{cbcomp}).  

Table \ref{table:basis} shows the orbital content of 500 dark matter particles taken from the inner region ($0.25r_{200}$) of the most massive relaxed cluster from the weak feedback
simulation at $z = 0$.  The orbital content as determined using the
Hernquist basis set for a number of different expansion coefficients is
shown in the top panel.  The orbital content derived using the
Clutton-Brock basis sets is shown below.   Varying the number of expansion
coefficients affects the orbital classifications at the level of a few
percent; a similar orbital content is found using both basis functions.

The choice of basis set is not
found to affect the reconstruction of the potential significantly over the
radial range we consider here.  Both basis sets reproduce the $N$-body potential with percent level accuracy
over this region (see Fig. \ref{cbcomp}).

\begin{table*}
\caption[A comparison of the SCF basis sets.]{Classifications of the orbits of
  500 dark matter particles taken from the inner region ($0.25r_{200}$) of the
  most massive relaxed cluster from the weak feedback simulation at $z = 0$.  Different numbers of expansion coefficients have been used to reconstruct the potential using the Hernquist basis set (top) and the Clutton-Brock basis set (bottom).} 
\centering %
\begin{tabular}{ c c c c c c c c} 
\hline 
(n,l) & Box  & Tube  & Irr    & resonant box & $x$-tube & $z$-tube  & Not classified \\
\hline

20,6 & 0.510 & 0.386 & 0.014        & 0.194        & 0.056  & 0.326 & 0.090  \\
20,4 & 0.528 & 0.376 & 0.006        & 0.194        & 0.034  & 0.340 & 0.090  \\
12,6 & 0.582 & 0.314 & 0.016        & 0.188        & 0.044  & 0.268 & 0.088    \\
12,4 & 0.532 & 0.370 & 0.009        & 0.162        & 0.042  & 0.328 & 0.088    \\
8,6  & 0.564 & 0.318 & 0.024        & 0.188        & 0.054  & 0.258 & 0.009  \\
8,4  & 0.484 & 0.400 & 0.006        & 0.144        & 0.030  & 0.366 & 0.110  \\

\hline 
20,6 & 0.582 & 0.338 & 0.008 & 0.238 & 0.038 & 0.298  & 0.072  \\
20,4 & 0.490 & 0.424 & 0.008 & 0.178 & 0.032 & 0.392  & 0.078  \\
12,6 & 0.458 & 0.428 & 0.012 & 0.182 & 0.029 & 0.398  & 0.102  \\
12,4 & 0.430 & 0.476 & 0.008 & 0.144 & 0.034 & 0.438 & 0.086  \\
8,6  & 0.422 & 0.468 & 0.008 & 0.186 & 0.014 & 0.452 & 0.102 \\
8,4 &  0.412 & 0.456 & 0.022 & 0.176 & 0.010 & 0.440 & 0.110 \\%

\hline

\end{tabular}
\label{table:basis} 
\end{table*}

\clearpage
\newpage


\end{appendix}

\label{lastpage}

\end{document}